\documentclass[a4paper,12pt]{article}
\pdfoutput=1
\usepackage{geometry}
\usepackage{amsmath,amssymb,amsfonts}
\usepackage{xcolor,graphicx,cite,soul}
\usepackage{array,booktabs,longtable}
\usepackage{caption,microtype}
\usepackage{subcaption}
\usepackage{hyperref}
\usepackage{datetime}
\usepackage{appendix}
\usepackage{multirow}
\usepackage{cancel}

\makeatletter
\providecommand{\tabularnewline}{\\}

\makeatother

\newcommand{\lsim}
{\;\raisebox{-.3em}{$\stackrel{\displaystyle <}{\sim}$}\;}
\newcommand{\gsim}
{\;\raisebox{-.3em}{$\stackrel{\displaystyle >}{\sim}$}\;}

\newcommand\al{\alpha}
\newcommand\be{\beta}
\newcommand\tb{\tan\beta}

\newcommand\CBA{c_{\beta - \alpha}}
\newcommand\SBA{s_{\beta - \alpha}}

\newcommand\ReDiag{\mathop{%
  \raise .5pt\hbox{[}%
  \widetilde{\mathrm{Re}}%
  \raise .5pt\hbox{]}}}
\newcommand\ReOffDiag{\mathop{%
  \raise .5pt\hbox{$\llbracket$}%
  \widetilde{\mathrm{Re}}%
  \raise .5pt\hbox{$\rrbracket$}}}

\newcommand\Mh{m_h}
\newcommand\MH{m_H}
\newcommand\MA{m_A}
\newcommand\MHp{m_{H^\pm}}

\newcommand\msq{m_{12}^{2}}

\newcommand\refeq[1]{Eq.~(\ref{#1})}
\newcommand\refeqs[1]{Eqs.~(\ref{#1})}
\newcommand\refta[1]{Tab.~\ref{#1}}
\newcommand\refse[1]{Sect.~\ref{#1}}
\newcommand\refses[1]{Sects.~\ref{#1}}
\newcommand\citere[1]{Ref.~\cite{#1}}
\newcommand\citeres[1]{Refs.~\cite{#1}}

\newcommand{\CP}{{\cal CP}}
\newcommand{\cp}{{\CP}}

\newcommand{\tev}{\,\, \mathrm{TeV}}
\newcommand{\gev}{\,\, \mathrm{GeV}}

\newcommand\HB{\texttt{HiggsBounds}}
\newcommand\HS{\texttt{HiggsSignals}}

\newcommand{\br}{\text{BR}}
\newcommand{\De}{\Delta}

\newcommand{\sig}{\sigma}

\def\reffi#1{\mbox{Fig.~\ref{#1}}}
\def\reffis#1{\mbox{Figs.~\ref{#1}}}

\def\Ga{\Gamma}
\def\ga{\gamma}

\def\la{\lambda}
\newcommand\kala{\ensuremath{\kappa_{\lambda}}}
\newcommand\laSM{\ensuremath{\lambda_{\mathrm{SM}}}}
\newcommand{\lahhh}{\ensuremath{\la_{hhh}}}
\newcommand{\lahhH}{\ensuremath{\la_{hhH}}}
\newcommand{\lahHH}{\ensuremath{\la_{hHH}}}
\newcommand{\lahAA}{\ensuremath{\la_{hAA}}}

\newcommand{\lahHpHm}{\ensuremath{\la_{hH^+H^-}}}

\newcommand{\inter}[2]{\ensuremath{[#1, #2]}}

\definecolor{Orange}{named}{orange}
\definecolor{Purple}{named}{purple}
\definecolor{Lightblue}{cmyk}{0.9,0.1,0.1,0.3}
\definecolor{dgelborange}{cmyk}{0.,0.3,0.5, 0.}
\definecolor{Lila}{rgb}{0.5,0.,1}
\definecolor{Darkgreen}{rgb}{0.,.7,0.2}

\graphicspath{{figs/}}
\allowdisplaybreaks
\begin{document}
\thispagestyle{empty}

\def\thefootnote{\fnsymbol{footnote}}

\begin{flushright}
\mbox{}
IFT--UAM/CSIC-22-032 \\
arXiv:2203.12684 [hep-ph]
\end{flushright}

\vspace{0.5cm}

\begin{center}

{\large\sc 
{\bf Triple Higgs Couplings in the 2HDM:\\[.5em]
  The Complete Picture}}\\  

\vspace{1cm}

{\sc
F.~Arco$^{1,2}$%
\footnote{email: Francisco.Arco@uam.es}%
, S.~Heinemeyer$^{2}$%
\footnote{email: Sven.Heinemeyer@cern.ch}%
~and M.J.~Herrero$^{1,2}$%
\footnote{email: Maria.Herrero@uam.es}%
}

\vspace*{.7cm}

{\sl
$^1$Departamento de F\'isica Te\'orica, 
Universidad Aut\'onoma de Madrid, \\ 
Cantoblanco, 28049, Madrid, Spain

\vspace*{0.1cm}

$^2$Instituto de F\'isica Te\'orica (UAM/CSIC), 
Universidad Aut\'onoma de Madrid, \\ 
Cantoblanco, 28049, Madrid, Spain

}

\end{center}

\vspace*{0.1cm}

\begin{abstract}
\noindent
The measurement of the triple Higgs coupling is one of the main tasks of
the (HL-)LHC and future lepton colliders. Similarly, triple Higgs
couplings involving BSM Higgs bosons are of high interest.
Within the framework of Two Higgs Doublet Models (2HDM) we
investigate the allowed ranges for all triple Higgs couplings involving at
least one light, SM-like Higgs boson. We present newly the allowed
ranges for 2HDM type~III and~IV and update the results within the
type~I and~II. 
We take into account theoretical constraints from unitarity and
stability, experimental constraints from 
direct BSM Higgs-boson searches, measurements of the rates of the
SM-like Higgs-boson at the LHC, as well as flavor observables and
electroweak precision data.
For the SM-type triple Higgs coupling w.r.t.\ its SM value,
$\lahhh/\laSM$,  we find allowed intervals of $\sim \inter{-0.5}{1.3}$
in type~I and $\sim \inter{0.5}{1.0}$ in the other Yukawa types.
These allowed ranges have important implications for the experimental
determination of this coupling at future collider experiments.
We find the coupling $\lahhH$ between $\sim -1.5$ and $\sim +1.5$
in the four Yukawa types.
For the triple Higgs couplings involving two heavy neutral
Higgs bosons, $\lahHH$ and 
$\lahAA$ we find values between $\sim -0.5$ and $\sim 16$, and
between $\sim -1$ and $\sim 32$ for $\lahHpHm$.
These potentially large values could lead to strongly enhanced
production of two Higgs-bosons at the HL-LHC or high-energy lepton colliders.

\end{abstract}


\def\thefootnote{\arabic{footnote}}
\setcounter{page}{0}
\setcounter{footnote}{0}

\newpage


\section{Introduction}
\label{sec:intro}

The discovery of a new scalar particle with a mass of $\sim 125 \gev$ by
ATLAS and CMS~\cite{Aad:2012tfa,Chatrchyan:2012xdj,Khachatryan:2016vau} 
-- within theoretical and experimental uncertainties -- is
consistent with the predictions of a Standard-Model~(SM) Higgs boson.
No conclusive evidence of physics beyond the~SM (BSM) has been found so
far at the LHC. 
However, the measurements of the Higgs-boson rates at the LHC, which are
known experimentally to a precision of roughly $\sim 10\%$, leave ample
room for BSM interpretations. Many models of BSM physics possess
extended Higgs-boson sectors. Consequently, one of the main tasks of the
LHC Run~3 and the HL-LHC is to determine whether 
the observed Higgs boson forms part of a Higgs sector of an extended
model.
A key element in the investigation of the Higgs-boson sector is the
measurement of the triple Higgs coupling of the SM-like Higgs boson,
\lahhh. The expected achievable precision at different future
colliders in the measurement of \lahhh\ depends on the value realized in
nature. 
In the case of an extended Higgs-boson sector, equally important are the
measurement of BSM triple Higgs-boson couplings.

One natural extension of the Higgs-boson sector of the SM is the ``Two Higgs
Doublet Model'' (2HDM) (for reviews see,
e.g.,~\citeres{Gunion:1989we,Aoki:2009ha, Branco:2011iw}). This model
possesses five physical Higgs bosons: the 
light and the heavy $\CP$-even $h$ and $H$, the $\CP$-odd $A$, and the
pair of charged Higgs bosons, $H^\pm$.
The ratio of the two vacuum expectation values, $\tb := v_2/v_1$,
defines the angle $\be$ that diagonalizes the $\CP$-odd and the charged
Higgs sector, while the independent angle $\al$ diagonalizes the
$\CP$-even Higgs sector.
For this work we assume that the
light $\CP$-even Higgs-boson $h$ is SM-like with a mass of 
$\Mh \sim 125 \gev$ with all other Higgs bosons assumed to be heavier.
To avoid flavor changing neutral currents (FCNC) at the 
tree-level, a $Z_2$~symmetry is imposed~\cite{Glashow:1976nt},
which is allowed to be softly broken by the parameter $\msq$.
The extension of the $Z_2$~symmetry to the fermion sector defines four
types of the 2HDM: 
type~I and~II, type~III (also called type~Y, or flipped)
and type~IV (also called type~X, or lepton specific)~\cite{Aoki:2009ha}.

In this paper we investigate the allowed ranges for
all triple Higgs couplings involving at 
least one light SM-like Higgs boson in all the four 2HDM types.
Concretely, these triple Higgs couplings are
\lahhh, \lahhH, \lahHH, \lahAA\ and \lahHpHm, extending and
completing our analysis in \citere{Arco:2020ucn}.
One important aspect of our explorations is to
find allowed parameter regions that lead to either
large non-SM triple Higgs boson couplings,
or to large deviations from unity in the ratio of the light
triple Higgs-boson coupling w.r.t.\ its SM value,
$\kala := \lahhh/\laSM$. 
Particularly, we explore scenarios up to relatively heavy
masses $\MH$, $\MA$ and $\MHp \lsim 1.6 \tev$, 
but not enforcing the so-called
\textit{alignment limit}, $\cos(\be-\al) \to 0$ (see,
e.g.,~\cite{Bernon:2015qea}). 
A related important aspect in the allowed ranges for the various triple Higgs
couplings is that they may affect the di-Higgs boson production rates at
current and future colliders. The production  of Higgs-boson pairs like
$hh$, $hH$, $HH$, $hA$, $hH^\pm$, $AA$ and $H^+H^-$ can be
significantly affected by the presence of sizable triple Higgs couplings
within the 2HDM, yet allowed by the present constraints. 
In particular, $e^+e^-$ colliders will be crucial to explore
deviations from the SM Higgs-boson self-coupling,   
as well as triple Higgs couplings to BSM Higgs bosons.
In the context of the 2HDM type~I and~II, we analyzed the effects from triple
Higgs couplings on the production of two neutral Higgs
bosons at $e^+e^-$ colliders in \citere{Arco:2021bvf}
(extended discussions can be found in
\citeres{Arco:2021ecv,Arco:2021zhb}).
Specifically,  in these previous works we
explored the sensitivity to BSM triple Higgs couplings
via the double Higgs production channels 
$e^+e^- \to h_ih_j \nu \bar \nu$ and $e^+e^- \to h_i h_j Z$
at possible future high-energy $e^+e^-$ colliders, such as the ILC
or CLIC. Further analyses
of triple Higgs couplings at $e^+e^-$ colliders
were presented in \citeres{Kon:2018vmv,Sonmez:2018smv}.
Recent reviews on triple Higgs couplings at $e^+e^-$
colliders can be found
in~\citeres{deBlas:2019rxi,DiMicco:2019ngk,Strube:2016eje,Roloff:2019crr}.

The allowed ranges of the triple Higgs couplings that we explore here
are restricted by theoretical constraints from unitarity and stability (we 
use~\citeres{Bhattacharyya:2015nca,Akeroyd:2000wc,Barroso:2013awa}
as implemented in our private code), 
as well as by experimental constraints from direct Higgs-boson searches
(we use 
\HB~\cite{Bechtle:2008jh,Bechtle:2011sb,Bechtle:2013wla,Bechtle:2015pma,Bechtle:2020pkv},
with data from
\citeres{CMS:2014afl,ATLAS:2018rnh,ATLAS:2019qdc,ATLAS:2018sbw,ATLAS:2020jqj,CMS:2019qcx,ATLAS:2020zms,CMS:2013fjq,CMS:2012jmc,ATLAS:2017ayi,ATLAS:2018uni,ATLAS:2018gfm}),
from the experimental measurements of the production and decay rates of
the Higgs boson at $\sim 125 \gev$ (we use
\HS~\cite{Bechtle:2013xfa,Bechtle:2014ewa,Bechtle:2020uwn}, where the 
experimental data are listed in \citere{higgssignals-www}),
from flavor observables (we use
\texttt{SuperIso}~\cite{Mahmoudi:2008tp,Mahmoudi:2009zz},
complemented with~\citeres{Li:2014fea,Cheng:2015yfu,Arnan:2017lxi} and
experimental data from
\citeres{Chen:2001fja,Aubert:2007my,Limosani:2009qg,Lees:2012ym,Lees:2012wg,Saito:2014das,Aaltonen:2013as,Aaij:2017vnw,Aaboud:2018mst,CMS:2019bbr,ParticleDataGroup:2020ssz}),
as well as from electroweak precision observables (EWPO) (we use~$S$, $T$
and~$U$~\cite{Peskin:1990zt,Peskin:1991sw}, complemented with
\cite{Grimus:2007if,Funk:2011ad} and bounds from
\cite{ParticleDataGroup:2020ssz}). 
To explore the 2HDM parameter space we use
\texttt{2HDMC}~\cite{Eriksson:2009ws}, which includes one-loop QCD corrections for the Higgs-boson decay widths. In the decays of neutral Higgses to quarks also two-loop QCD corrections are included. Furthermore, for all decays of a Higgs boson to quarks the leading logarithmic corrections to all orders are implemented by using the running $\overline{\rm MS}$ quark masses in the couplings, see \citere{Eriksson:2009ws} for more details.
The analysis of the values of the triple Higgs couplings has
been performed with our private code, which is based on the tree-level formulas for these couplings as given in the ``physical basis'' in terms of our chosen input parameters, see the appendix of \citere{Arco:2020ucn}.

Our analysis extends the work presented in \citere{Arco:2020ucn} in several
ways. In \citere{Arco:2020ucn} we focused on the 2HDM type~I and~II,
with the then available constraints. While the theoretical constraints
remain effectively the same, there have been important updates in the
experimental constraints. Particularly, we can now apply the full set of
available LHC Higgs-boson rate measurements, especially the STXS
measurements via \texttt{\HS} into our evaluation. This leads to
somewhat tighter limits on $\cos(\be-\al)$ (see below) and
correspondingly to smaller allowed intervals for the various triple
Higgs couplings, particularly in type~II. More importantly, we now
extend our analysis to the full set of 2HDM types. In this way we
provide a direct comparison of the four types w.r.t.\ the various
theoretical and experimental constraints.

This also constitutes one of the main differences
between our new study 
and  previous studies on constraints in the 2HDM, from LHC physics
\cite{Sirunyan:2018koj,Aad:2019mbh,Kling:2020hmi,Abouabid:2021yvw},  EWPO
\cite{Bertolini:1985ia,Hollik:1986gg,Grimus:2007if}, flavor physics
\cite{Enomoto:2015wbn,Atkinson:2022pcn} and global fits
\cite{Bernon:2015qea,Arbey:2017gmh,Haller:2018nnx,Kraml:2019sis}.
With the fully updated results for the allowed ranges
of the triple
Higgs couplings presented here,  one could then explore the 
sensitivities to those couplings at future $e^+e^-$ colliders.  Such an
analysis,  extending our first proposal in
\citeres{Arco:2021bvf,Arco:2021ecv,Arco:2021zhb}
is left for future work.  
        
Our paper is organized as follows. In \refse{sec:model} we briefly review
the details of the 2HDM and fix our notation.  We also discuss
the theoretical and experimental constraints applied to
our sampling of the 2HDMs. The four 2HDM types are compared to each
other in \refse{sec:comparison} in several selected benchmark planes,
where we discuss in detail the impact of the various constraints on them. 
In \refse{sec:thc-anal} we define specific planes for each of the four
types, exhibiting large effects on the triple Higgs couplings. We
analyze the maximum deviations of \lahhh\ from the SM  
that are still allowed taking into account all constraints. We also discuss
the values that can be reached for the other triple Higgs couplings
involving at least one $h$. Our conclusions are given in
\refse{sec:conclusions}.


\section{The Model and the constraints}
\label{sec:model}

In this section we give a brief description of the 2HDM to fix our
notation. We also review the theoretical and experimental constraints,
which are in general the same as in
\citere{Arco:2020ucn}, but where details of the experimental constraints
have been updated. 


\subsection{The 2HDM}
\label{sec:2hdm}

We assume the $\cp$ conserving 2HDM (see
\citeres{Gunion:1989we,Aoki:2009ha, Branco:2011iw} for reviews).
The potential can be written as:
\begin{align}
V &= m_{11}^2 (\Phi_1^\dagger\Phi_1) + m_{22}^2 (\Phi_2^\dagger\Phi_2) - \msq (\Phi_1^\dagger
\Phi_2 + \Phi_2^\dagger\Phi_1) + \frac{\la_1}{2} (\Phi_1^\dagger \Phi_1)^2 +
\frac{\la_2}{2} (\Phi_2^\dagger \Phi_2)^2 \nonumber \\
&\quad + \la_3
(\Phi_1^\dagger \Phi_1) (\Phi_2^\dagger \Phi_2) + \la_4
(\Phi_1^\dagger \Phi_2) (\Phi_2^\dagger \Phi_1) + \frac{\la_5}{2}
[(\Phi_1^\dagger \Phi_2)^2 +(\Phi_2^\dagger \Phi_1)^2]  \;.
\label{eq:scalarpot}
\end{align}
\noindent
The two $SU(2)_L$ doublets are denoted as $\Phi_1$ and $\Phi_2$, 
\begin{eqnarray}
\Phi_1 = \left( \begin{array}{c} \phi_1^+ \\ \frac{1}{\sqrt{2}} (v_1 +
    \rho_1 + i \eta_1) \end{array} \right) \;, \quad
\Phi_2 = \left( \begin{array}{c} \phi_2^+ \\ \frac{1}{\sqrt{2}} (v_2 +
    \rho_2 + i \eta_2) \end{array} \right) \;,
\label{eq:2hdmvevs}
\end{eqnarray}
where $v_1, v_2$ are the two real vacuum expectation values (vevs)
acquired by the fields 
$\Phi_1, \Phi_2$, respectively, and they satisfy the 
relation $v = \sqrt{(v_1^2 +v_2^2)}$ where $v\simeq246\gev$ is the SM
vev. We furthermore define $\tb := v_2/v_1$.
The eight degrees of freedom above, $\phi_{1,2}^\pm$, $\rho_{1,2}$ and
$\eta_{1,2}$, give rise to three Goldstone bosons, $G^\pm$ and $G^0$,
and five massive physical scalar fields: two $\cp$-even scalar fields,
$h$ and $H$, one $\cp$-odd one, $A$, and one charged pair, $H^\pm$.
Here the mixing angle $\al$ diagonalizes the $\CP$-even scalar bosons,
whereas the angle $\be$ diagonalizes the $\cp$-odd and the charged scalar
bosons. 

A $Z_2$ symmetry is imposed to avoid the occurrence of tree-level
FCNC. This symmetry  is softly broken by the parameter $\msq$ in
the Lagrangian. The extension of the $Z_2$ symmetry to the Yukawa
sector of the model forbids tree-level FCNCs. 
This results in four variants of 2HDM, 
depending on the $Z_2$ parities of the 
fermions, where the corresponding coupling to fermions are
listed in \refta{tab:types} and \refta{tab:coupling}.

\begin{table}[htb!]
\begin{center}
\begin{tabular}{lccc} 
\hline
  & $u$-type & $d$-type & leptons \\
\hline
Type~I 									& $\Phi_2$ & $\Phi_2$ & $\Phi_2$ \\
Type~II 								& $\Phi_2$ & $\Phi_1$ & $\Phi_1$ \\
Type~III/Flipped/Y 				& $\Phi_2$ & $\Phi_1$ & $\Phi_2$ \\
Type~IV/Lepton-specific/X	& $\Phi_2$ & $\Phi_2$ & $\Phi_1$ \\
\hline
\end{tabular}
\caption{Allowed fermion couplings in 
the four 2HDM types.}
\label{tab:types}
\end{center}
\end{table}

We will study the 2HDM in the so-called ``physical basis'', where the
free parameters in \refeq{eq:scalarpot} can be re-expressed in terms of
the following set:
\begin{equation}
c_{\be-\al} \; , \quad \tb \;, \quad v \; ,
\quad \Mh\;, \quad \MH \;, \quad \MA \;, \quad \MHp \;, \quad \msq \;,
\label{eq:inputs}
\end{equation}
which we take here as input parameters.
From now on we use sometimes the short-hand notation $s_x = \sin(x)$,
$c_x = \cos(x)$. 
In our analysis we will identify the lightest $\cp$-even Higgs boson,
$h$, with the observed Higgs boson at $\sim 125 \gev$.

The couplings of the extended Higgs sector to SM particles within the
2HDM are different than in the SM.  In particular,  the couplings of the
lightest Higgs boson are modified 
w.r.t.\ the SM Higgs-coupling predictions due to the mixing in the Higgs
sector.  The corresponding 2HDM Lagrangian is given by:
\begin{eqnarray}
	\mathcal{L} &=&-\sum_{f=u,d,l}\frac{m_f}{v}\left[\xi_h^f\bar{f}fh + \xi_H^f\bar{f}fH +i \xi_A^f\bar{f}\gamma_5fA \right] \nonumber \\
	&& -\left[\frac{\sqrt{2}}{v}\bar{u}\left(m_{u}V_{\mathrm{CKM}}\xi_{A}^{u}P_{L}+V_{\mathrm{CKM}}m_{d}\xi_{A}^{d}P_{R}\right)dH^{+}+\frac{\sqrt{2}m_{l}}{v}\xi_{A}^{l}\bar{\nu}P_{R}lH^{+}+\mathrm{h.c.}\right] \nonumber \\
&&+\sum_{h_i=h,H,A}	\left[   g m_W \xi_{h_i}^W  W_\mu W^\mu h_i + \frac{1}{2} g m_Z \xi_{h_i}^Z  Z_\mu Z^\mu h_i\right] .
\label{eq:Lag-xi}
\end{eqnarray}
Here $m_{f,f'}$, $m_W$ and $m_Z$ are the fermion masses, the $W$ mass
and the $Z$ mass, respectively. The factors in the couplings  to
fermions,  $\xi_{h,H,A}^f$, and to gauge-bosons, $\xi_{h,H,A}^V$,    
are summarized in \refta{tab:coupling}.

\begin{table}
\begin{center}
\begin{tabular}{c|c|c|c|c}
 & Type~I  & Type~II & Type~III/Flipped/Y & Type~IV/Lepton-specific/X\tabularnewline
\hline 
$\xi_{h}^{u}$ & $s_{\beta-\alpha}+c_{\beta-\alpha}\cot\beta$ & $s_{\beta-\alpha}+c_{\beta-\alpha}\cot\beta$ & $s_{\beta-\alpha}+c_{\beta-\alpha}\cot\beta$ & $s_{\beta-\alpha}+c_{\beta-\alpha}\cot\beta$\tabularnewline
$\xi_{h}^{d}$ & $s_{\beta-\alpha}+c_{\beta-\alpha}\cot\beta$ & $s_{\beta-\alpha}-c_{\beta-\alpha}\tan\beta$ & $s_{\beta-\alpha}-c_{\beta-\alpha}\tan\beta$ & $s_{\beta-\alpha}+c_{\beta-\alpha}\cot\beta$\tabularnewline
$\xi_{h}^{l}$ & $s_{\beta-\alpha}+c_{\beta-\alpha}\cot\beta$ & $s_{\beta-\alpha}-c_{\beta-\alpha}\tan\beta$ & $s_{\beta-\alpha}+c_{\beta-\alpha}\cot\beta$ & $s_{\beta-\alpha}-c_{\beta-\alpha}\tan\beta$\tabularnewline
$\xi_h^V$ & $s_{\be-\al}$ & $s_{\be-\al}$ & $s_{\be-\al}$ & $s_{\be-\al}$\tabularnewline\hline
$\xi_{H}^{u}$ & $c_{\beta-\alpha}-s_{\beta-\alpha}\cot\beta$ & $c_{\beta-\alpha}-s_{\beta-\alpha}\cot\beta$ & $c_{\beta-\alpha}-s_{\beta-\alpha}\cot\beta$ & $c_{\beta-\alpha}-s_{\beta-\alpha}\cot\beta$\tabularnewline
$\xi_{H}^{d}$ & $c_{\beta-\alpha}-s_{\beta-\alpha}\cot\beta$ & $c_{\beta-\alpha}+s_{\beta-\alpha}\tan\beta$ & $c_{\beta-\alpha}+s_{\beta-\alpha}\tan\beta$ & $c_{\beta-\alpha}-s_{\beta-\alpha}\cot\beta$\tabularnewline
$\xi_{H}^{l}$ & $c_{\beta-\alpha}-s_{\beta-\alpha}\cot\beta$ & $c_{\beta-\alpha}+s_{\beta-\alpha}\tan\beta$ & $c_{\beta-\alpha}-s_{\beta-\alpha}\cot\beta$ & $c_{\beta-\alpha}+s_{\beta-\alpha}\tan\beta$\tabularnewline
$\xi_H^V$ & $c_{\be-\al}$ & $c_{\be-\al}$ & $c_{\be-\al}$ & $c_{\be-\al}$\tabularnewline\hline
$\xi_{A}^{u}$ & $-\cot\beta$ & $-\cot\beta$ & $-\cot\beta$ & $-\cot\beta$\tabularnewline
$\xi_{A}^{d}$ & $\cot\beta$ & $-\tan\beta$ & $-\tan\beta$ & $\cot\beta$\tabularnewline
$\xi_{A}^{l}$ & $\cot\beta$ & $-\tan\beta$ & $\cot\beta$ & $-\tan\beta$\tabularnewline
$\xi_A^V$ & 0 & 0  & 0 & 0 \tabularnewline
\end{tabular}
\caption{Relevant factors appearing in the couplings of the Higgs 
bosons to fermions,  $\xi_{h,H,A}^f$,  and to gauge-bosons,
$\xi_{h,H,A}^V$,  according to \refeq{eq:Lag-xi},
in the four types of the 2HDM considered here.}
\label{tab:coupling}
\end{center}
\end{table}

In this paper we focus on the couplings of the
lightest $\cp$-even Higgs boson with the other BSM bosons, concretely
\lahhh,  \lahhH, \lahHH\ and \lahAA.
We define these $\la_{h_i h_j h_k}$ couplings such that the Feynman rules
are given by: 
\begin{equation}
	\begin{gathered}
		\includegraphics{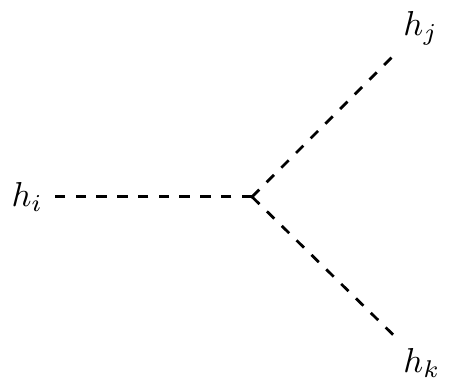}
	\end{gathered}
	=- i\, v\, n!\; \la_{h_i h_j h_k}\,,
\label{eq:lambda}
\end{equation}
where $n$ is the number of identical particles in the vertex. 
Explicit expressions for the couplings $\la_{hh_ih_j}$ in terms of our
input parameters in \refeq{eq:inputs} can be found in
the Appendix of \citere{Arco:2020ucn}. 
Following the convention in \refeq{eq:lambda} the light Higgs
triple coupling $\lahhh$ has the same normalization as $\laSM$  in the
SM, i.e. $-6iv\laSM$ with 
$\laSM=\Mh^2/2v^2\simeq0.13$.
We furthermore define $\kala := \lahhh/\laSM$.

An important limit of the 2HDM is reached for $\CBA \to 0$, the
so-called \textit{alignment limit}. 
In particular, if $\CBA=0$ one recovers all the
interactions of the SM Higgs boson for the $h$ state.
However, also in the alignment limit one can still have BSM physics
related to the extended Higgs sector, like $hHH$ or $ZHA$ interactions, 
for example.


\subsection{Experimental and theoretical constraints}
\label{sec:constraints}

In this subsection we briefly summarize the various theoretical and
experimental constraints considered in our scans, with an emphasis on
differences w.r.t.\ the constraints used in \citere{Arco:2020ucn}.

\begin{itemize}

\item {\bf Constraints from electroweak precision data}\\
For ``pure'' Higgs-sector extensions of the SM, 
constraints from the electroweak precision observables (EWPO)
can be parametrized well in terms of the oblique parameters $S$, $T$ and
$U$~\cite{Peskin:1990zt,Peskin:1991sw}.
In the 2HDM the most constraining EWPO is the
$T$~parameter~\cite{Grimus:2007if,Funk:2011ad}.  
It requires either $\MHp \approx \MA$ or $\MHp \approx \MH$.  
In \citere{Arco:2020ucn} we explored three scenarios: 
(A)\;$\MHp = \MA$ with independent $\MH$,
(B)\;$\MHp = \MH$ with independent $\MA$,
and (C)\;$\MHp = \MA = \MH$.
In the central section of this work, \refse{sec:comparison}, we will
focus on scenario~C with $m := \MHp = \MA = \MH$.
In the following \refse{sec:thc-anal} we will analyze and compare both
scenarios, the complete degenarate scenario~C and the non-fully
degenerate scenario~A, allowing also for a comparison of these scenarios.
From the technical side the 2HDM parameter space is explored with the code
\texttt{2HDMC-1.8.0}~\cite{Eriksson:2009ws}, where the predictions
for the triple Higgs couplings are analyzed with our private code. 

\item {\bf Theoretical constraints}\\
The important theoretical constraints come from
tree-level perturbartive unitarity and the stability of the vacuum.
These constraints are ensured by an explicit test of the underlying Lagrangian
parameters \cite{Bhattacharyya:2015nca,Akeroyd:2000wc,Barroso:2013awa},
  see also \citere{Arco:2020ucn} for more details. It should be noted
that $ \msq$ is a free input parameter in our study,  but we
have also analyzed specific choices of $ \msq$ that turn out to be
interesting for the present study.  Concretely,  
the parameter space allowed by the two mentioned theoretical constraints
can be enlarged, in 
particular to higher values of the BSM Higgs masses by the particular condition, which we have applied in our analysis in some cases,
\begin{equation}
  \msq = \frac{\MH^2\cos^2\al}{\tb}~.
  \label{eq:m12special}
\end{equation}
In some other cases of our analysis we have applied an
alternative  condition on $m_{12}^2$, which can be obtained by enforcing  
the stability condition $\la_3+\la_4-|\la_5|+\sqrt{\la_1\la_2}=0$.
This can be written as:
\begin{equation}
	\msq = \frac{1}{2}\frac{\Mh^2\MH^2\sin(2\beta)}{\Mh^2 s^2_{\be-\al}+\MH^2 c^2_{\be-\al}}
	\simeq \frac{1}{2}\frac{\Mh^2\MH^2\sin(2\beta)}{\Mh^2 +\MH^2 c^2_{\be-\al}}\,.
	\label{eq:m12special2}
\end{equation}
It is interesting to notice that both of the equations above go to
the same expression in the alignment limit:
\begin{equation}
	\msq=m_H^2\sin\beta\cos\beta.
	\label{eq:m12AL}
\end{equation}

\item {\bf Constraints from direct searches at colliders}\\
The exclusion limits at the $95\%$ confidence level (CL) of all
relevant searches for BSM Higgs bosons are included in the public code
\HB\,\texttt{v.5.9}~\cite{Bechtle:2008jh,Bechtle:2011sb,Bechtle:2013wla,Bechtle:2015pma,Bechtle:2020pkv},
including Run~2 data from the LHC.
Each parameter point in the 2HDM (or any other model) gives a set of
theoretical predictions for the Higgs-boson sector.
\HB\ determines which is the most 
sensitive channel for this parameter point and then determines, based on
this most sensitive channel, whether the point is allowed or not at the
$95\%$~CL. 
As input \HB\ requires some specific predictions from the model,
like branching ratios or Higgs couplings, that we computed with the
help of \texttt{2HDMC}~\cite{Eriksson:2009ws}.

\item {\bf Constraints from the SM-like Higgs-boson properties}\\
Any model beyond the SM has to accommodate the SM-like Higgs boson,
with mass and signal strengths as measured at the LHC (within
theoretical and experimental uncertainties). 
In our scans the compatibility of the $\cp$-even scalar $h$ with a mass
of $125.09\gev$ with the LHC measurements of rates 
is checked with the code
\HS\,\texttt{v.2.6.1}~\cite{Bechtle:2013xfa,Bechtle:2014ewa,Bechtle:2020uwn}. 
This code provides a
statistical $\chi^2$ analysis of the SM-like Higgs-boson predictions of
a certain model w.r.t.\ the LHC measurement of Higgs-boson rates
and masses. As for the BSM Higgs searches, the predictions of the 2HDM
have been obtained with \texttt{2HDMC}~\cite{Eriksson:2009ws}. 
As in \citere{Arco:2020ucn}, in this work we will require that for a
parameter point of the 2HDM to be allowed, the corresponding $\chi^2$ is
within $2\,\sig$ ($\De\chi^2 = 6.18$) of
the SM fit: $\chi_\mathrm{SM}^2=85.76$ with 107 observables.

Many of the recent LHC Higgs rate measurements are now given in terms of
``STXS observables''. 
As an important update w.r.t.\ our previous analysis in \citere{Arco:2020ucn} 
the {\tt 2HDMC} output can now allow the application of the STXS
observables (as more recently implemented in \HS). 
This results in substantially stronger limits on, in particular, $\CBA$,
especially in the 2HDM type~II. 
This leads to substantially smaller allowed intervals of the triple
Higgs couplings in some cases.

\item {\bf Constraints from flavor physics}\\
Constraints from flavor physics can be very significant
in the 2HDM mainly due to the presence of the charged Higgs-boson.
Various flavor observables, e.g.\ rare $B$~decays, 
$B$~meson mixing parameters, $\br(B \to X_s \gamma)$, but also 
LEP constraints on $Z$ decay partial widths
etc., are sensitive to charged Higgs boson exchange. Consequently, they
can provide effective constraints on the available 
parameter space~\cite{Enomoto:2015wbn,Arbey:2017gmh}. 
In this work we take into account the most important constraints,  given
by the decays $B \to X_s \gamma$ and $B_s \to \mu^+ \mu^-$. 
We consider the following experimental values from \cite{ParticleDataGroup:2020ssz},  
with $\mathrm{BR}(B\to X_s\ga)=(3.49\pm0.19) \times 10^{-4}$ 
(averaged value from \cite{Chen:2001fja,Aubert:2007my,Limosani:2009qg,Lees:2012ym,Lees:2012wg,Saito:2014das}) and $\mathrm{BR}(B_s\to\mu^+\mu^-)=(2.9\pm0.4) \times 10^{-9}$ (averaged value from \cite{Aaltonen:2013as,Aaij:2017vnw,Aaboud:2018mst,CMS:2019bbr}).
We employ
the code \texttt{SuperIso4.0}~\cite{Mahmoudi:2008tp,Mahmoudi:2009zz}
where again the model input is given by \texttt{2HDMC}.  We have modified the
code to include the Higgs-Penguin type corrections in
$B_s \to \mu^+ \mu^-$~\cite{Li:2014fea,Arnan:2017lxi,Cheng:2015yfu},
which were not included in the original version of
\texttt{SuperIso}. These corrections can be relevant for the present
work since precisely these Higgs-Penguin contributions are the ones
containing the effects from triple Higgs couplings in $B_s \to \mu^+ \mu^-$.   

\end{itemize}



\section{Comparison of the four 2HDM types}
\label{sec:comparison}

In this section we will compare the four 2HDM types w.r.t.\ the various
constraints, as described in the previous section. As discussed in
\refse{sec:constraints},  in order to simplify our analysis,
we set in this section all the heavy Higgs-boson masses
to be equal, $m := \MHp = \MA = \MH$. 
Based on the analysis in \citere{Arco:2020ucn} we define three 
benchmark planes for this comparison.  In
order  to leave some allowed parameter space  by the most constraining
flavor observables at low $\tan \beta$,  $B \to X_s \ga$ and
$B \to \mu \mu$,  specially for the types~II and~III,  whenever we have
to fix $m$ we choose moderately heavy values for this parameter.
Concretely,  in our benchmark planes we  
set $m = 550 \gev$ or leave $m$ as a free parameter.  Similarly,
whenever we have to fix the value of $\CBA$ in our plots we choose a
moderately small value for this parameter in order to get some allowed
parameter space imposing the LHC constraints.  Concretely, we choose
$\CBA=0.01, 0.02$ or leave it as a free parameter.
Furthermore, in the benchmark scenarios with a
fixed value of $\tb$ we set it to relatively low values,
where the four 2HDM-types manifest some allowed parameter space.
The particular non-vanishing fixed value for $m_{12}^2$ in our
scenarios is not as relevant as the others, regarding the experimental
constraints,  but we set it in our benchmark planes (in this and
the following section) within the explored
interval $\inter{0}{(2\times10^6\sim1400^2) \gev^2}$ to get a wide
allowed region of the parameter space after applying the theoretical
constraints.  Concretely,  the three 
benchmark scenarios chosen for this section are defined by:

\begin{enumerate}

\item
  $m \equiv \MHp = \MH = \MA = 550 \gev$,
  $\msq=60000\gev^2$ \\
  free parameters: $\CBA$, $\tb$
  
\item
  $m \equiv \MHp = \MH = \MA=550\gev$,
  $\CBA=0.02$,\\
  free parameters: $\msq$, $\tan\beta$.
  
\item
  $\tan\beta=3.0$,
  $\CBA=0.01$,\\
  free parameters: $\msq$, $m \equiv \MHp = \MH = \MA$.

\end{enumerate}

The results for the three benchmark scenarios~1,~2,~3 are shown in
\reffis{fig:bench-cba-tb}, \ref{fig:bench-m12-tb} and
\ref{fig:bench-m12-m}, respectively. Each figure is split into two
subfigures: in subfigure~(A) we focus on the various constraints.
We show the results for type~I, II, III
and~IV in the left, second, third and right column, respectively.
Concerning the first rows in \reffis{fig:bench-cba-tb}(A),
\ref{fig:bench-m12-tb}(A) and \ref{fig:bench-m12-m}(A), the areas
permitted by the Higgs-boson rate measurements, as evaluated with \HS,
are shown as dark (light) yellow regions allowed at the 
$1\,(2)\,\sig$ level, corresponding to a $\De\chi^2 = 2.30 (6.18)$
w.r.t.\ the SM value.
The areas, allowed at the 95\% CL by the (BSM) Higgs-boson searches at
LHC with \HB\ are indicated as blue regions. The 
small letters shown on the various parts of the edges indicate the
channel that is responsible (via the \HB\ selection) for the
respective part of the exclusion bounds. The letters correspond to the
following channels:\\
\begin{itemize}

\item[(a)] $pp\to h\to\gamma\gamma$~\cite{CMS:2014afl}

\item[(b)] $pp\to H\to hh\to bbbb$~\cite{ATLAS:2018rnh}

\item[(c)] $pp\to H\to hh\to bb/\tau\tau/WW/\gamma\gamma$~\cite{ATLAS:2019qdc}

\item[(d)] $pp\to H\to VV$~\cite{ATLAS:2018sbw}

\item[(e)] $pp\to H^\pm tb\to tbtb$~\cite{ATLAS:2020jqj}

\item[(f)] $gg\to A\to Zh\to llbb$~\cite{CMS:2019qcx}

\item[(g)] $pp\to H\to\tau\tau$ and $pp\to A\to\tau\tau$~\cite{ATLAS:2020zms}

\item[(h)] $pp\to h\to ZZ\to llll$~\cite{CMS:2013fjq}

\item[(i)] $pp\to H\ (\mathrm{VBF})/HW/HZ/Htt$
  with $H\to\gamma\gamma$~\cite{ATLAS:2017ayi}

\item[(j)] $pp\to h\tau\tau$~\cite{CMS:2012jmc}

\item[(k)] $pp\to AW/AZ/Att$ with $A\to\gamma\gamma$~\cite{ATLAS:2017ayi}

\item[(l)] $pp\to H\to hh\to bb/\tau\tau$~\cite{ATLAS:2018uni}

\item[(m)] $pp\to H^\pm tb\to \tau\nu_\tau tb$~\cite{ATLAS:2018gfm}

\end{itemize}
The areas allowed by both, Higgs rate measurements and BSM Higgs-boson
searches at the 95\% CL are shown as dotted areas in the first rows of
\reffis{fig:bench-cba-tb}(A),
\ref{fig:bench-m12-tb}(A) and \ref{fig:bench-m12-m}(A).
In the second rows of \reffis{fig:bench-cba-tb}(A),
\ref{fig:bench-m12-tb}(A) and \ref{fig:bench-m12-m}(A) we show
the restrictions from flavor physics. The regions allowed by $B \to X_s \ga$
($B_s \to \mu\mu$) are given by the pink (teal) area.
The parameter space allowed by both constraints is shown as
dotted area.
The third rows of \reffis{fig:bench-cba-tb}(A),
\ref{fig:bench-m12-tb}(A) and \ref{fig:bench-m12-m}(A) indicate the
restrictions from unitarity (light green) and stability (light pink), see
\refse{sec:constraints} for details.
The parameter space allowed by both types of constraints is shown as
dotted area.
The violet solid line follows \refeq{eq:m12special},  whereas the yellow
dashed line satisfies \refeq{eq:m12special2}.
Since the Higgs potential is identical in all four types,  the
constraints from unitarity and stability are identical. We show them
for all four types individually to have all constraints for one type
collected in one column.
The fourth rows of \reffis{fig:bench-cba-tb}(A),
\ref{fig:bench-m12-tb}(A) and \ref{fig:bench-m12-m}(A) indicate
the regions allowed by all constraints in the respective scenario, shown
as dotted area, with a solid black, solid blue, dotted pink or dotted
orange line around for the Yukawa types~I, II, III and~IV, respectively.

The subfigure~(B) in \reffis{fig:bench-cba-tb},
\ref{fig:bench-m12-tb} and \ref{fig:bench-m12-m}
then present the results for the various triple Higgs
couplings in these benchmark planes, where in each plot the four
regions allowed in the four types are indicated together. Here we show
$\kala := \lahhh/\laSM$ in the upper left, $\lahhH$ in the upper
right, $\lahHH$ in the lower left and $\lahHpHm = 2 \lahAA$
(the latter equality holds in our scenarios because of $\MA = \MHp$) 
in the lower right plots of \reffis{fig:bench-cba-tb}(B),
\ref{fig:bench-m12-tb}(B) and \ref{fig:bench-m12-m}(B). 
Here it should be kept in mind that the values for
the various triple Higgs couplings are displayed for a qualitative
comparison of the four 2HDM types. An analysis of the largest
deviation from the SM or the largest possible values in the four types
will be performed with ``optimized'' planes in the following sections.

\begin{figure}[t!]
\centering
\includegraphics[width=0.24\textwidth]{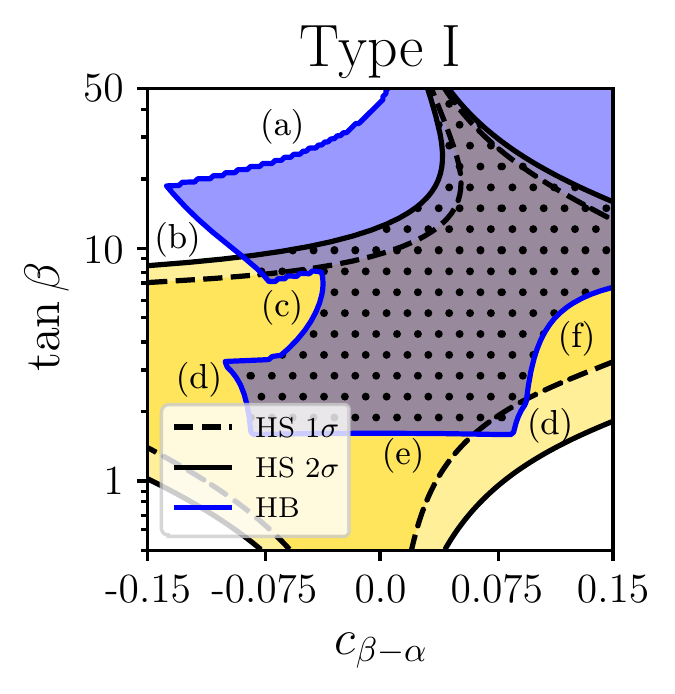}
\includegraphics[width=0.24\textwidth]{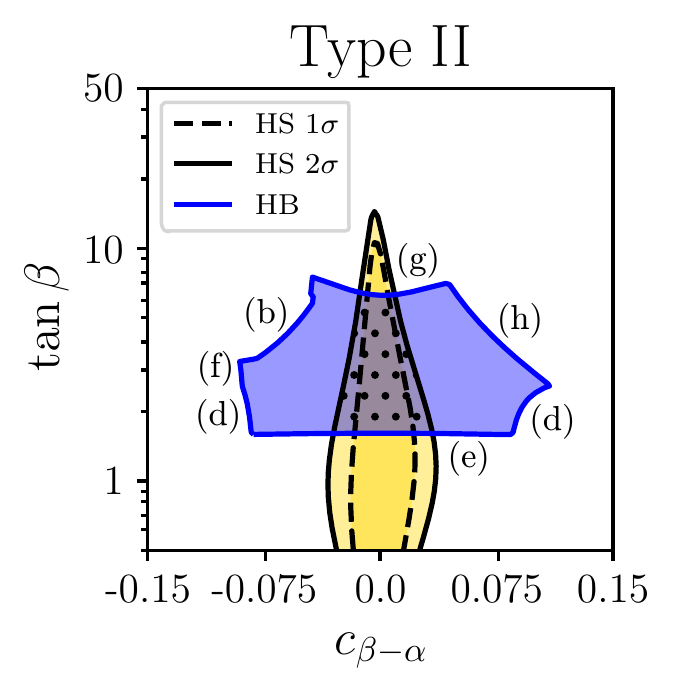}
\includegraphics[width=0.24\textwidth]{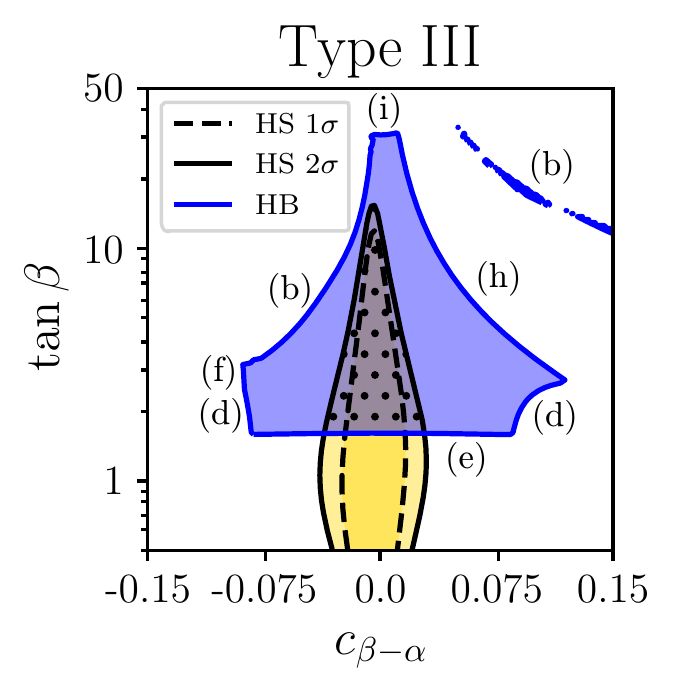}
\includegraphics[width=0.24\textwidth]{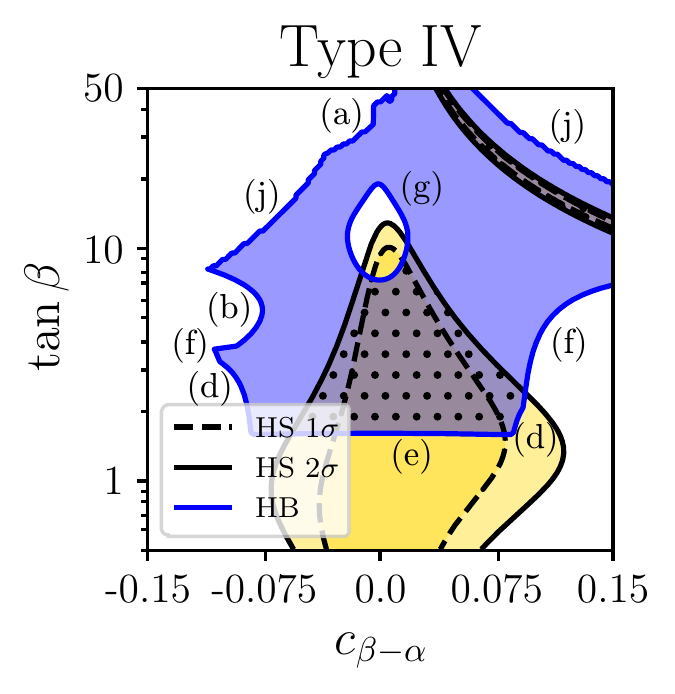}

\includegraphics[width=0.24\textwidth]{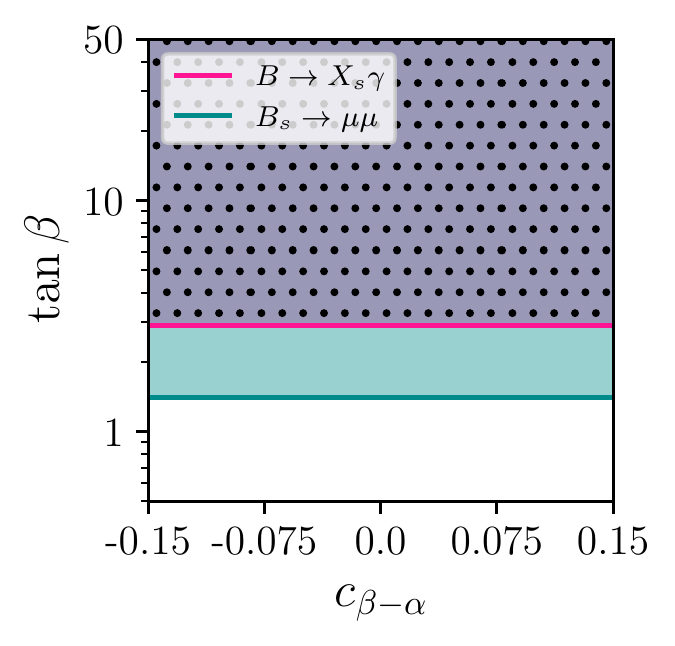}
\includegraphics[width=0.24\textwidth]{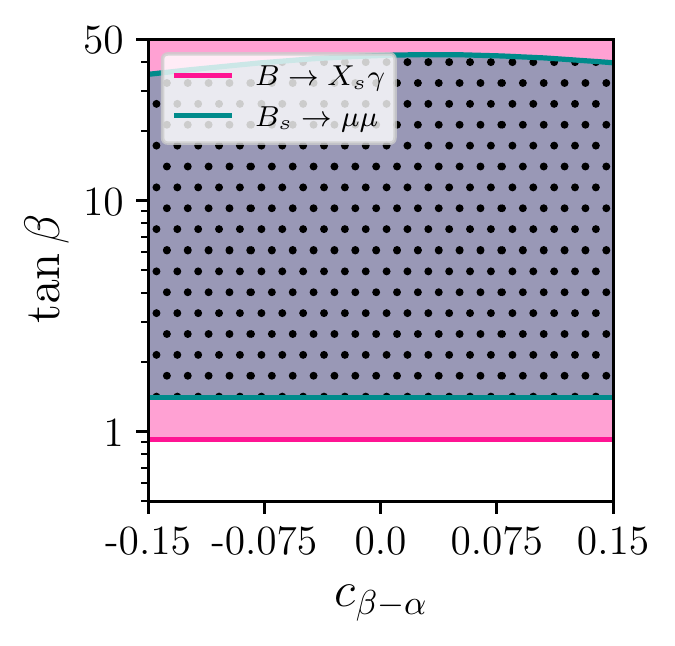}
\includegraphics[width=0.24\textwidth]{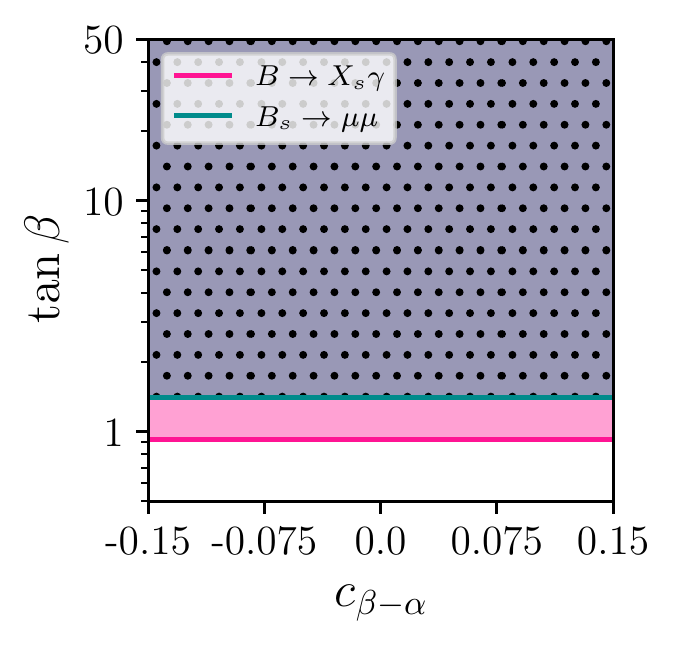}
\includegraphics[width=0.24\textwidth]{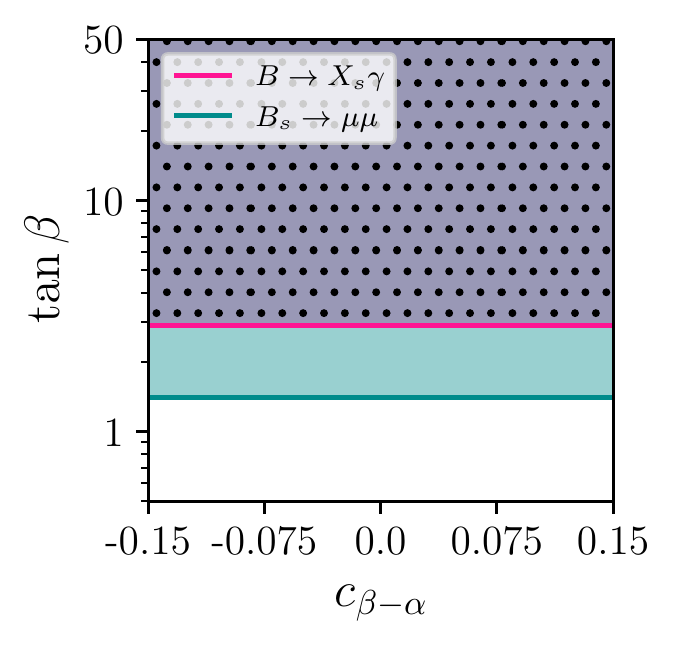}

\includegraphics[width=0.24\textwidth]{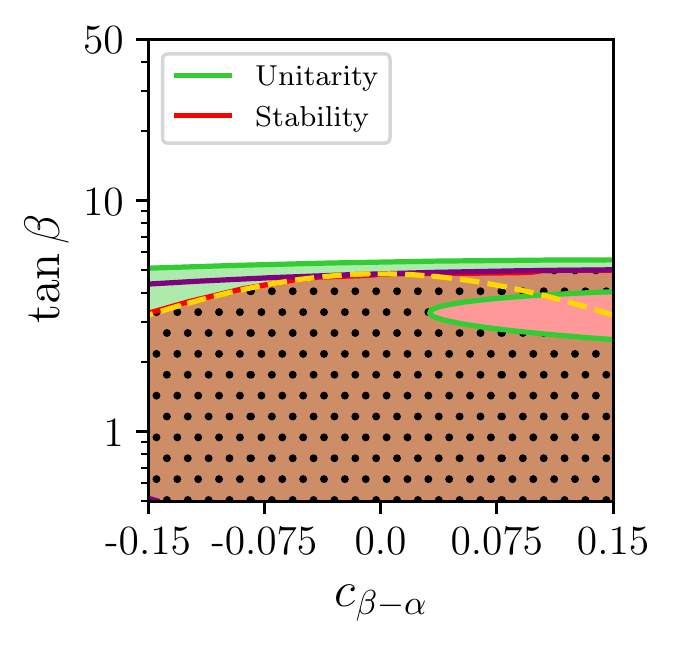}
\includegraphics[width=0.24\textwidth]{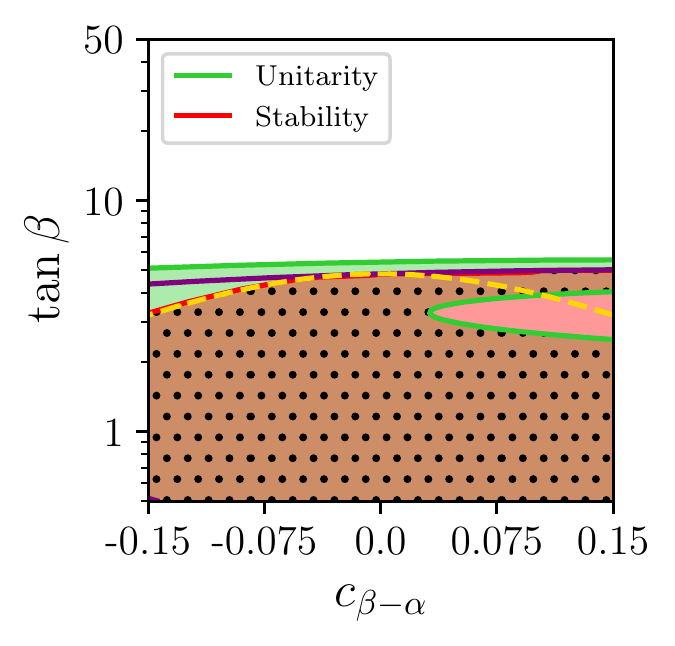}
\includegraphics[width=0.24\textwidth]{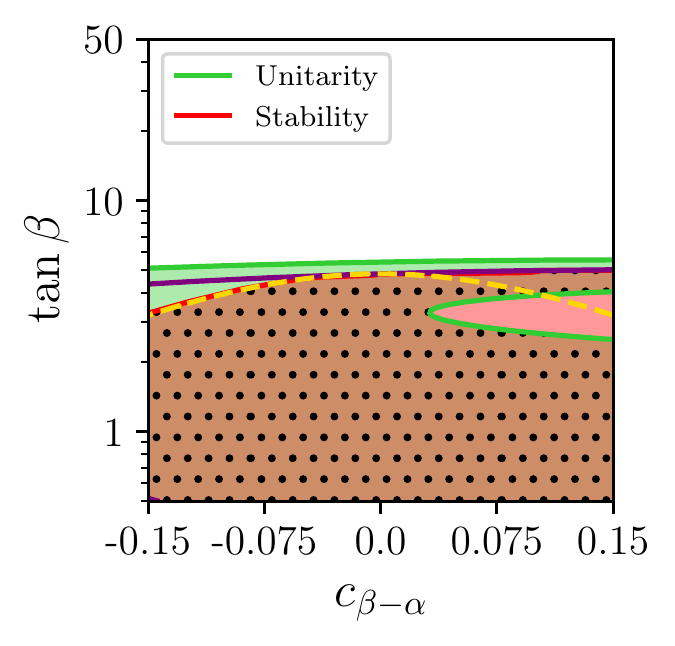}
\includegraphics[width=0.24\textwidth]{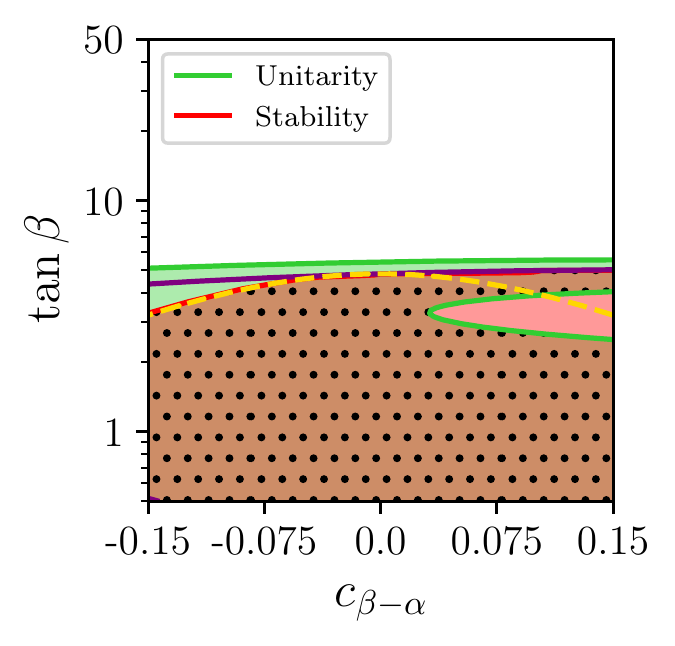}

\includegraphics[width=0.24\textwidth]{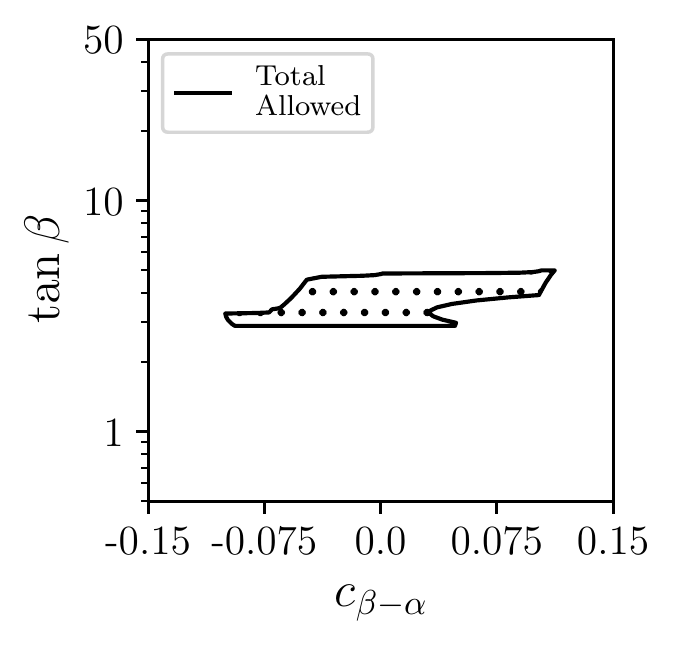}
\includegraphics[width=0.24\textwidth]{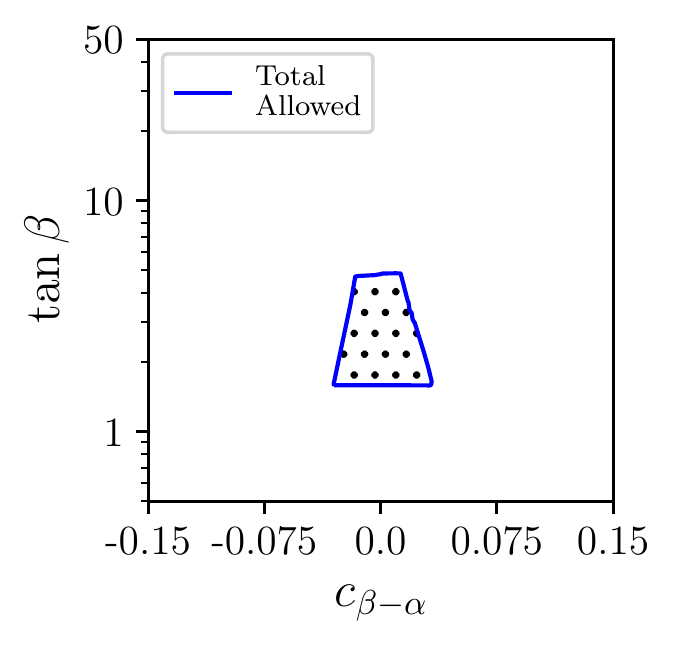}
\includegraphics[width=0.24\textwidth]{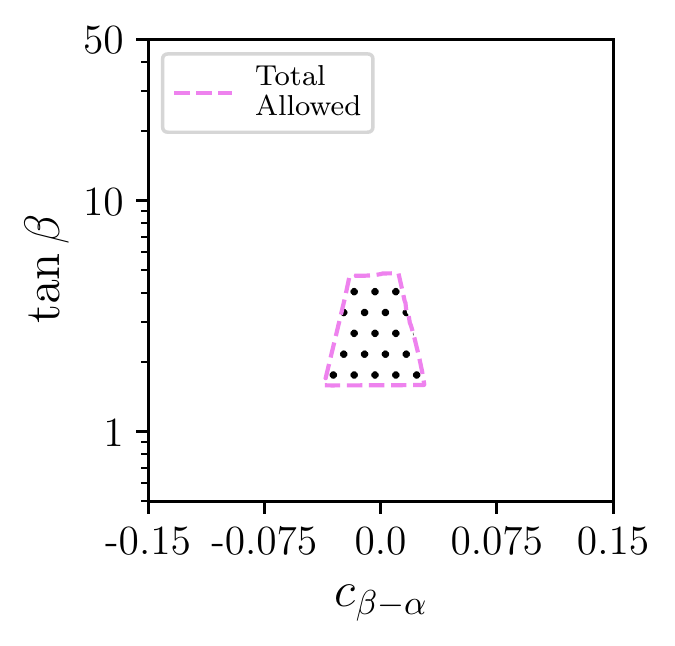}
\includegraphics[width=0.24\textwidth]{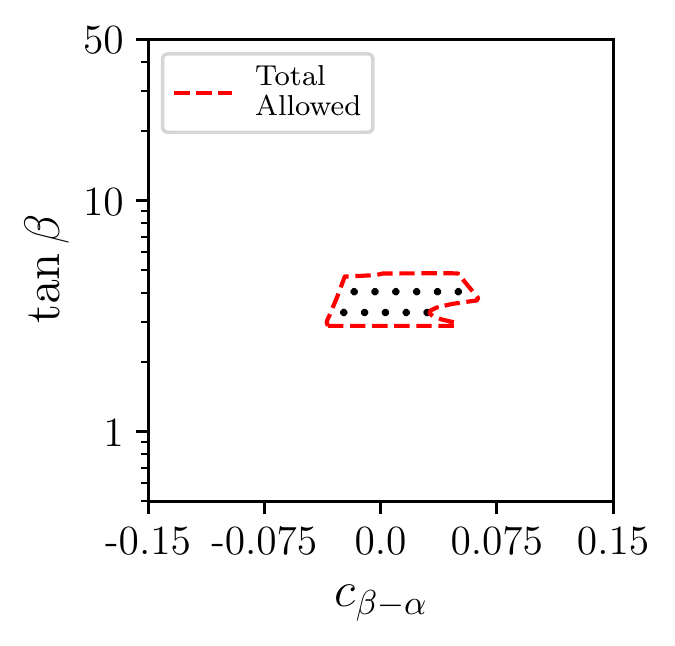}
\caption{\textbf{(A)} Allowed regions from the restrictions on the
  parameter space in benchmark 
  scenario~1 in the $\CBA$--$\tb$ plane
  with $m = 550 \gev$ and $\msq = 60000 \gev^2$. The results
for type~I, II, III and~IV in the left, second, third and right
column, respectively. The upper, second and third row show the
restrictions from \HB/\HS, the flavor observables and from
unitarity/stability, respectively. The fourth row indicates
the regions allowed by all constraints in the respective scenario.}
\end{figure}

\begin{figure}[t!]\ContinuedFloat
\centering
\includegraphics[height=0.25\textheight]{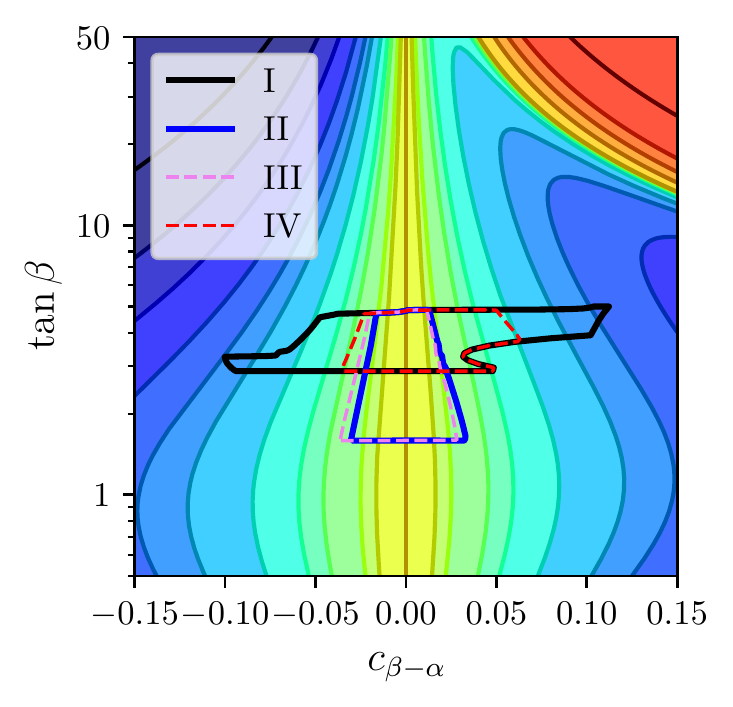}
\includegraphics[height=0.25\textheight]{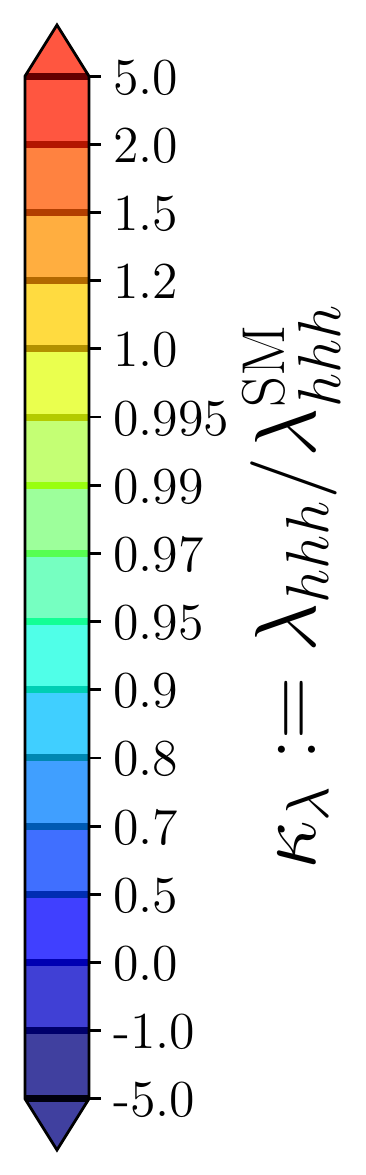}
\includegraphics[height=0.25\textheight]{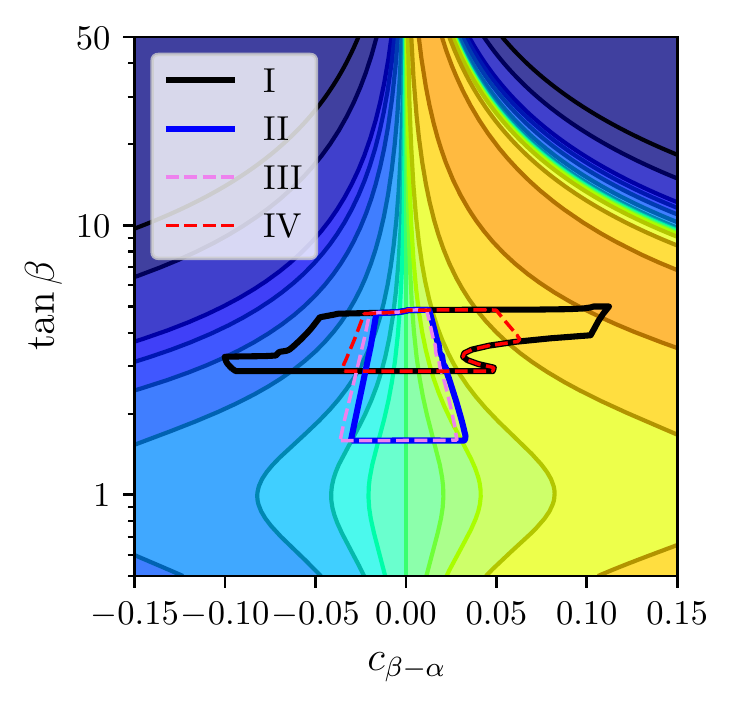}
\includegraphics[height=0.25\textheight]{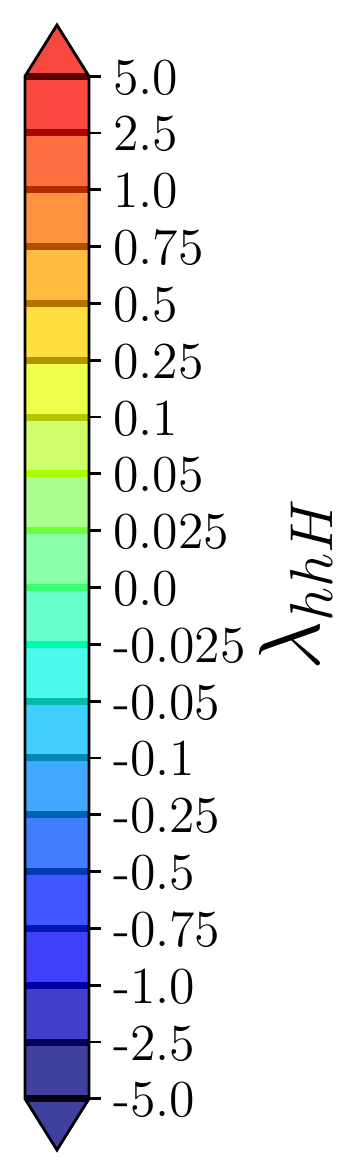}

\includegraphics[height=0.25\textheight]{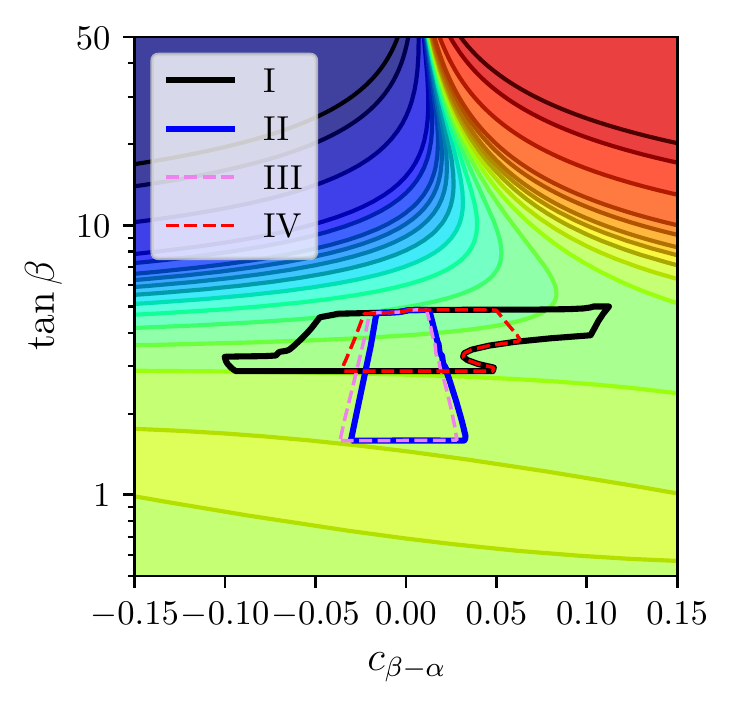}
\includegraphics[height=0.25\textheight]{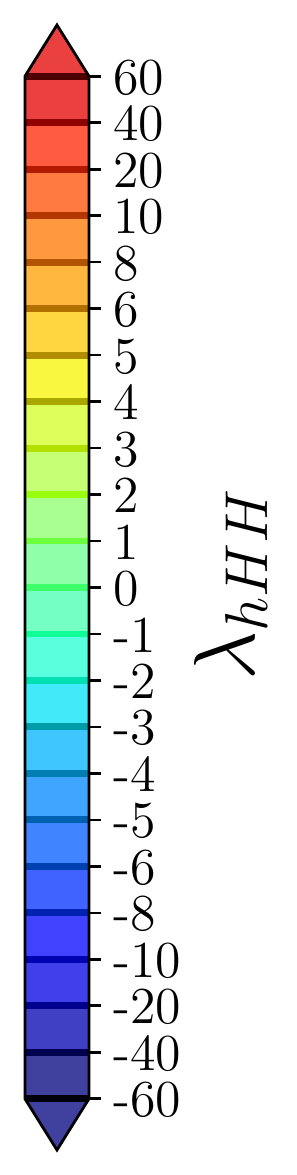}
\includegraphics[height=0.25\textheight]{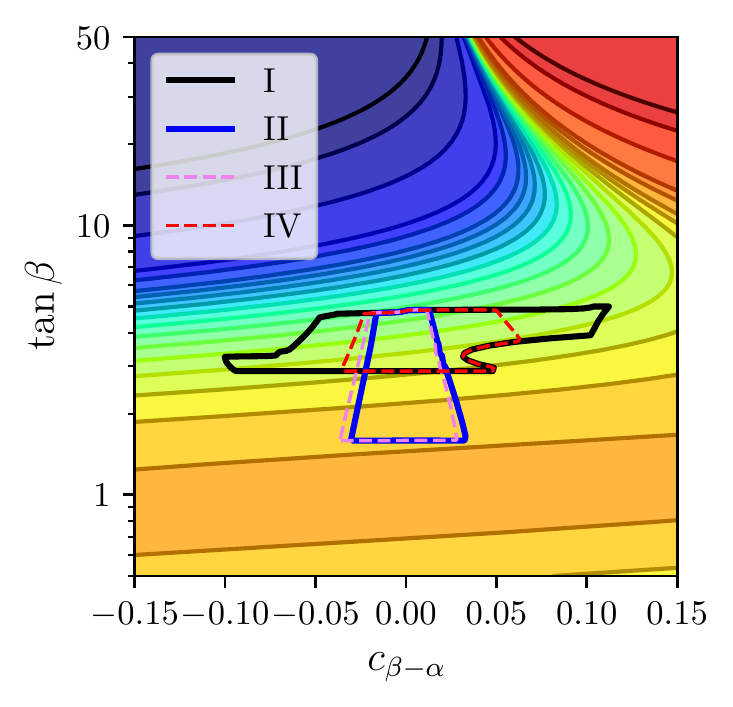}
\includegraphics[height=0.25\textheight]{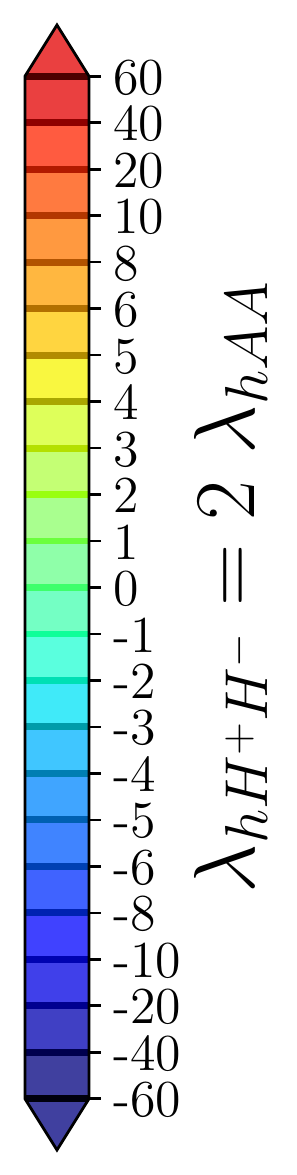}
\caption{\textbf{(B)} Triple Higgs couplings in benchmark scenario~1
  in the $\CBA$--$\tb$ plane
  with $m = 550 \gev$ and $\msq = 60000 \gev^2$. Shown are
  $\kala := \lahhh/\laSM$ (upper left), $\lahhH$ (upper right),
  $\lahHH$ (lower left) and $\lahHpHm = 2 \lahAA$ (lower
  right). Indicated by the interior of lines are the allowed regions
  for type~I (solid black), type~II (solid blue), type~III (dashed
  pink), type~IV (dashed red).}
\label{fig:bench-cba-tb}
\end{figure}

We start our comparison in \reffi{fig:bench-cba-tb}(A) with benchmark
scenario~1, i.e.\ the $\CBA-\tb$ plane with $m = 550 \gev$ and
$\msq = 60000 \gev^2$. The largest differences between the four 2HDM
types can be observed in the first row, where we show the restrictions
from the LHC data
based on the BSM Higgs-boson searches, obtained via \HB, and on the
Higgs-boson rate measurements, obtained via \HS.
Concerning the latter, very
roughly speaking, one observes that type~I has the ``largest
allowed'' parameter space, and type~II resembles type~III. Both can be
explained by the couplings of the various Higgs bosons to fermions as
specified in \refta{tab:coupling} as follows. Overall, it
can be observed that the  parameter space is strongly constrained
for $\CBA$ to be close to the alignment limit, such that $h$
behaves sufficiently SM-like. In particular, the $2\sig$ allowed regions
for the Yukawa type~II and~III (2nd and 3rd column) are substantially
smaller compared to type~I (left column). This is caused by an
enhancement of the coupling of $h$ to $b$-quark (see
\refta{tab:coupling}) in these two types. For type~IV the restrictions
are caused by the enhanced coupling of the $h$ to $\tau$-leptons
(which is also present in type~II). 
As $\tb$ increases the types~II, III and~IV are forced to be very
close to the alignment limit to agree with the experimental data.  
For type~I the constraints are weaker, specially for $\tb>3$, where we
can accommodate inside the $2\sigma$ region values for $\CBA$ 
between -0.35 and 0.25
when $\tb\sim6$.  For very large values of $\tb$ the
restrictions in type~I depend strongly on the chosen value of $\msq$,
as has been discussed in \citere{Arco:2020ucn}. 
In this benchmark scenario, having a fixed value of $\msq$, 
(even) in the alignment limit there is an upper
limit on $\tb$. This is caused by the charged Higgs contribution to
$\Ga(h \to \ga\ga)$. The $hH^+H^-$ coupling has a contribution that
scales with $\msq\,\tb$, such that for fixed $\msq$ extremely large
loop contributions and thus extremely large values of $\br(h \to \ga\ga)$
are reached, which are in disagreement with the LHC measurements.
On the other hand, in all
four types the region allowed by Higgs-boson rate measurements extends
to very large values of $\tb$ for $\CBA = 0$ and $\msq = 0$.
It is interesting to observe 
that in the type~IV analysis a new allowed branch appears in the
upper right 
part of the plot which corresponds to $\xi_h^d = -\xi_h^l = 1$, known as the
\textit{wrong sign} Yukawa region.  
The explicit expression for $\tb$ in this limit, only valid if $\CBA>0$,
is given by  
\begin{equation}
	\tb=\frac{1+\SBA}{\CBA}=\frac{\CBA}{1-\SBA}.
	\label{eq:wrongsign}
\end{equation}

We now turn to the regions allowed by BSM Higgs-boson searches,
shown in blue.
The various exclusion bounds are directly related to the Higgs-boson
couplings in the respective Yukawa type, as summarized in
\refta{tab:coupling}. The coupling of the heavy Higgs bosons to
top-quarks in all four types become large for small
$\tb$. Consequently, all four types possess a lower limit for
$\tb \sim 1.5$ (with the value of $\MHp = 550 \gev$ fixed) from the charged
Higgs-boson searches, channel~(e). For slightly larger $\tb$ and the
largest allowed $\CBA$ values, the search for $H \to VV$
(channel~(d)) becomes relevant, which is then superseded by the
channel~(f) $gg \to A \to Zh$. However, depending on the type, other
channels take over for larger $\tb$.  First the channels~(c) and~(b), via
the decay $H \to hh$,  become important. Most relevant,  however,  is
the search for $b\bar b \to H/A \to \tau\tau$,  which becomes important
for larger $\tb$ in type~II, where the production and decay both scale
with $\tb$.  Also for type~IV this channel becomes important, but only
for intermediate $\tb$, since the production channel here scales with
$1/\tb$, and only an ``island'' is excluded by channel~(g).
In type~I,  which can extend to larger $\tb$ values than the others,  a
different channel becomes relevant, $h \to \ga\ga$~(a),  see the
discussion on the Higgs signal rates above.  In types~II and~III for
larger $\tb$ and larger positive $\CBA$ also the channel~(h),
$h\to ZZ\to llll$ restricts the allowed parameter space. 
In type~III at very large $\tb$ the channel~(i),
$H \to \ga\ga$ becomes important due to an enhanced $HH^+H^-$
coupling.  Finally, in type~IV the channel~(j), $h \to \tau\tau$
restricts the allowed parameter space due to the enhanced Higgs
coupling to leptons in this Yukawa type.

The restrictions from flavor physics are discussed in the second row of
\reffi{fig:bench-cba-tb}(A). Again type~II and type~III strongly resemble
each other, and type~I is very similar to type~IV.
In general, all the four types of the 2HDM exhibit an excluded area
at low $\tb$ values, $\tb \lsim 1$. The most constraining observable in
this low $\tb$ region is $\br(B \to X_s \ga)$ in the  types~I and~IV, 
and $\br(B_s \to \mu\mu)$ in the types~II and~III.
These similarities in the allowed areas
of types~I and~IV, on one hand, and those of types~II and~III, on
the other hand, are due to the dominant loop effects in
flavor observables involving the $H^\pm$. Its couplings to
fermions, given in terms of $\xi_{A}^{u}$ and $\xi_{A}^{d}$, are the same
in type~I and type~IV as well as type~II and type~III.  
Furthermore,
the case of type~II shows a peculiarity at large $\tb$, where a region
appears constrained from $B_s \to \mu\mu$.
This is due to the large contributions
from the Higgs-penguin loops that are mediated by the neutral Higgs
bosons. These contributions are enhanced maximally in the Yukawa
type~II due to the involved coupling factors, $\xi_{h,H,A}^{d}$
and $\xi_{h,H,A}^{l}$, which all grow with $\tb$. 

The third row of \reffi{fig:bench-cba-tb}(A) discusses the restrictions
from unitarity and stability,  which is identical in all four types.
Both constraints disallow values of $\tb\gtrsim 5$,  since
$\la_1$, $\la_3$,  $\la_4,$ and $\la_5$,  present in \refeq{eq:scalarpot},
grow  with $\tb$, making it complicated to
fulfill the theoretical constraints.
It can also be seen how \refeq{eq:m12special},  plotted in solid purple, 
follows the boundary of the allowed region by the stability conditions
for $\CBA>0$. 
On the other hand, for negative values of $\CBA$,
\refeq{eq:m12special2}, plotted in dashed yellow, marks this boundary. 

The overall allowed parameter regions in the $\CBA$-$\tb$ plane in the
four 2HDM types is summarized in the last row of \reffi{fig:bench-cba-tb}(A)
as the interior of solid lines (in black, blue, pink and red).
In the benchmark scenario~1 the allowed region
extends more in $\CBA$ for type~I, whereas it it extends further down in
$\tb$ for types~II and~III. These allowed regions are now contrasted
with the predictions of the various triple Higgs couplings in
\reffi{fig:bench-cba-tb}(B).
In the upper left plot the prediction for
$\kala := \lahhh/\laSM$ is shown. By definition one finds $\kala = 1$
in the alignment limit, $\CBA = 0$.  Larger deviations from unity are
found for larger $|\CBA|$, and consequently, type~I naturally features
larger deviations from the SM. The situation is similar for $\lahhH$,
as shown in the upper right plot. This coupling goes to zero in the
alignment limit, and larger positive (negative) values are found for
larger positive (negative) values of $\CBA$. Consequently, also for
this coupling type~I allows for the largest values of $|\lahhH|$
(reached for $\tb = 3$ in this benchmark plane).
The situation is reversed for the trilinear couplings involving two
heavy Higgs bosons, as shown in the lower row of
\reffi{fig:bench-cba-tb}, with $\lahHH$ on the left and
$\lahHpHm = 2 \lahAA$ on the right. Larger variations of these couplings are
found (in the allowed regions) for a variation of $\tb$
(this pattern changes for $\tb$ values somewhat higher than in the
allowed regions). Consequently, the largest values are found 
in Yukawa types~II and~III in the lowest allowed $\tb$ region
in this benchmark  plane.
On the other hand, it is interesting to note that the
behavior of $\lahHpHm$ in the region $\tb \gsim 10$ correlates
with the parameter space allowed by \HS\ in 
Yukawa type~I. As discussed above, this is due to the charged Higgs
contribution to $\Ga(h \to \ga\ga)$. Also the other three types would
exhibit the same feature, but other constraints already constrain the
allowed parameter space to lower $\tb$ and smaller $|\CBA|$.

\begin{figure}[t!]
\centering
\includegraphics[width=0.225\textwidth]{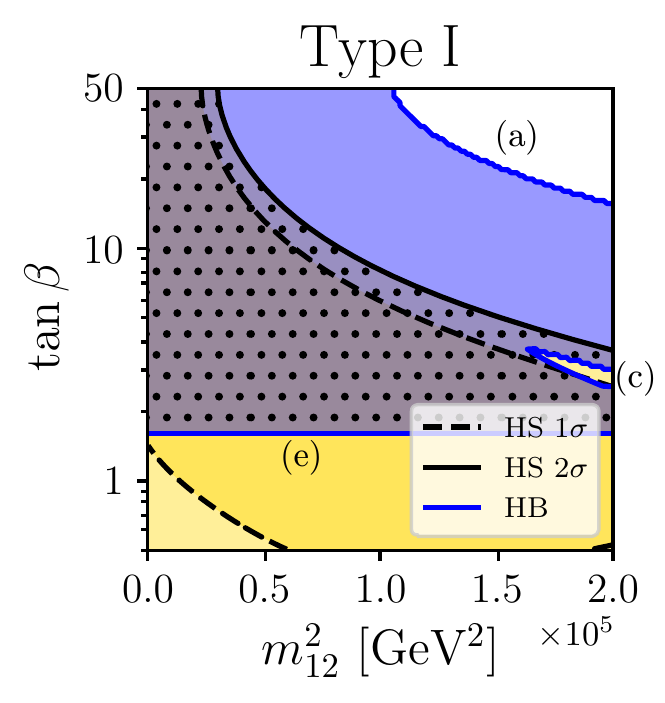}
\includegraphics[width=0.225\textwidth]{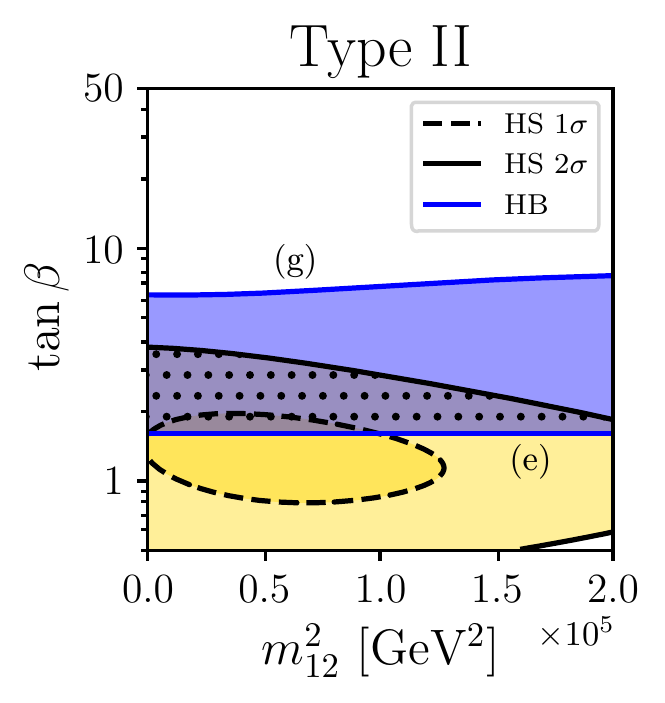}
\includegraphics[width=0.225\textwidth]{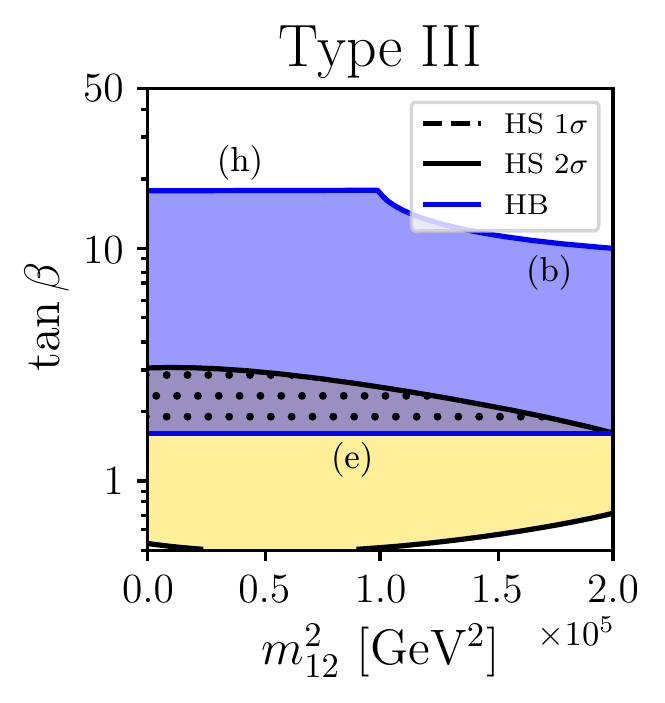}
\includegraphics[width=0.225\textwidth]{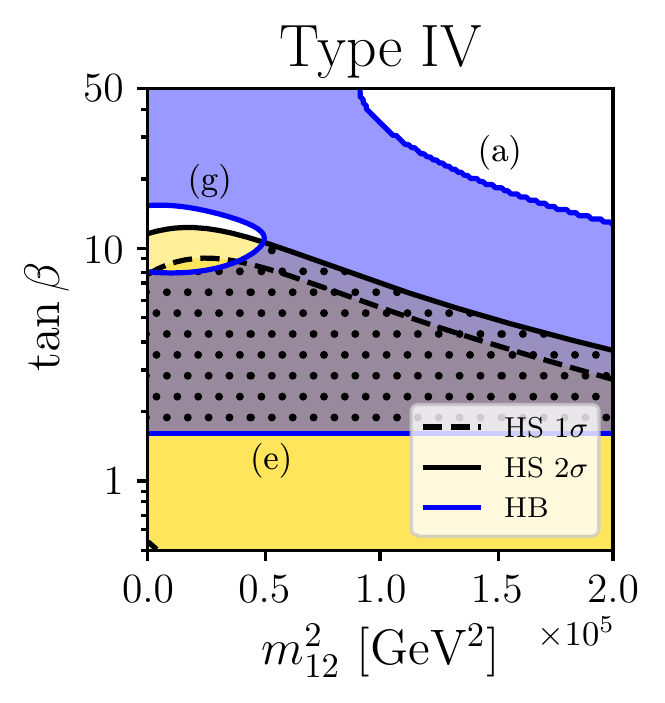}

\includegraphics[width=0.225\textwidth]{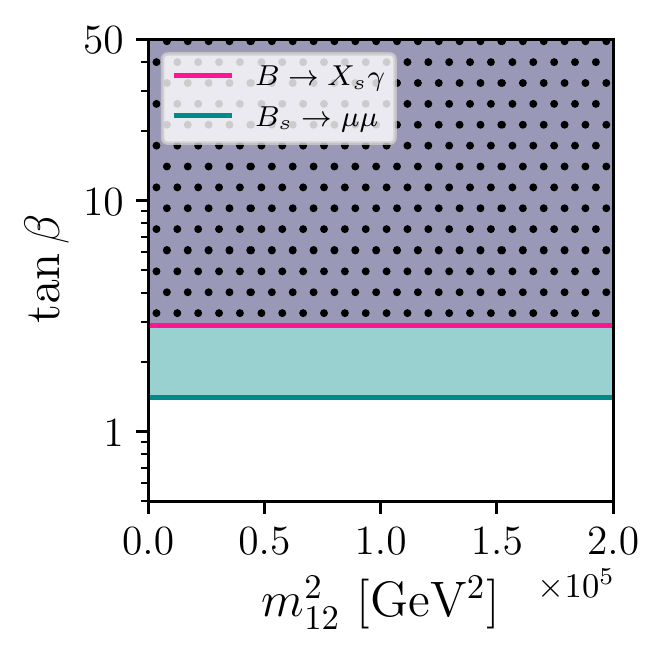}
\includegraphics[width=0.225\textwidth]{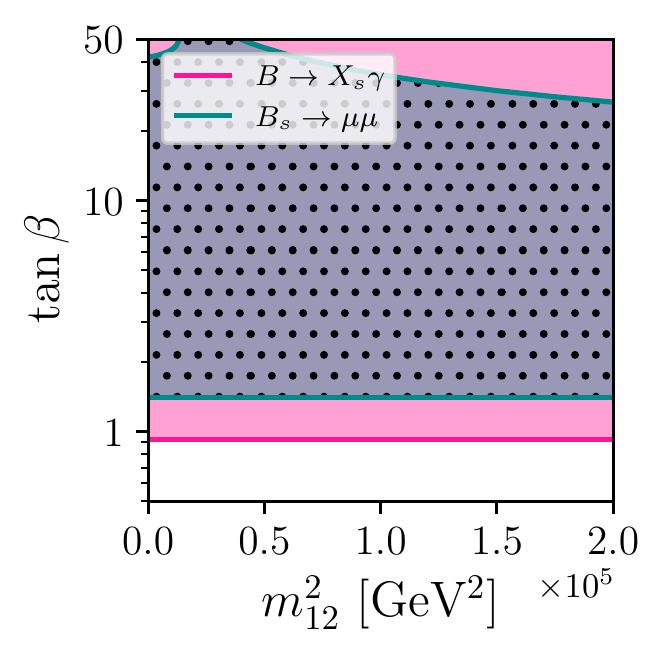}
\includegraphics[width=0.225\textwidth]{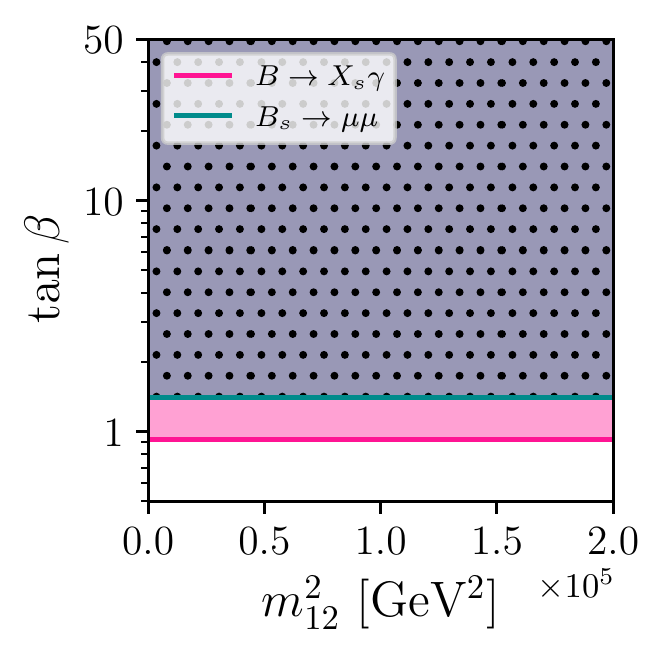}
\includegraphics[width=0.225\textwidth]{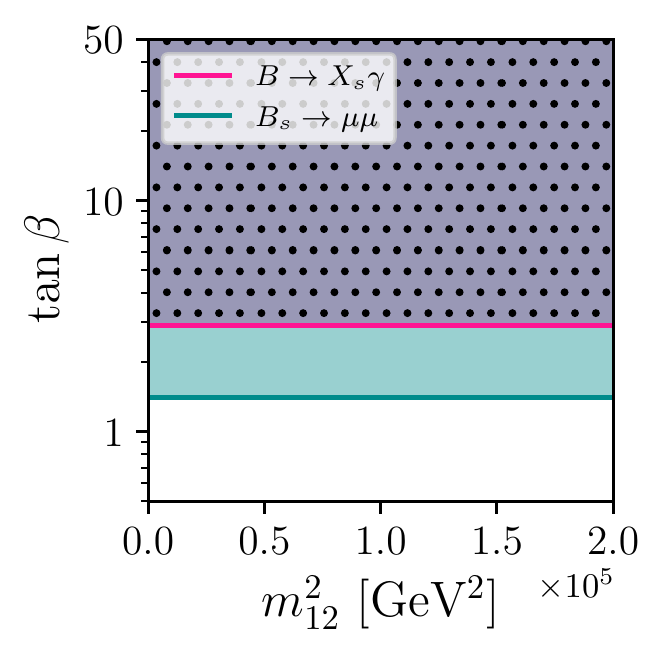}

\includegraphics[width=0.225\textwidth]{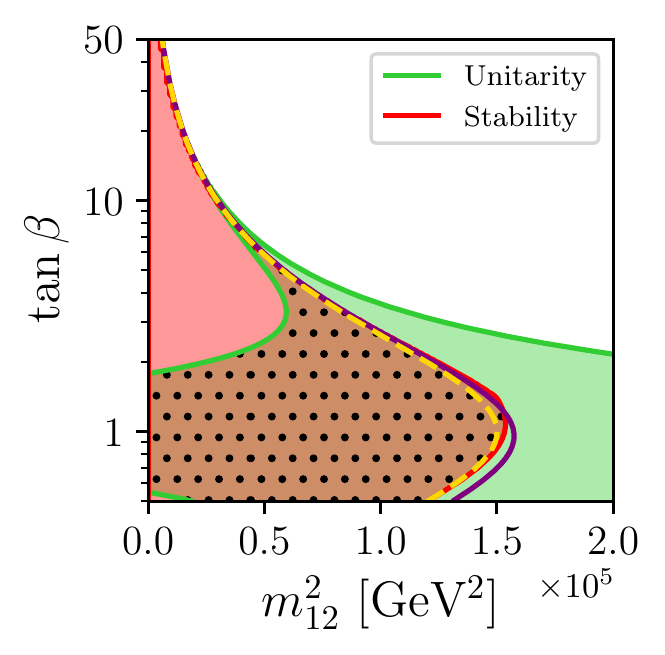}
\includegraphics[width=0.225\textwidth]{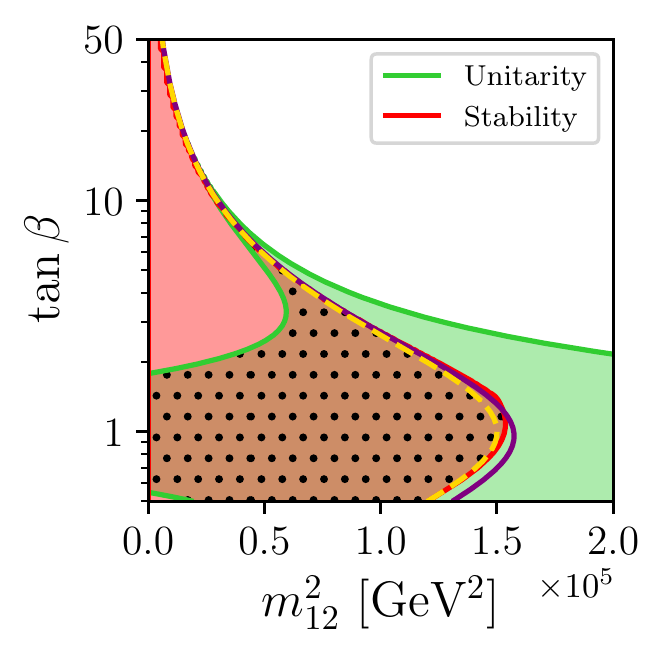}
\includegraphics[width=0.225\textwidth]{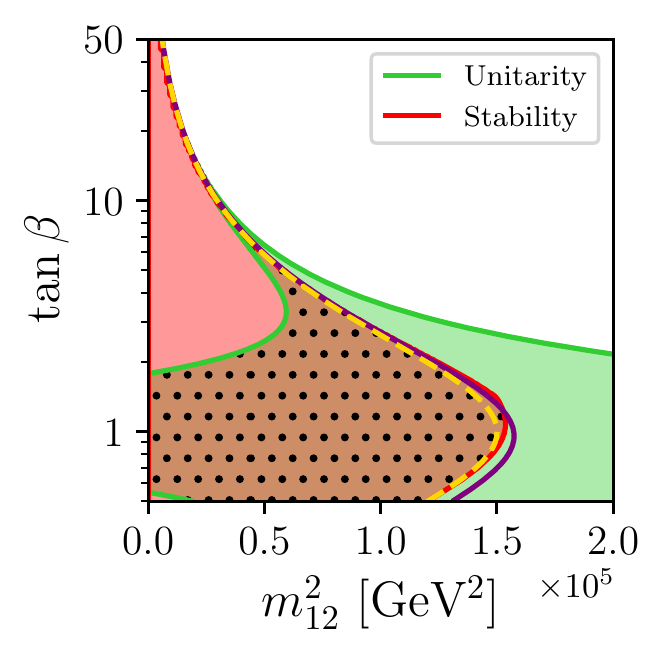}
\includegraphics[width=0.225\textwidth]{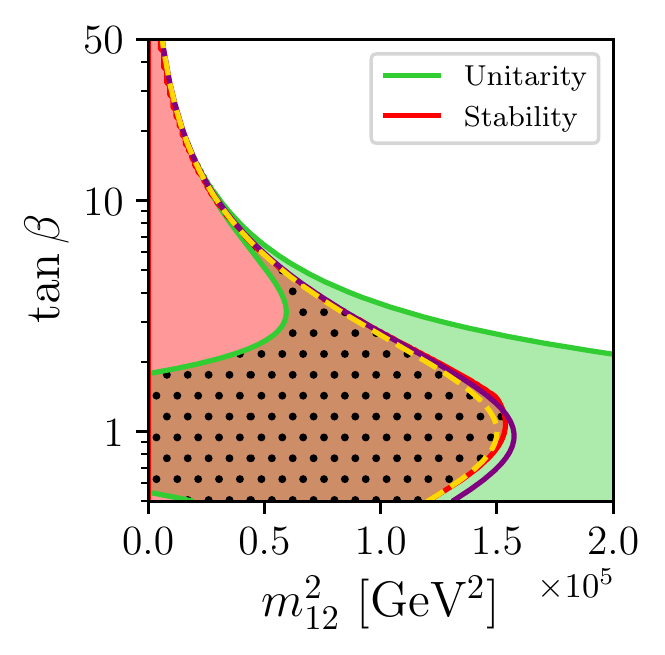}

\includegraphics[width=0.225\textwidth]{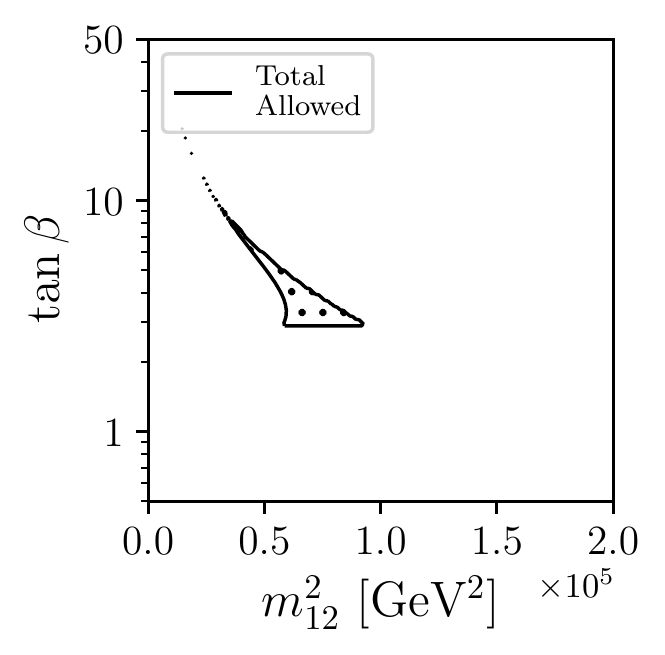}
\includegraphics[width=0.225\textwidth]{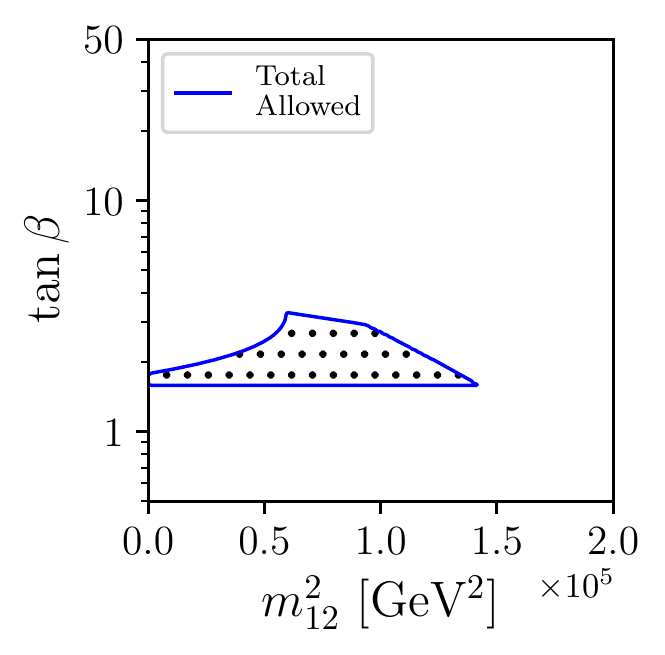}
\includegraphics[width=0.225\textwidth]{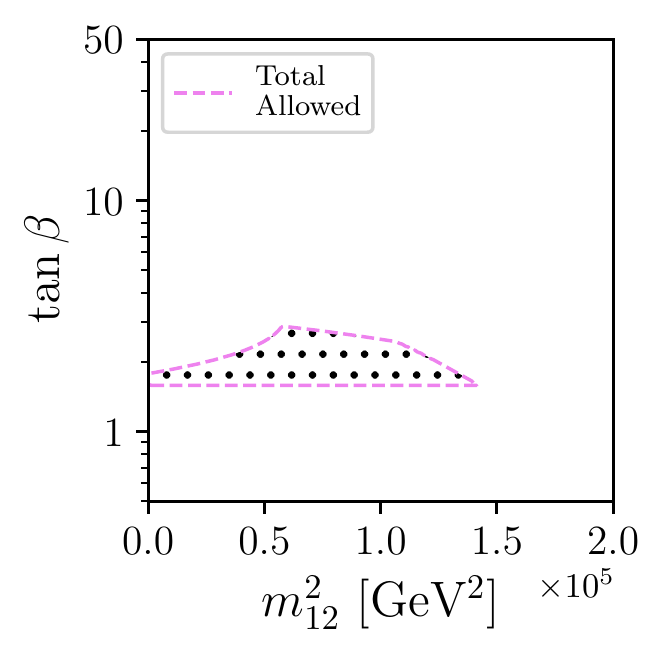}
\includegraphics[width=0.225\textwidth]{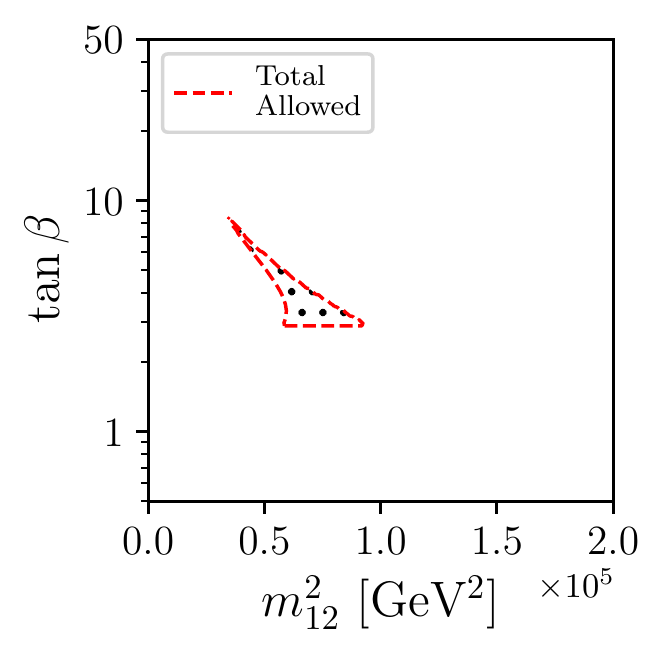}
\caption{\textbf{(A)} Allowed regions from the restrictions on the
  parameter space in benchmark 
  scenario~2 in the $\msq$--$\tb$ plane
  with $m = 550 \gev$ and $\CBA = 0.02$. The results
for type~I, II, III and~IV in the left, second, third and right
column, respectively. The upper, second and third row show the
restrictions from \HB/\HS, the flavor observables and from
unitarity/stability, respectively. The fourth row indicates
the regions allowed by all constraints in the respective scenario.}
\end{figure}
\begin{figure}[t!]\ContinuedFloat
\centering
\includegraphics[height=0.25\textheight]{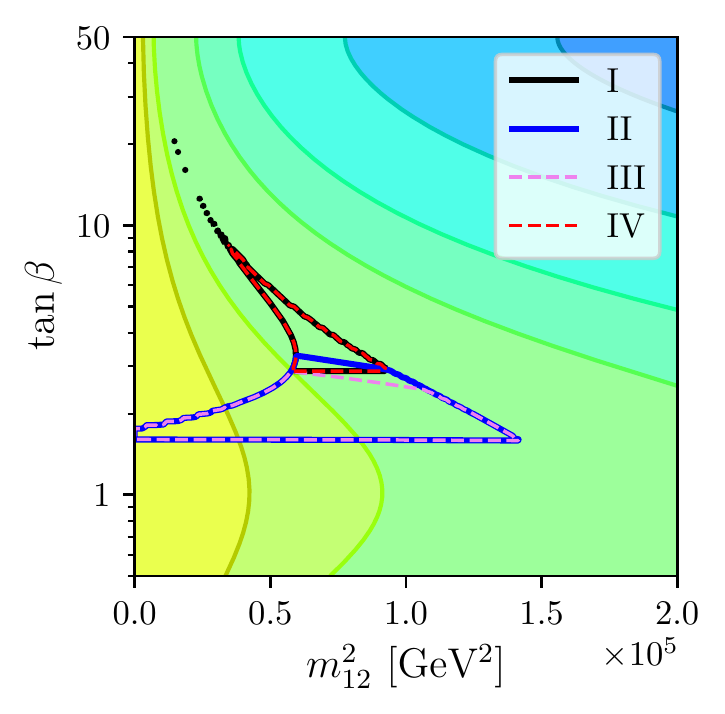}
\includegraphics[height=0.25\textheight]{h1h1h1_colorbar_vertical}
\includegraphics[height=0.25\textheight]{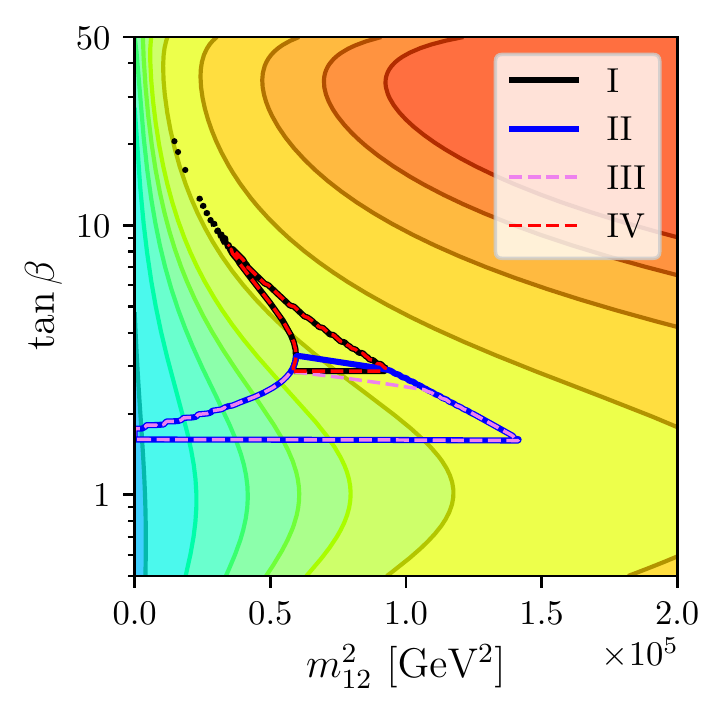}
\includegraphics[height=0.25\textheight]{h1h1h2_colorbar_vertical}

\includegraphics[height=0.25\textheight]{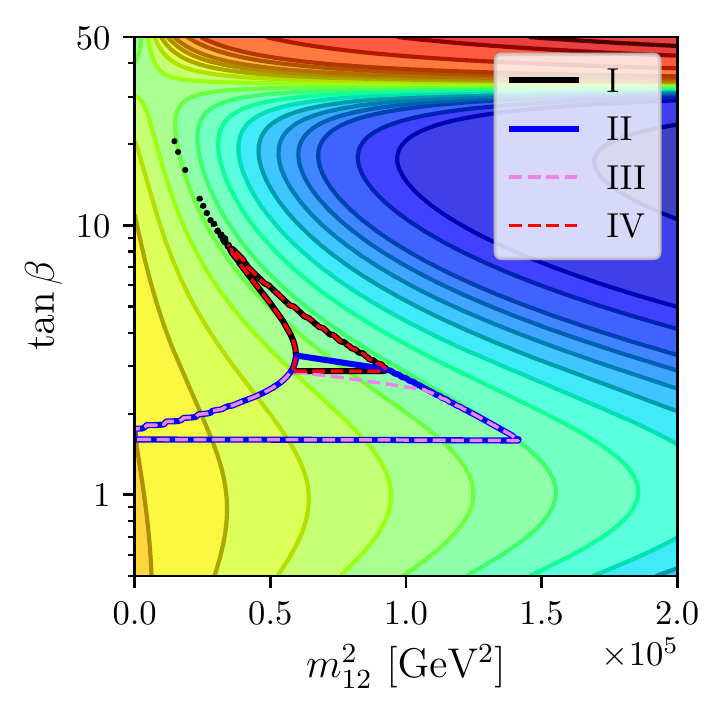}
\includegraphics[height=0.25\textheight]{h1h2h2_colorbar_vertical}
\includegraphics[height=0.25\textheight]{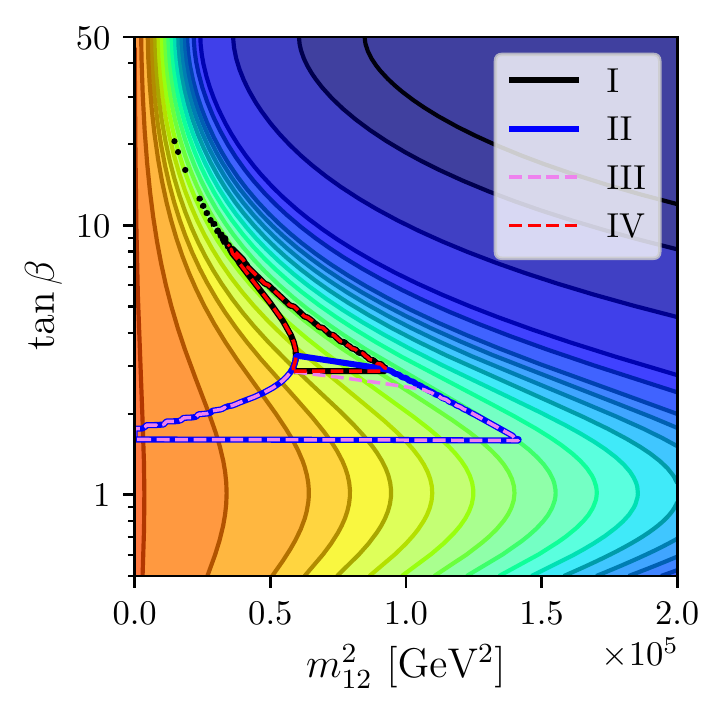}
\includegraphics[height=0.25\textheight]{h1HpHm_colorbar_vertical}
\caption{\textbf{(B)} Triple Higgs couplings in benchmark scenario~2
  in the $\msq$--$\tb$ plane
  with $m = 550 \gev$ and $\CBA = 0.02$. Shown are
  $\kala := \lahhh/\laSM$ (upper left), $\lahhH$ (upper right),
  $\lahHH$ (lower left) and $\lahHpHm = 2 \lahAA$ (lower
  right). Indicated by the interior of lines are the allowed regions
  for type~I (solid black), type~II (solid blue), type~III (dashed
  pink), type~IV (dashed red).}
\label{fig:bench-m12-tb}
\end{figure}

\medskip
The next set of comparisons of the four 2HDM types, in benchmark
scenario~2, in the $\msq$--$\tb$ plane is presented in
\reffi{fig:bench-m12-tb}. The overall mass scale is fixed to $m = 550 \gev$,
and $\CBA = 0.02$, i.e.\ the decoupling limit is explicitly excluded
from this benchmark. As in the first benchmark scenario, the
parameter space allowed by the Higgs-boson rate measurements, shown in
yellow in the first row of \reffi{fig:bench-m12-tb}(A) is largest for
type~I and similar for type~II and~III. In all four types the lowest
$\tb$ values of $\tb \sim 0.5$ are allowed, where in type~II and~III the
largest $\msq$ values shown in combination with very low $\tb$ are
excluded, which can be traced back to $h \to \ga\ga$.
Going to larger $\tb$, the
upper limit in type~I, and largely also in type~IV, is given by the
charged Higgs-boson contribution to $\Ga(h \to \ga\ga)$, see the
discussion of benchmark~1. In type~II and~III the upper limit is
encountered already for lower $\tb$ values, where the enhancement of the
$hb\bar b$ coupling becomes stronger in these two types.

Concerning the searches for BSM Higgs bosons, at low $\tb$ the same
pattern as in benchmark~1 is observed. The coupling of the heavy Higgs
bosons to top-quarks in all four types become large for small
$\tb$. Consequently, all four types possess a lower limit for
$\tb \sim 1.5$ (and $\MHp = 550 \gev$) from the search for charged
Higgs-boson searches, channel~(e). However, the four types differ
substantially in their upper $\tb$ limits. In type~I all couplings of
the heavy Higgs bosons to SM fermions decrease with increasing $\tb$,
yielding a large allowed parameter space. The limit then comes from the
too large rate in $\br(h \to \ga\ga)$, channel~(a). For intermediate
$\tb$ and large $\msq$ also the channel~(c), $H \to hh$, plays a minor
role. The situation is completely different in type~II, where $\tb \gsim 6$
is excluded from the ``classical'' search channel $H/A \to \tau\tau$ for
this Yukawa type. In type~III the situation is again different. For
smaller $\msq$ at $\tb \sim 19$ the channel~(h), $h \to ZZ \to llll$,
becomes important. For these large values of $\tb$ the $hb\bar b$
coupling is reduced substantially in type~III and,  becomes~0 for
$\xi_{h}^{d}(\mbox{type III}) = \SBA - \CBA \tb = 0$, see
\refta{tab:coupling}. For the
chosen value of $\CBA = 0.02$ this is reached for
$\tb \sim 50$. Thus, an increase in $\tb$ yields a decrease of
$\Ga(h \to b \bar b)$ and correspondingly an increase of
$\br(h \to ZZ \to llll)$, where the experimental bound is reached for
$\tb \sim 19$. Going to
larger $\msq$ the $H \to hh$ channel~(b) takes over. Type~IV, because
of its Yukawa structure, is restricted at high $\tb$ from $\br(h \to \ga\ga)$,
channel~(a). However, for small $\msq$, as in benchmark~1, for
intermediate $\tb$ values the $H/A \to \tau\tau$ channel becomes strong,
where the same interplay as described for benchmark~1 takes
place. Consequently, also in benchmark~2 type~IV exhibits a ``hole'' in
the allowed parameter space at $\tb \sim 10$. Overall, the lower limits
on $\tb$ are set by the charged Higgs-boson searches, which are
effectively the same in the four types. On the other hand, the upper
limits are given by the Higgs-boson rate measurements, resulting in
higher $\tb$ limits in type~I and~IV, and in quite low limits in type~II
and~III.

The restrictions from flavor physics are discussed in the second row of
\reffi{fig:bench-m12-tb}(A). Again type~II and type~III strongly resemble
each other, and type~I is very similar to type~IV in the low $\tan\beta$ 
region,  again because the coupling of $H^\pm$ to quarks is identical in
both cases.   
For types~I and~IV,  $B\to X_s\gamma$ disallows $\tb < 3$, whereas 
for types~II and~III $B_s\to\mu\mu$ is the most constraining observable
setting the limit on $\tb \gsim 1$.
In addition,  for type~II we see again a disallowed region for large
$\tb$ and $\msq$,  originating from the effect of the Higgs
mediated penguin diagrams in $B_s\to\mu\mu$.

The third row of \reffi{fig:bench-m12-tb}(A) shows the restrictions
from unitarity and stability, which are identical in all four types.
The largest allowed range for $\msq$ occurs at $\tb \sim 1$, where 
this parameter can reach values from 0 up to $1.5 \times 10^5 \gev^2$. 
For larger values of $\tb$ the region allowed by the unitarity constraints
narrows drastically, closing in to \refeq{eq:m12special} and
\refeq{eq:m12special2}, 
plotted in solid purple and dashed yellow respectively. 
Notice that these two equations provide contour lines in this plane that are at 
the boundaries of the allowed region by the stability constraints which is also 
quite narrow at large $\tb$,  as can be seen in this figure.
If a value for $\CBA$ further from the alignment limit was chosen, 
the narrow region allowed by unitarity  shrinks even further and would separate from the allowed
region by the stability conditions.  In this case,  only
\refeq{eq:m12special} will enter in the extremely narrow allowed region by unitarity.
Furthermore, 
\refeq{eq:m12special2} is very close to upper bounds to $\msq$ set 
by the theoretical constraints for all $\tb$ values.  
Negative values of $\msq$ are disallowed by the condition
that requires the minimum of the potential to be a global minimum.

The overall allowed parameter regions in the $\msq$-$\tb$ plane in
benchmark~2 in the four 2HDM types are summarized in the last row of
\reffi{fig:bench-m12-tb}(A).
According to our discussion, the regions are similar for type~I and~IV,
as well as for type~II and~III. In Yukawa types~I and~IV the regions
extend for intermediate $\msq$ from $\tb \sim 3$ to $\tb \sim 10$.
Conversely, in type~II and~III the allowed regions extend from
$\msq = 0$ to $\msq \sim 150000 \gev^2$ and from $\tb \sim 1.7$ to
$\tb \lsim 3.5$. 
This complementarity results in equally complementary results for the
tripe Higgs couplings, shown in \reffi{fig:bench-m12-tb}(B). For $\kala$
only a value $\neq 1$ is allowed in types~I and~IV, although the
difference never exceeds~1\%. In types~II and~III, reaching to small
$\msq$, also $\kala = 1$  is almost  reached.  However, due to the choice
$\CBA = 0.02$, i.e.\ very close to the decoupling limit, $\kala$ is
bound to be close to unity. 
Correspondingly, for $\lahhH$ only relatively small values are found. 
In type~I and~IV values between~0.1 and~0.25 are found. In type~II
and~III, which allow to go to small $\msq$ and lower $\tb$ values,
also smaller $\lahhH$ are realized, which can become even negative.
Larger values of triple Higgs couplings are possible for $\lahHH$,
$\lahAA$ and $\lahHpHm$. However, the overall structure remains as for
the other triple Higgs couplings. The contours of the allowed regions
for type~I and type~IV somewhat follow the iso-contours of the three
remaining triple Higgs couplings, while the allowed regions for
types~II and~III show larger allowed ranges for $\msq$ in a lower $\tb$
region. Values of $\sim 2$ and $\sim 4$ are found for $\lahHH$ and
$\lahAA = \lahHpHm/2$, respectively, in types~I and~IV. Values up to
$\sim 5$ and $\sim 10$, respectively, are found in types~II and~III,
where the largest values are found for $\msq = 0$. 
As in benchmark~1, it is interesting to note that the $\lahHpHm$ coupling for
large $\tb$ correlates with the parameter space allowed by \HS\ in
Yukawa type~I and~IV (due to the charged Higgs
contribution to $\Ga(h \to \ga\ga)$). 

\medskip
We finish our comparison of the four 2HDM Yukawa types with benchmark
scenario~3, shown in \reffi{fig:bench-m12-m}. In this scenario the 
input parameters are fixed to $\CBA = 0.01$ and $\tb = 3$, and the
comparison is performed in the mass plane $\msq$--$m$. As discussed
above, the angles 
have been chosen to find larger regions in the parameter space in all
four types that are in agreement with the constraints. As will become
clear, in such a case the masses, contrary to the angles, play a
very similar role in the four Yukawa types.
As before, we start the discussion with the restrictions coming from the
Higgs-boson rate measurements at the LHC. In all four types the allowed
region goes from low $m$ and $\msq$ to $m \sim 800 \gev$ (depending
somewhat on the Yukawa type) for the largest analyzed $\msq$ values.
The allowed regions from BSM Higgs boson searches exhibit a richer
structure for $m \lsim 500 \gev$, but very roughly allow points with
$m \lsim 350 \gev$, with the exception of type~III, where values down to
$m \sim 200 \gev$ are allowed. This is mainly due to the absence of the
$H/A \to \tau\tau$ channel~(g) in this Yukawa type. The other relevant
channels in all four types are $H \to hh$~(b),(l) and $h \to \ga\ga$~(a).

\begin{figure}[t!]
\centering
\includegraphics[width=0.225\textwidth]{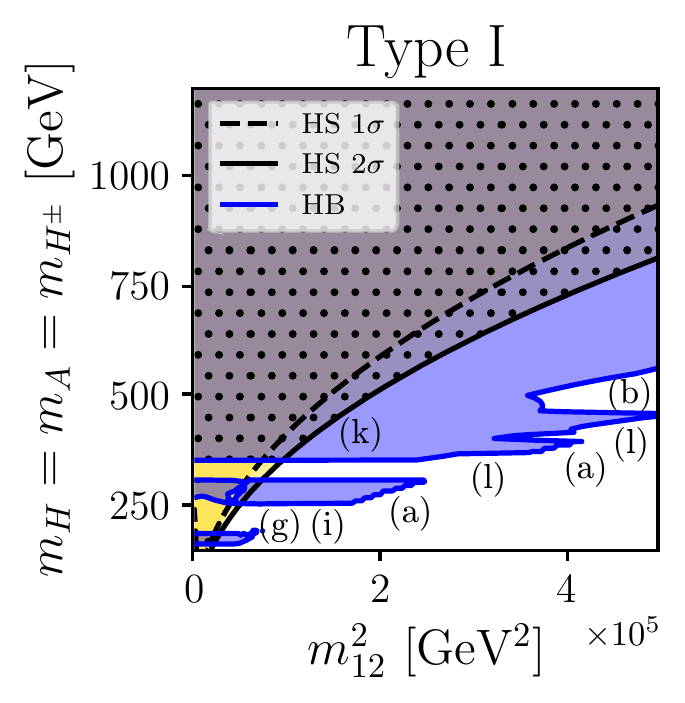}
\includegraphics[width=0.225\textwidth]{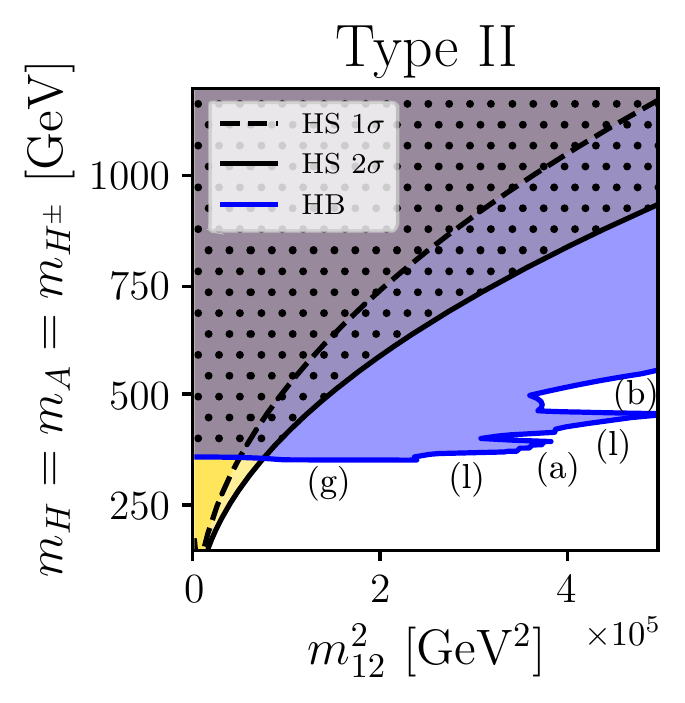}
\includegraphics[width=0.225\textwidth]{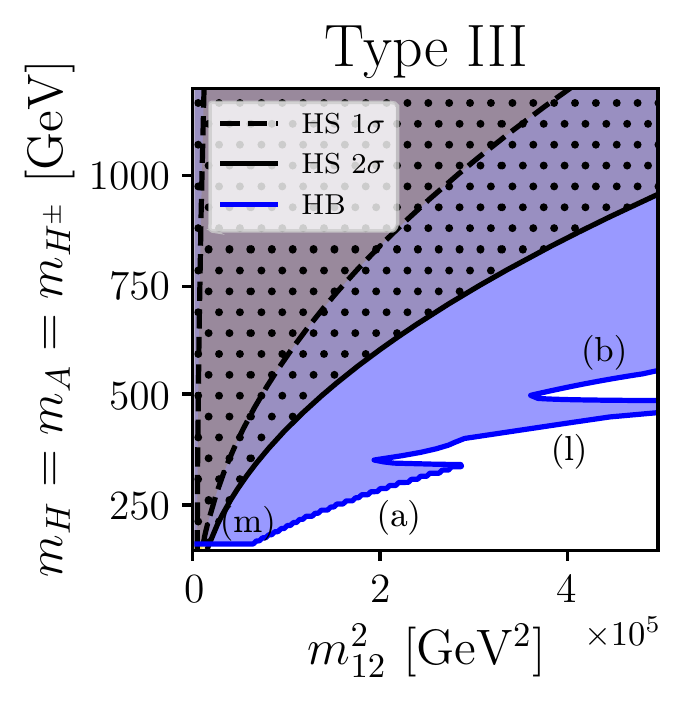}
\includegraphics[width=0.225\textwidth]{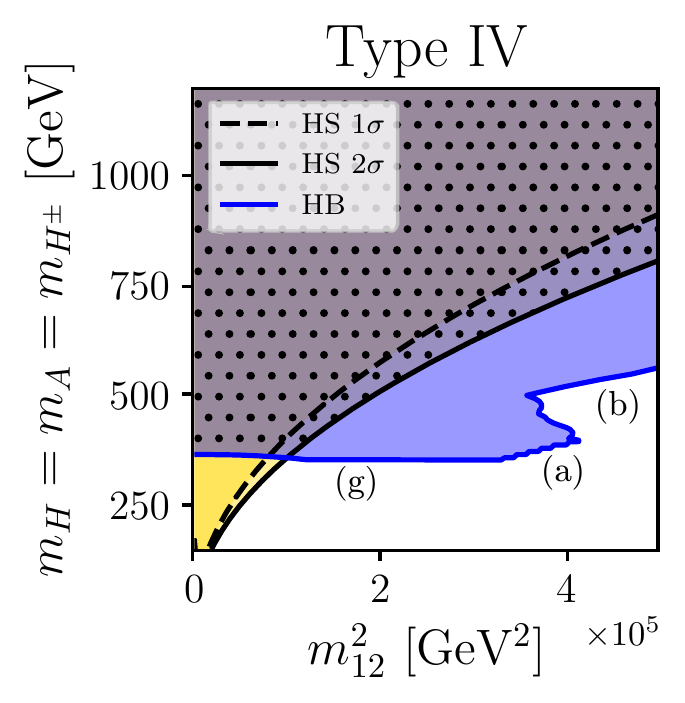}

\includegraphics[width=0.225\textwidth]{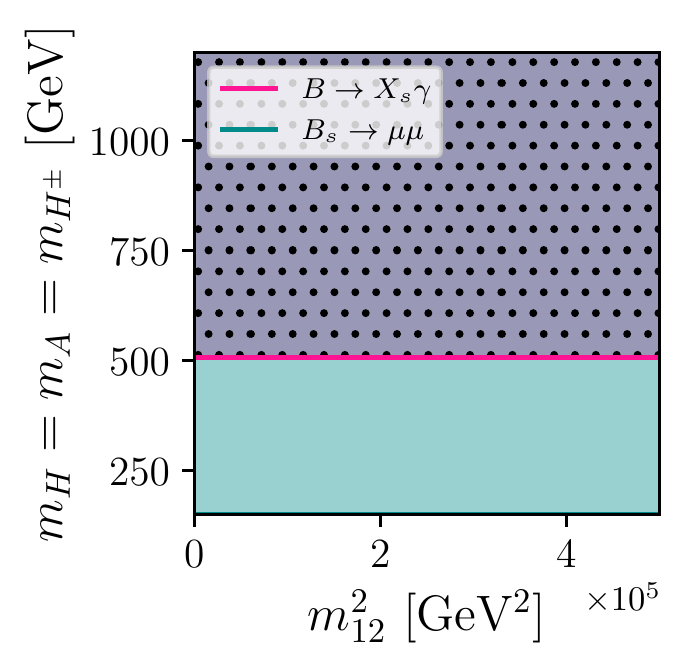}
\includegraphics[width=0.225\textwidth]{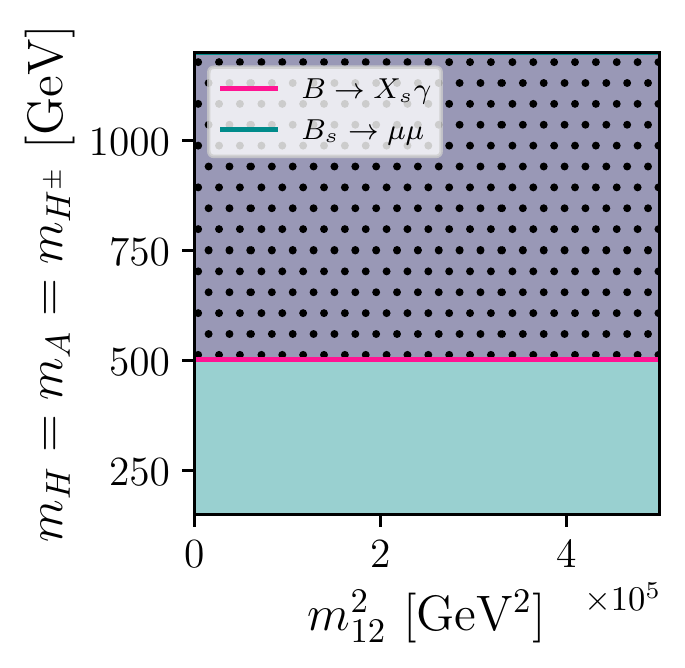}
\includegraphics[width=0.225\textwidth]{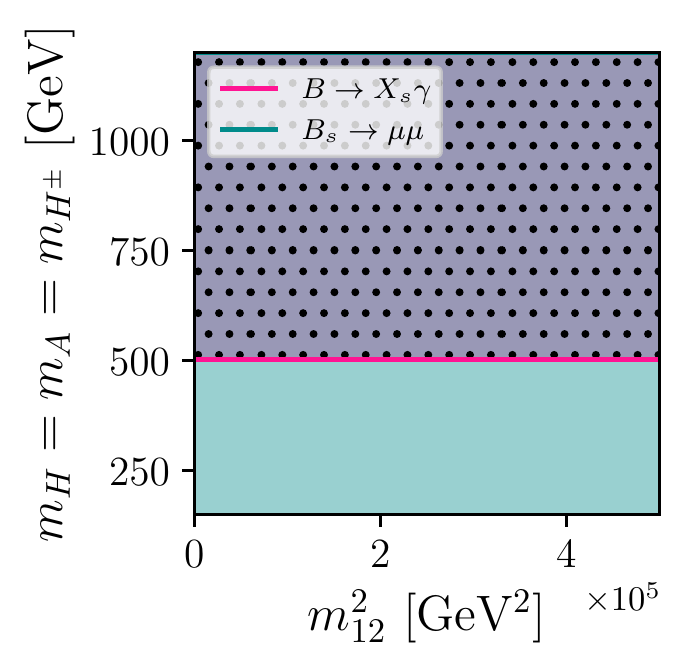}
\includegraphics[width=0.225\textwidth]{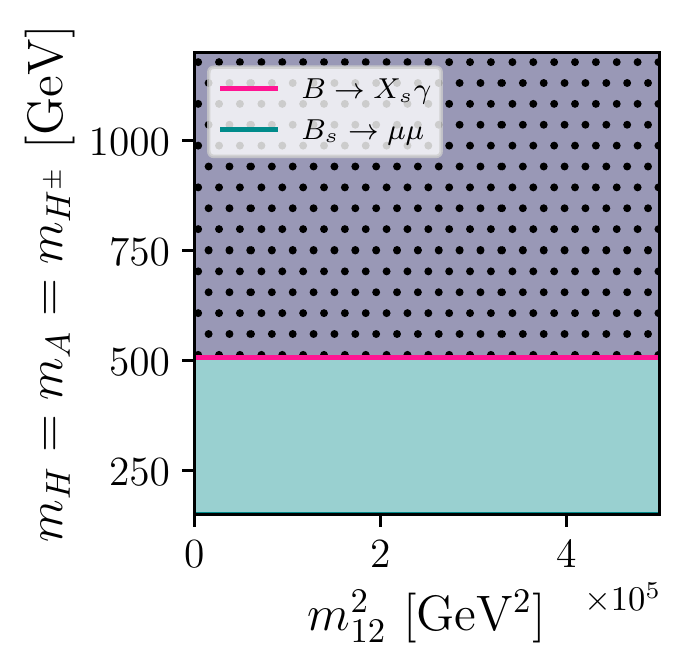}

\includegraphics[width=0.225\textwidth]{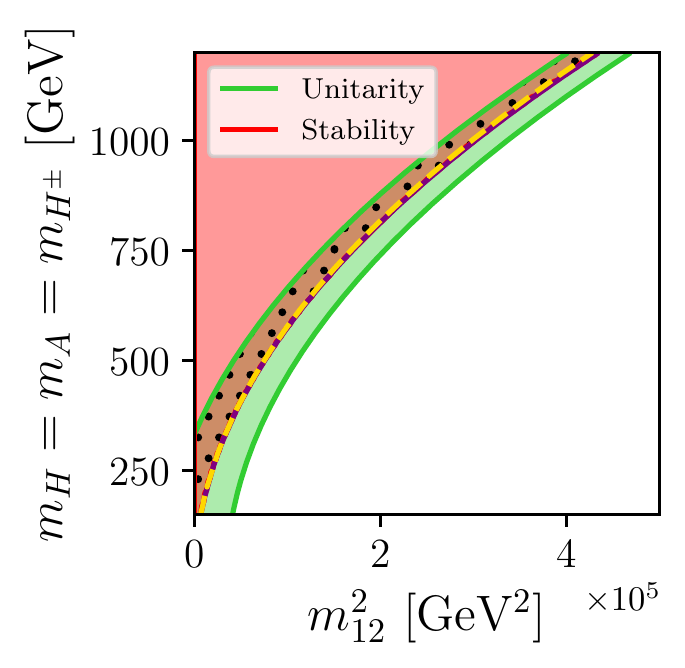}
\includegraphics[width=0.225\textwidth]{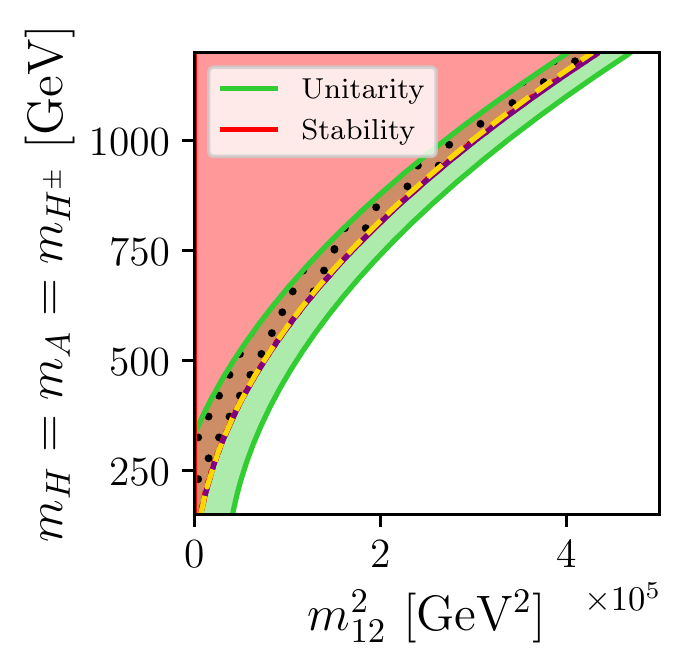}
\includegraphics[width=0.225\textwidth]{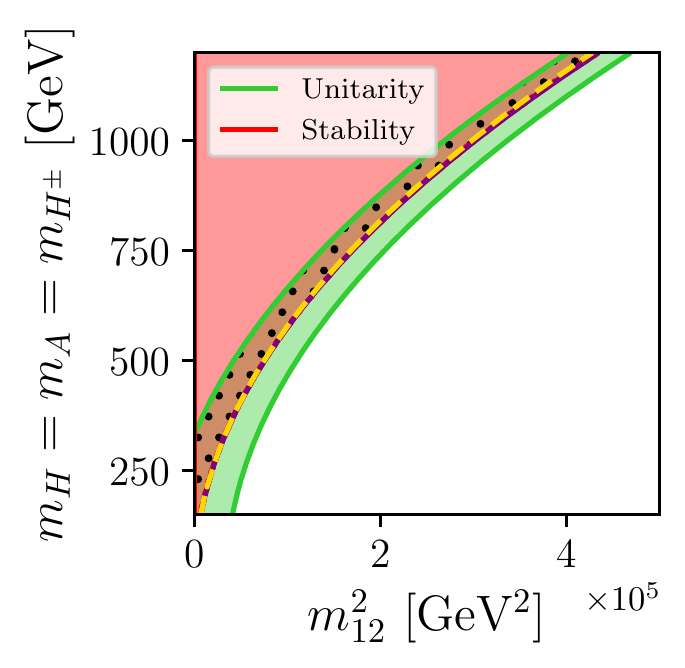}
\includegraphics[width=0.225\textwidth]{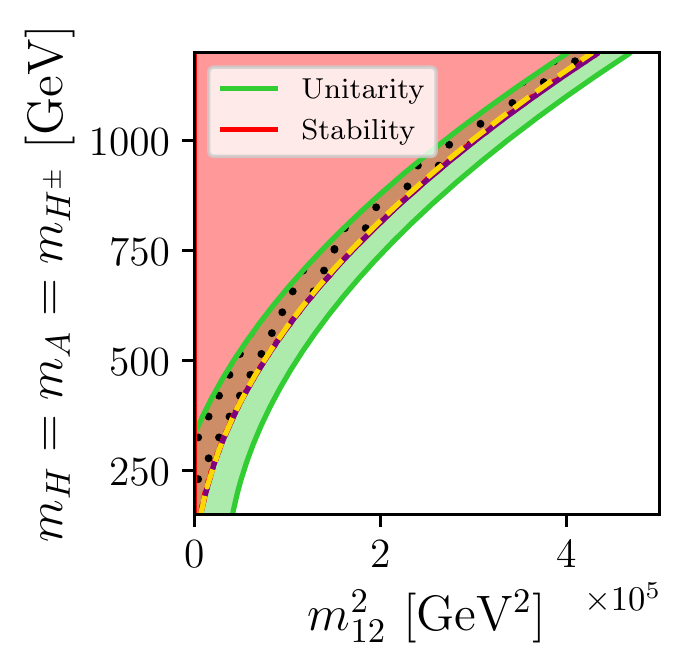}

\includegraphics[width=0.225\textwidth]{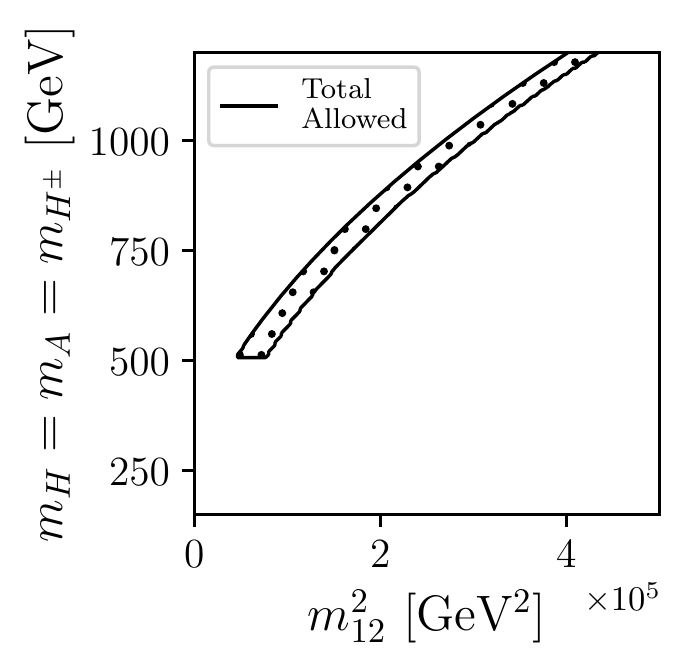}
\includegraphics[width=0.225\textwidth]{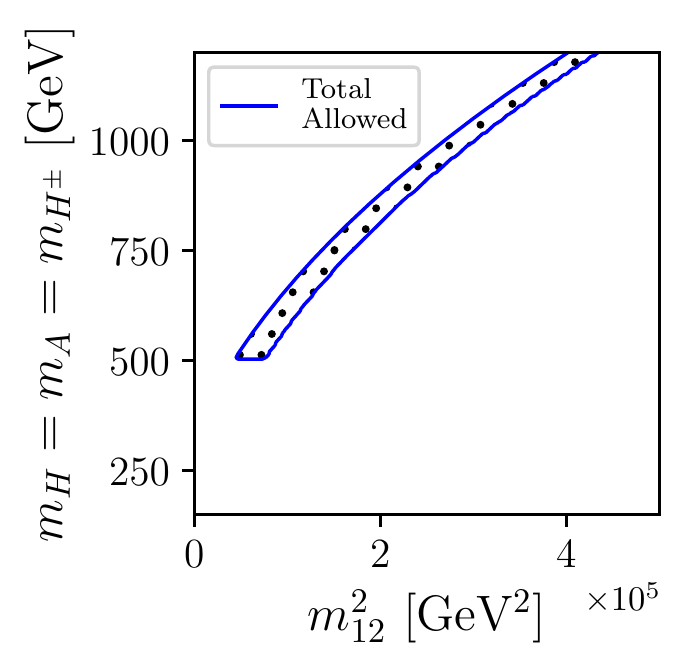}
\includegraphics[width=0.225\textwidth]{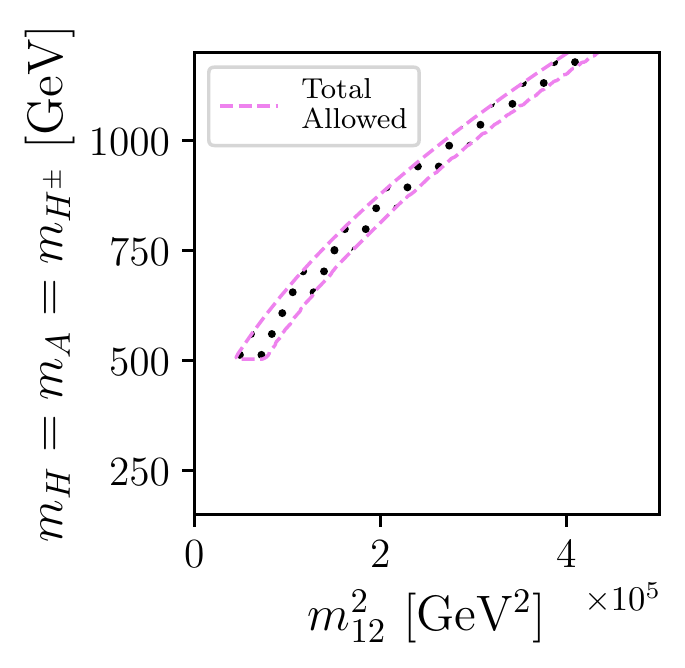}
\includegraphics[width=0.225\textwidth]{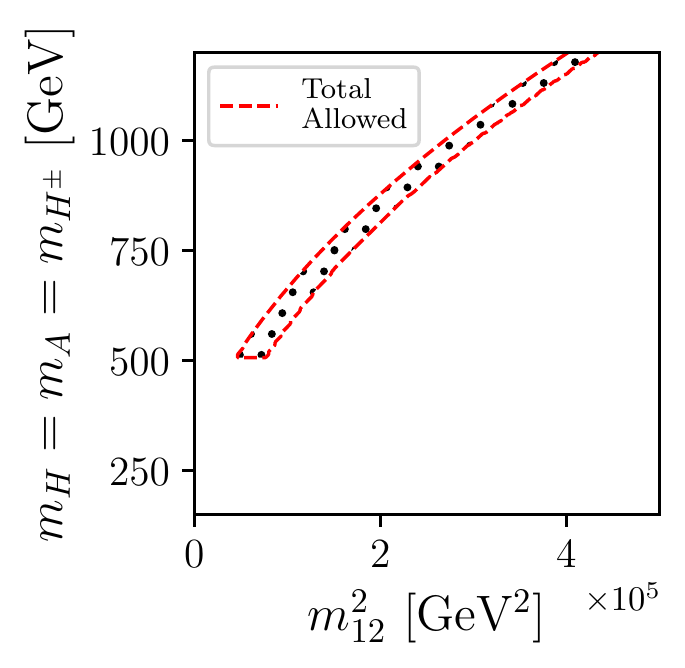}
\caption{\textbf{(A)} Allowed regions from the restrictions on the
  parameter space in benchmark 
  scenario~3 in the $\msq$--$m$ plane
  with $\tb = 3$ and $\CBA = 0.01$. The results
for type~I, II, III and~IV in the left, second, third and right
column, respectively. The upper, second and third row show the
restrictions from \HB/\HS, the flavor observables and from
unitarity/stability, respectively. The fourth row indicates
the regions allowed by all constraints in the respective scenario.}
\end{figure}
\begin{figure}[t!]\ContinuedFloat
\centering
\includegraphics[height=0.245\textheight]{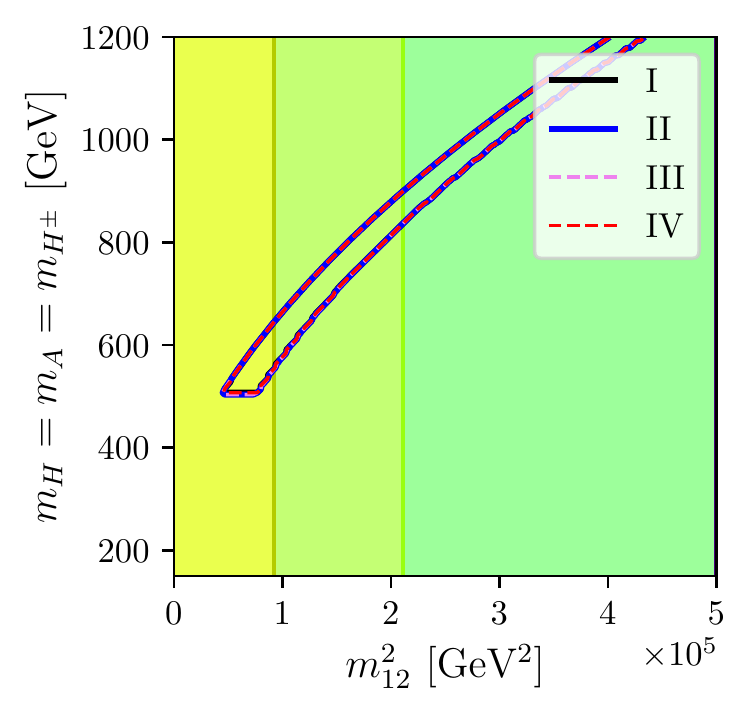}
\includegraphics[height=0.245\textheight]{h1h1h1_colorbar_vertical}
\includegraphics[height=0.245\textheight]{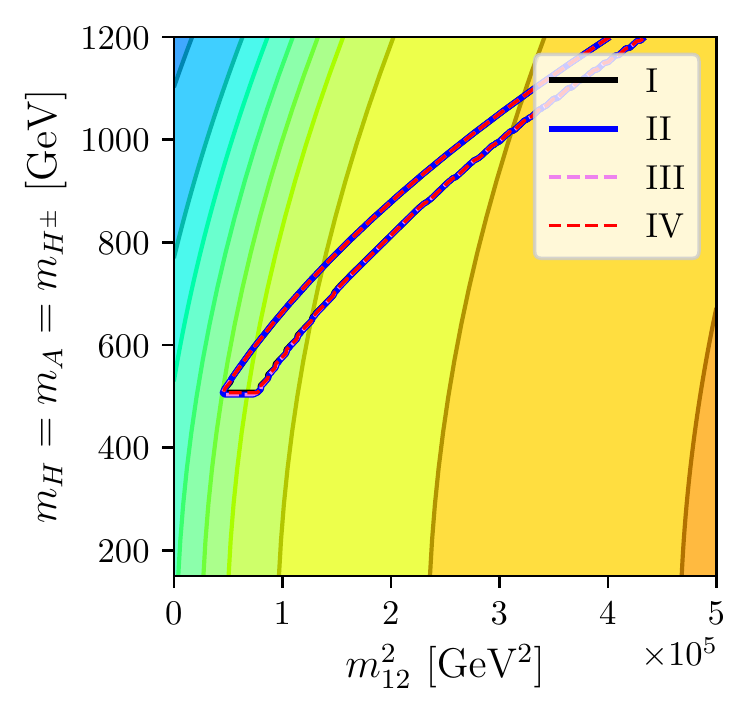}
\includegraphics[height=0.245\textheight]{h1h1h2_colorbar_vertical}

\includegraphics[height=0.245\textheight]{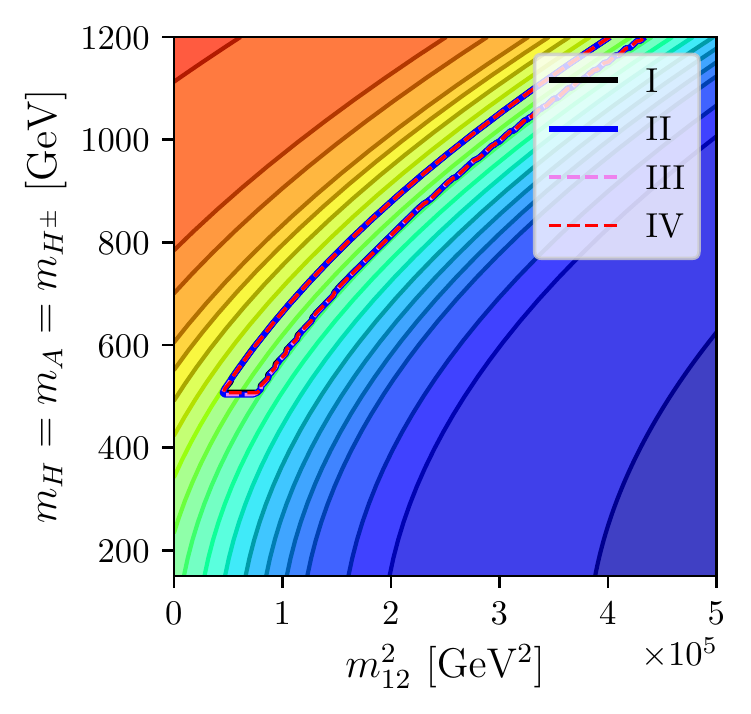}
\includegraphics[height=0.245\textheight]{h1h2h2_colorbar_vertical}
\includegraphics[height=0.245\textheight]{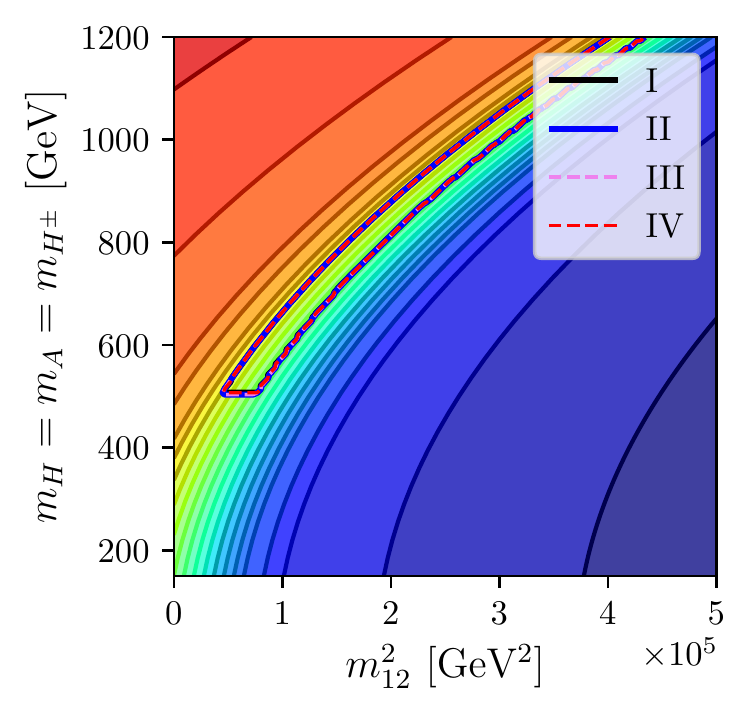}
\includegraphics[height=0.245\textheight]{h1HpHm_colorbar_vertical}
\caption{\textbf{(B)} Triple Higgs couplings in benchmark
  scenario~3 in the $\msq$--$m$ plane
  with $\tb = 3$ and $\CBA = 0.01$. Shown are
  $\kala := \lahhh/\laSM$ (upper left), $\lahhH$ (upper right),
  $\lahHH$ (lower left) and $\lahHpHm = 2 \lahAA$ (lower
  right). Indicated by the interior of lines are the allowed regions
  for type~I (solid black), type~II (solid blue), type~III (dashed
  pink), type~IV (dashed red).}
\label{fig:bench-m12-m}
\end{figure}

The restrictions from flavor physics are discussed in the second row of
\reffi{fig:bench-m12-m}(A). All four types are very similar to each other.
For the chosen value of $\tb$, the $\br(B\to X_s\ga)$ bound on
$m \lesssim 500 \gev$ occurs for the same value 
of the common heavy Higgs mass, even though 
the couplings of the heavy Higgs bosons are
different in types~I and~IV as compared to types~II and~III.
On the other hand, the value chosen 
for $\tb$ makes $\br(B_s\to\mu\mu)$ save for all four types in the
whole plane. 

The third row of \reffi{fig:bench-m12-m}(A) shows the restrictions
from unitarity and stability, which by definition are identical in all
four types. 
The unitarity constraints
only allow a narrow
region with nearly constant width around \refeq{eq:m12special} and
\refeq{eq:m12special2}.
The stability constraint further reduces the width of the allowed strip
where the lower border is then given by \refeq{eq:m12special} and
\refeq{eq:m12special2}.

Since we have a 
small value for $\CBA$,  both equations are very close.  This narrow
corridor goes from very low values of $m$ and $\msq$ and it goes
to very large values of these parameters,  even outside the figure limits
on this plane.
This plot demonstrates that for values of $\tb$ not much 
larger than~1, if $m$ increases, $\msq$ can not be arbitrary but 
it must increase accordingly to satisfy the unitarity and the stability
requirements of the theory. 

It is in fact the unitarity/stability constraints that restrict the
parameter space most.
Since this is identical in all four types, and
also the other restrictions turn out to be very similar for $\CBA$
and $\tb$ fixed to moderate values, the overall allowed parameter space
is effectively identical in types~I,~II,~III and~IV,  as can can be seen in the
fourth row of \reffi{fig:bench-m12-m}(A).  It should be noted that
the final allowed narrow corridors in these plots all end at
approximately $m=\MHp=\MA=\MH=500 \gev$, where this lower limit on
the heavy Higgs boson masses arises  from the
flavor constraints on $\MHp$.

The possible values of the triple Higgs couplings in this benchmark
plane~3 can be seen in \reffi{fig:bench-m12-m}(B). Since
$\CBA = 0.01$ is very close to the alignment limit, $\kala \sim 1$ 
is reached in the four Yukawa types, where the largest deviation of up
to $\sim 2\%$ are reached for the largest $\msq$
values. For $\lahhH$ values between $\sim 0.025$ and $\sim 0.35$ 
are found. Similarly, the values reached for $\lahHH$ and
$\lahAA = \lahHpHm/2$ do not exceed $\sim 2$, where the allowed region
follows the iso-contour lines of these triple Higgs couplings.


\clearpage
\newpage
\section{Analysis of the triple Higgs couplings}
\label{sec:thc-anal}

In this section we analyze numerically which intervals (or extreme
values) of the various triple Higgs couplings are still allowed, taking
into account all experimental and theoretical constraints as discussed
in \refse{sec:constraints}. In the case of $\lahhh$ this is relevant to
judge correctly which collider option may be needed to perform a precise
experimental determination. For the triple Higgs couplings involving one
or two heavy Higgs bosons the analysis will indicate in which processes
large effects, 
e.g.\ possibly enhanced production cross sections, can be expected due to
large triple Higgs couplings (following the strategies discussed in
\citeres{Arco:2021bvf,Arco:2021ecv,Arco:2021zhb}). 

The evaluation has been performed in all four 2HDM types, focusing
first on the ``simplest'' scenario~C with fully degenerate heavy
Higgs-boson masses $m$.  In the final part of this section, showing the
complete picture,  we also discuss the alternative scenario~A with non
fully degenerate masses,  namely,  assuming  $\MHp=\MA$ and $\MH$ as
independent and generically different mass parameters.   

The results  for scenario~C in the following three subsections will be
shown in different benchmark planes, which are 
chosen in each scenario individually.
In some benchmark planes the particular values of the
other parameters are chosen such as to maximize the 
deviations of $\lahhh$ from it SM value (the plots below show
$\kala := \lahhh/\laSM$)).%
\footnote{It should be noted that this is a tree-level analysis. It was
shown that one-loop~\cite{lahhh1L} and even two-loop corrections to
$\lahhh$ can substantially enhance their values~\cite{Bahl:2022jnx}.}%
~Other benchmark planes are chosen such as to
maximize the (absolute) size of the triple Higgs couplings involving one
or two heavy Higgs bosons. The plots below show the triple Higgs
couplings as defined in \refeq{eq:lambda}.

The present analysis in the 2HDM type~I has changed only slightly
w.r.t.\ \citere{Arco:2020ucn} and we briefly update the
corresponding results in \reffis{fig:I-1} - \ref{fig:I-4}.
Concerning the 2HDM type~II, the constraints in particular from the
Higgs-boson rate measurements have tightened in a relevant way
w.r.t.\ \citere{Arco:2020ucn}, affecting in particular the allowed ranges
for $\CBA$. Furthermore, only one scenario with
$m \equiv \MHp = \MH = \MA$ had been investigated in our previous work
\citere{Arco:2020ucn}. 
Consequently, we update our analysis from this previous work
analyzing the triple Higgs couplings in several additional planes.
The results for the 2HDM types~III and~IV are new and complete
the triple Higgs-boson coupling analysis in the 2HDM.
The results in type~II and~III turned out to be very
similar. Consequently, we analyze these two types together, as shown in
\reffis{fig:II-III-1} - \ref{fig:II-III-4}.  
The results for type~IV are presented in \reffis{fig:IV-1} -
\ref{fig:IV-4}.  

The figures are organized as follows.
The upper rows (the upper row for type~I and~IV,
the upper two for type~II and~III)
summarize the constraints in each
benchmark plane: the first, second and third plots correspond to
the constraints (with the same color coding) as shown in the first,
second and third row of \reffis{fig:bench-cba-tb}(A),
\ref{fig:bench-m12-tb}(A) and \ref{fig:bench-m12-m}(A), i.e.\
the constraints from Higgs rate measurements and BSM Higgs boson
searches, from flavor observables and from unitarity/stability,
respectively. The corresponding right plots in
\reffis{fig:I-1} - \ref{fig:IV-4} 
show the overall allowed region, depicted as dotted areas.
The lower rows of \reffi{fig:I-1} - \ref{fig:IV-4} present
the result for the triple Higgs couplings: the first, second, third and
fourth plot show the predictions for $\kala$, $\lahhH$, $\lahHH$ and
$2 \lahAA = \lahHpHm$, respectively. The overall allowed regions is
indicated by a solid black (type~I),
solid blue (type~II), dashed pink (type~III) and
dashed red line (type~IV).


\newpage

\subsection{Triple Higgs couplings in the 2HDM type I}
\label{sec:typeI}

The benchmark planes for the 2HDM type~I had been defined in
\citere{Arco:2020ucn} as:

\begin{itemize}

\item[I-1:]
  $m \equiv \MHp = \MH = \MA = 1000 \gev$,
  $\msq$ fixed via \refeq{eq:m12special},\\
  free parameters: $\CBA$, $\tb$

\item[I-2:]
  $m \equiv \MHp = \MH = \MA = 650 \gev$, $\tb = 7.5$,\\
  free parameters: $\CBA$, $\msq$

\item[I-3:]
  $\msq$ fixed via \refeq{eq:m12special}, $\tb = 10$,\\
  free parameters: $\CBA$, $m$
  
\item[I-4:]
   $\CBA = 0.1$, 
  $\msq$ fixed by \refeq{eq:m12special},\\
  free parameters $m\equiv \MH = \MA = \MHp$,$\tb$.
  
\end{itemize}

\begin{figure}[t]
\centering
\includegraphics[width=0.24\textwidth]{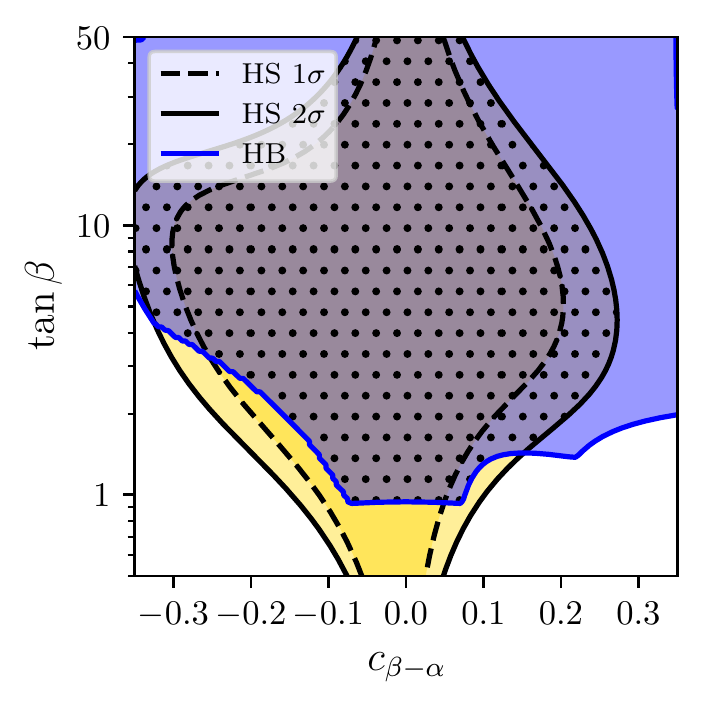}
\includegraphics[width=0.24\textwidth]{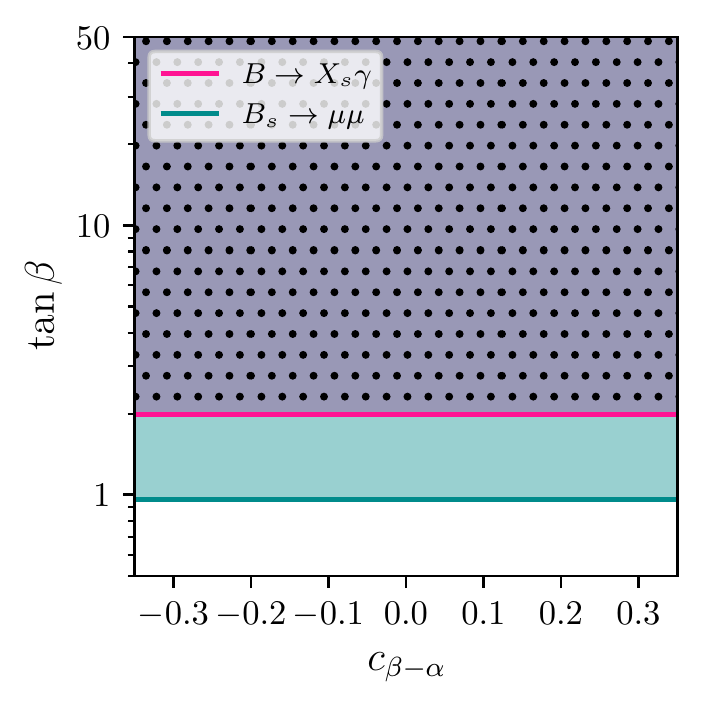}
\includegraphics[width=0.24\textwidth]{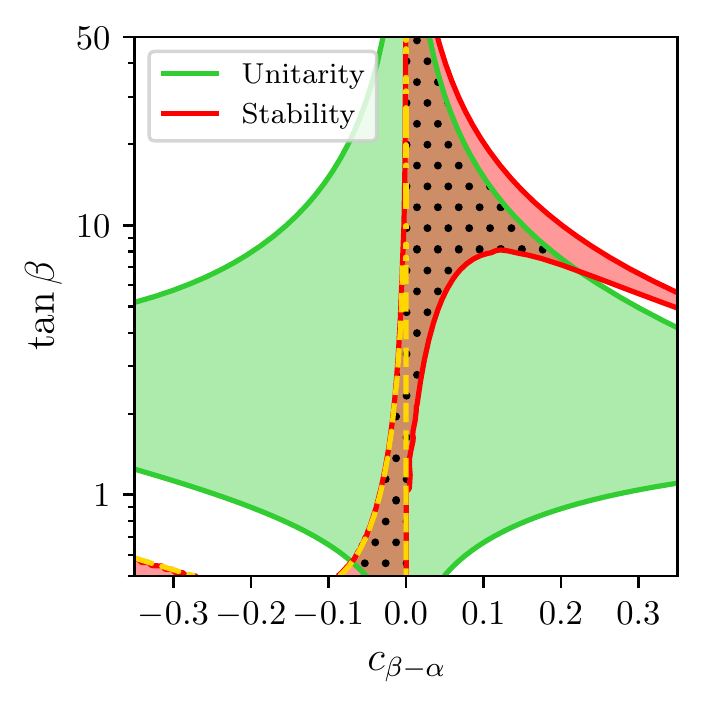}
\includegraphics[width=0.24\textwidth]{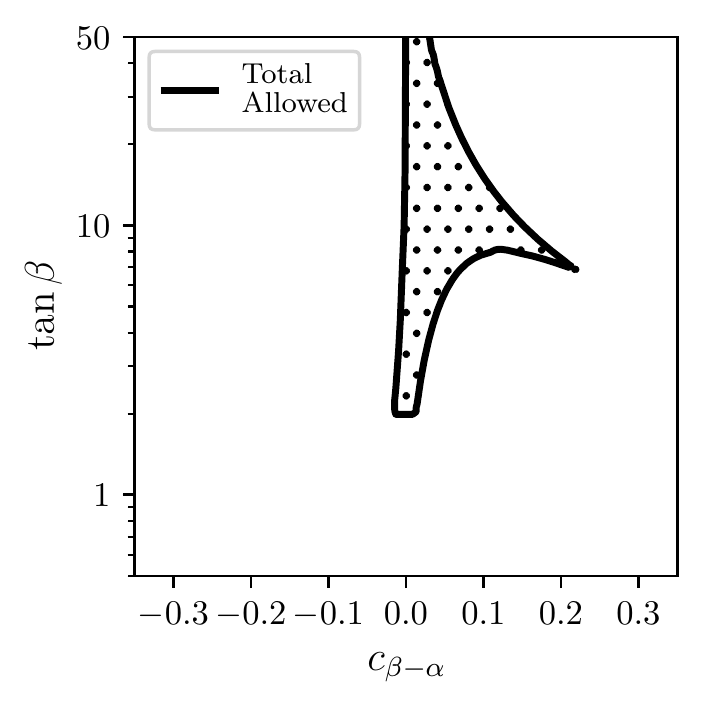}\\[1em]

\includegraphics[width=0.24\textwidth]{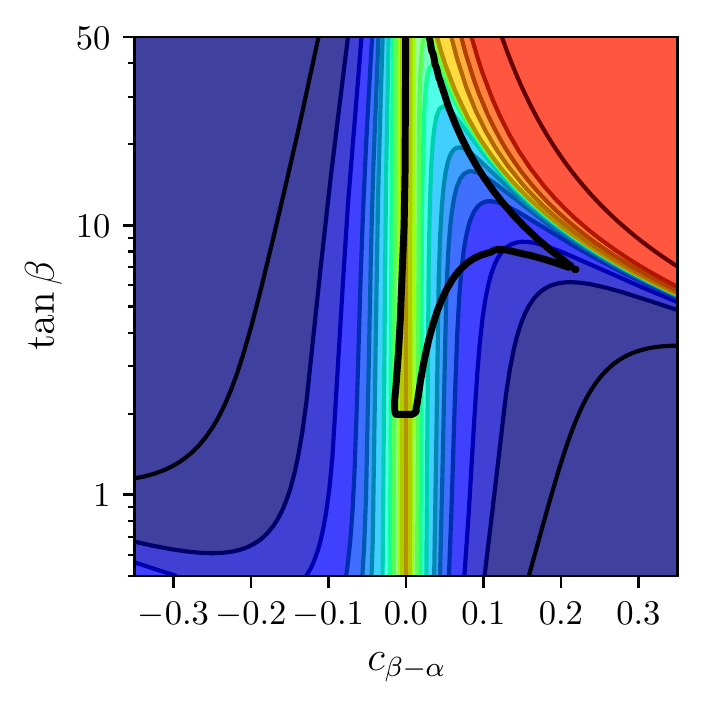}
\includegraphics[width=0.24\textwidth]{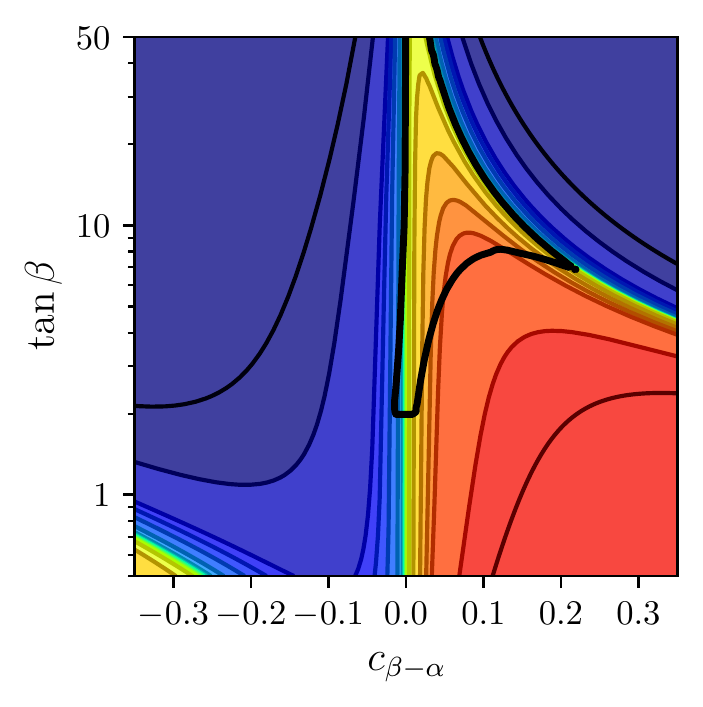}
\includegraphics[width=0.24\textwidth]{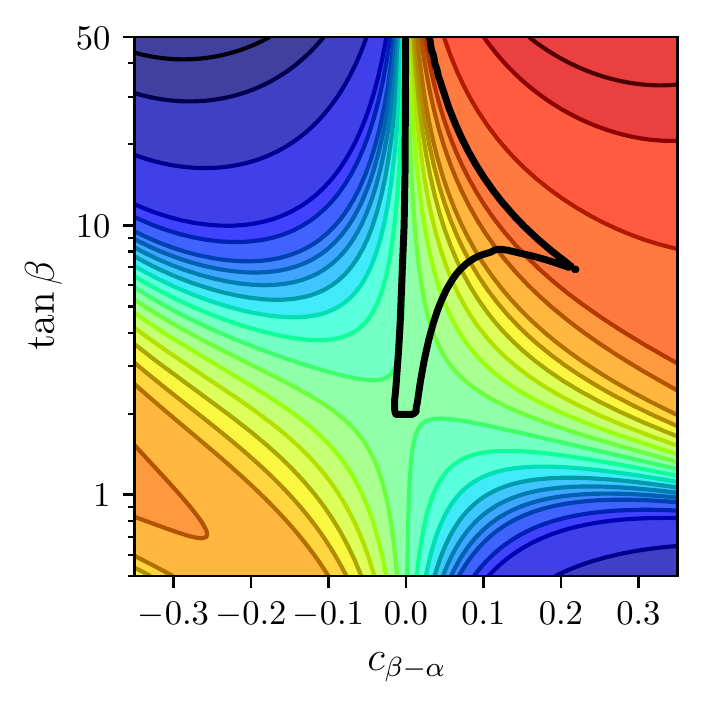}
\includegraphics[width=0.24\textwidth]{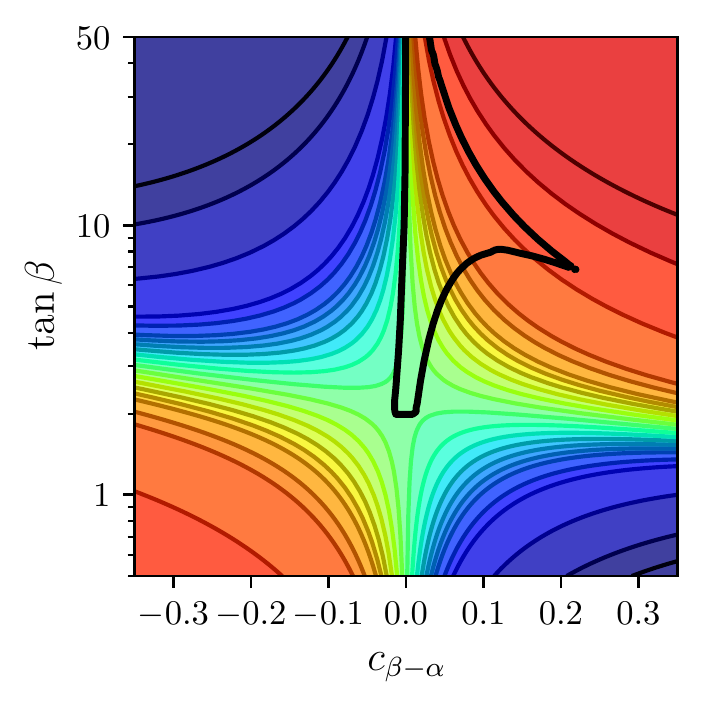}

\includegraphics[width=0.24\textwidth]{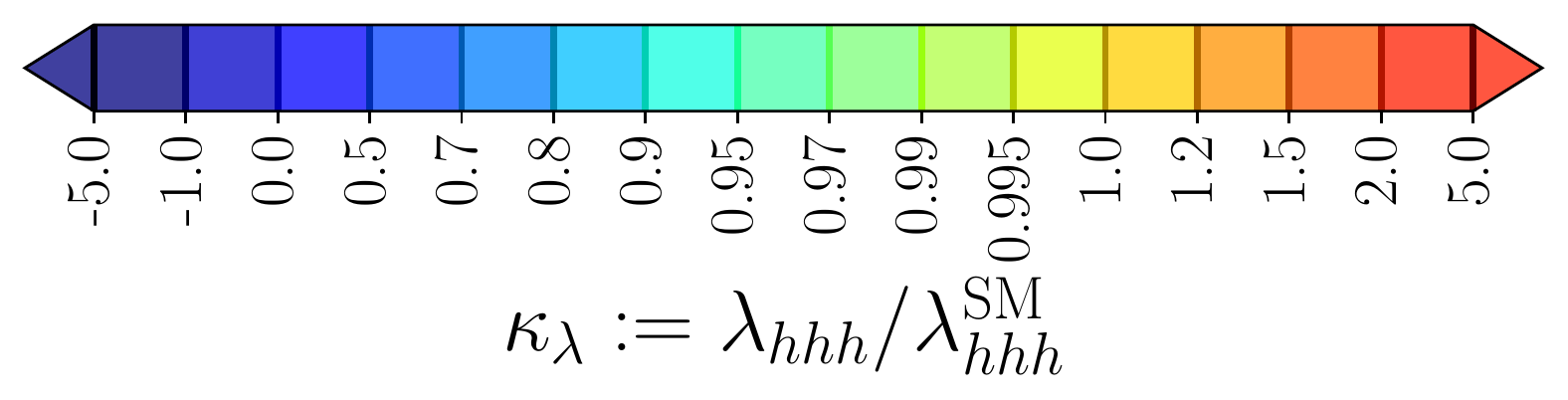}
\includegraphics[width=0.24\textwidth]{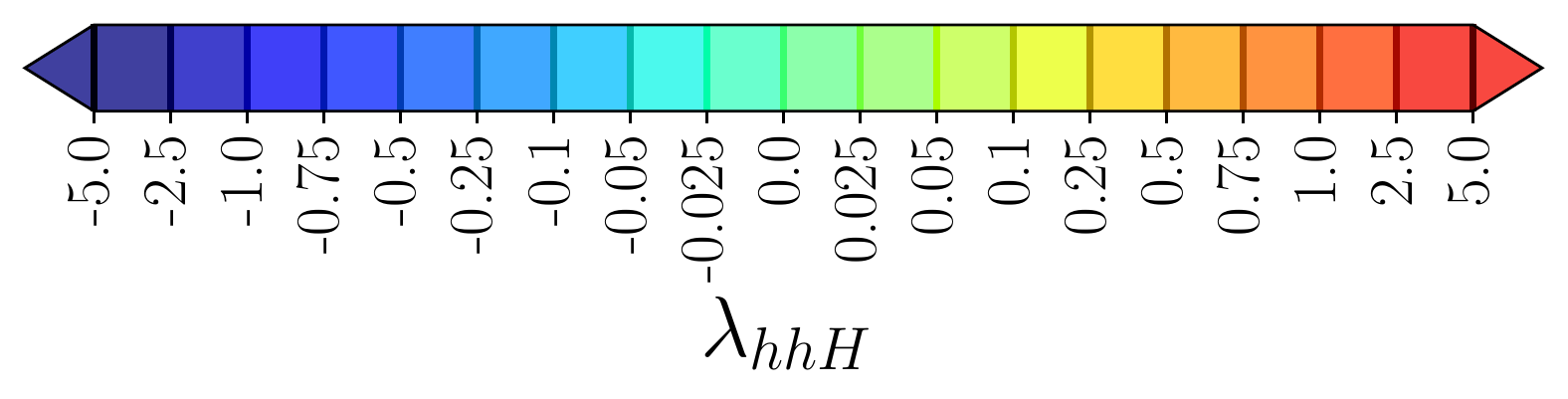}
\includegraphics[width=0.24\textwidth]{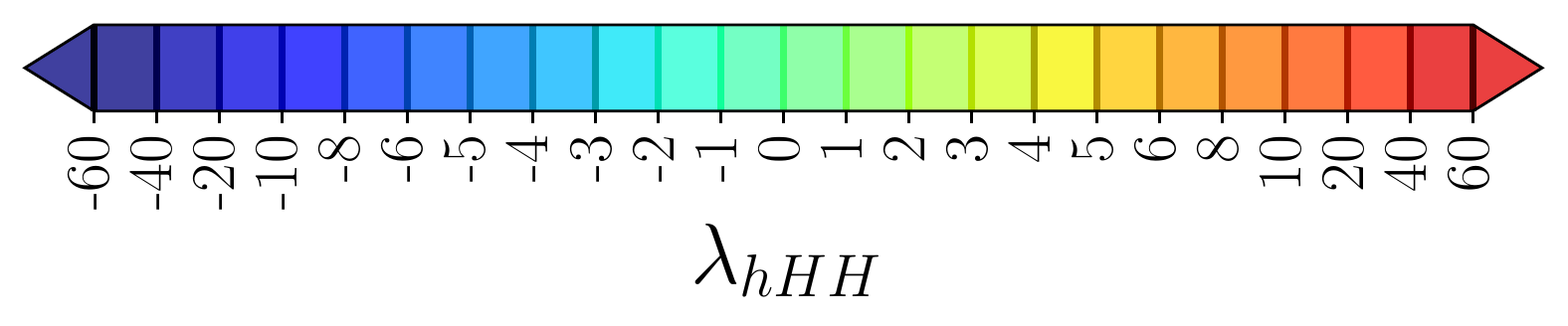}
\includegraphics[width=0.24\textwidth]{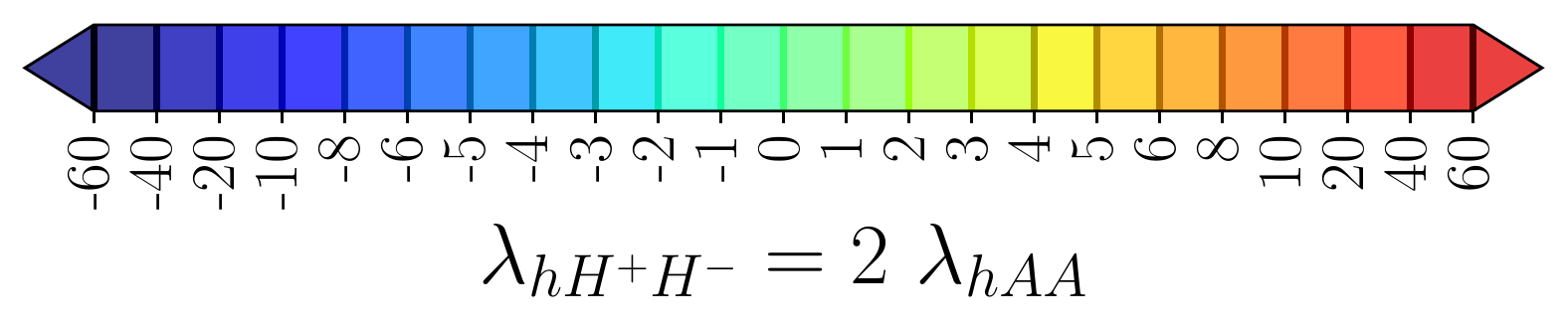}\\[1em]
\caption{Allowed areas (dotted regions) from the various constraints (upper row) and triple Higgs couplings (lower row)
  for the benchmark scenario I-1 in the $\CBA$--$\tb$ plane
  with $\msq$ fixed via \refeq{eq:m12special} and $m = 1000 \gev$.}
\label{fig:I-1}
\end{figure}

\begin{figure}[h!]
\centering
\includegraphics[width=0.24\textwidth]{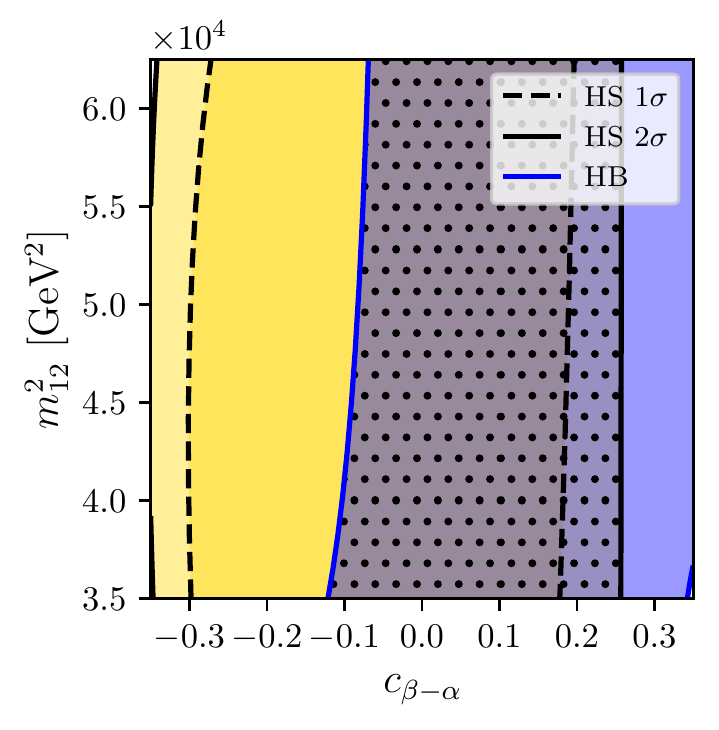}
\includegraphics[width=0.24\textwidth]{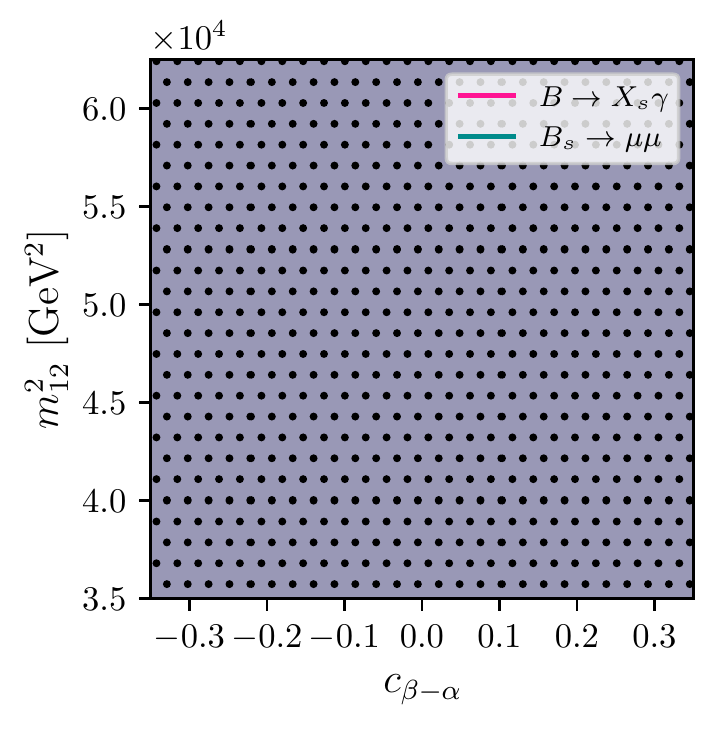}
\includegraphics[width=0.24\textwidth]{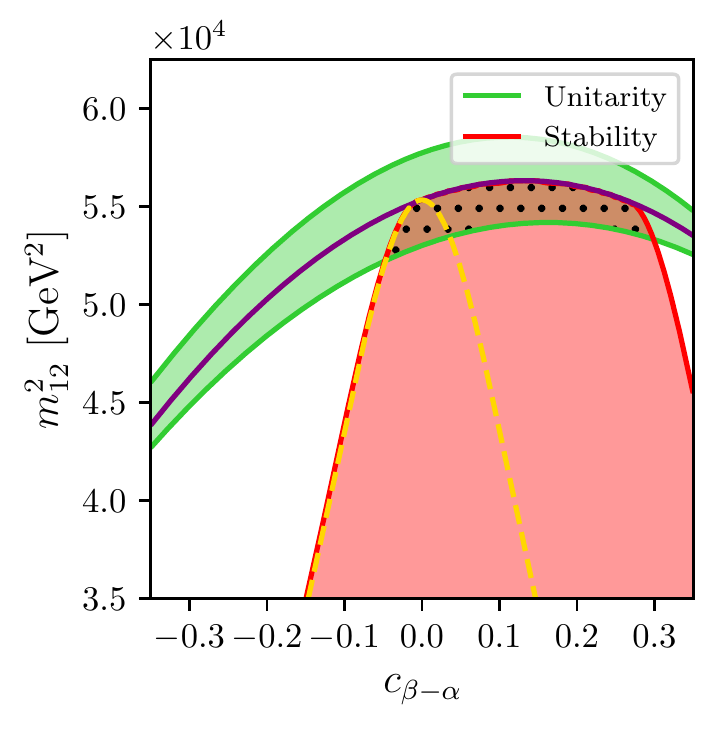}
\includegraphics[width=0.24\textwidth]{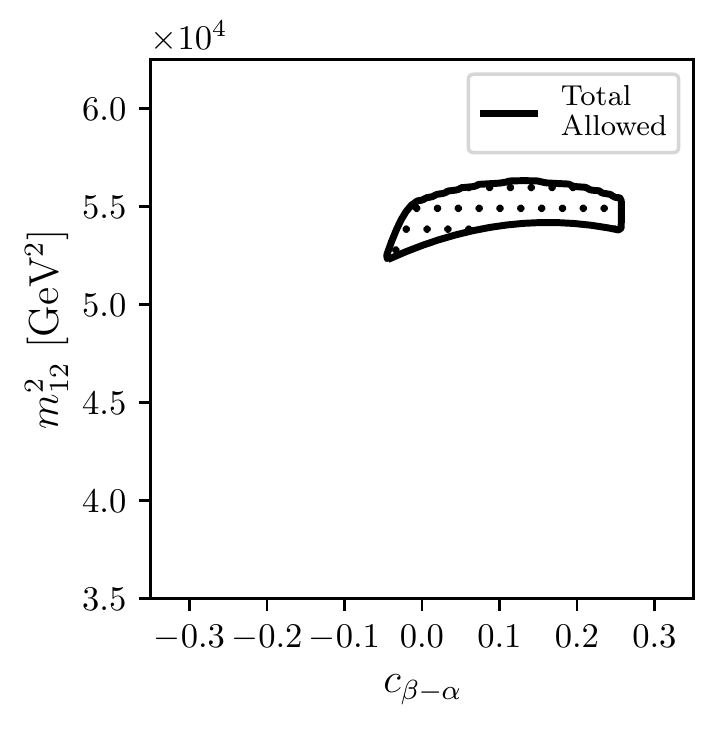}\\[1em]

\includegraphics[width=0.24\textwidth]{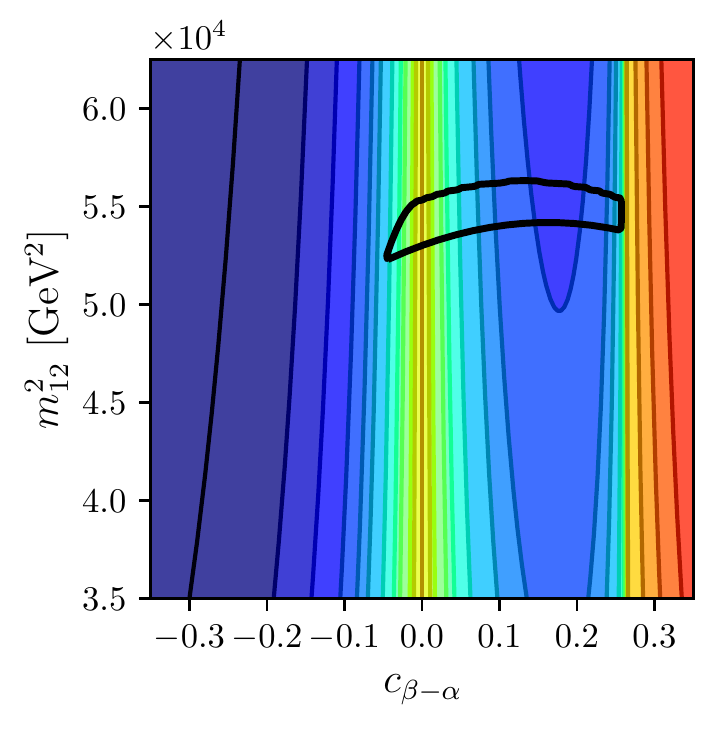}
\includegraphics[width=0.24\textwidth]{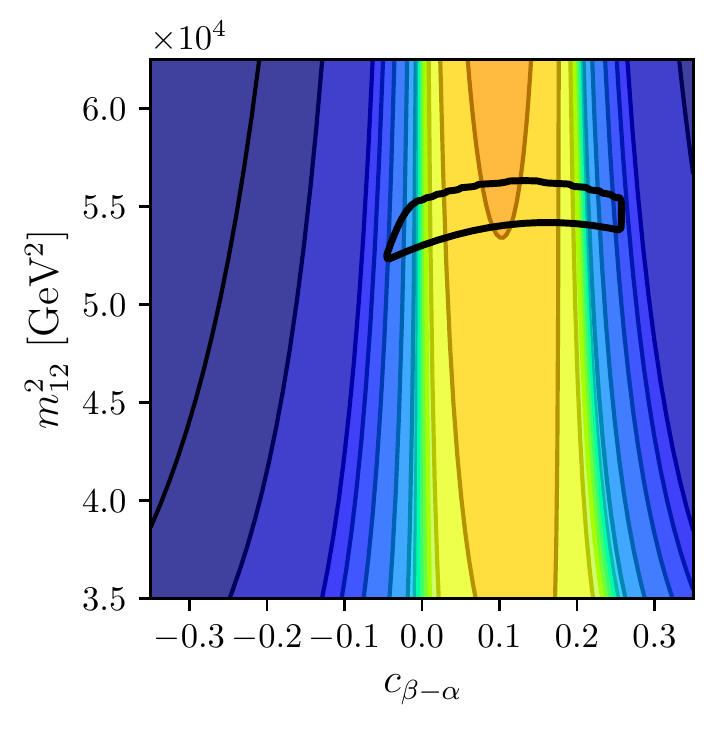}
\includegraphics[width=0.24\textwidth]{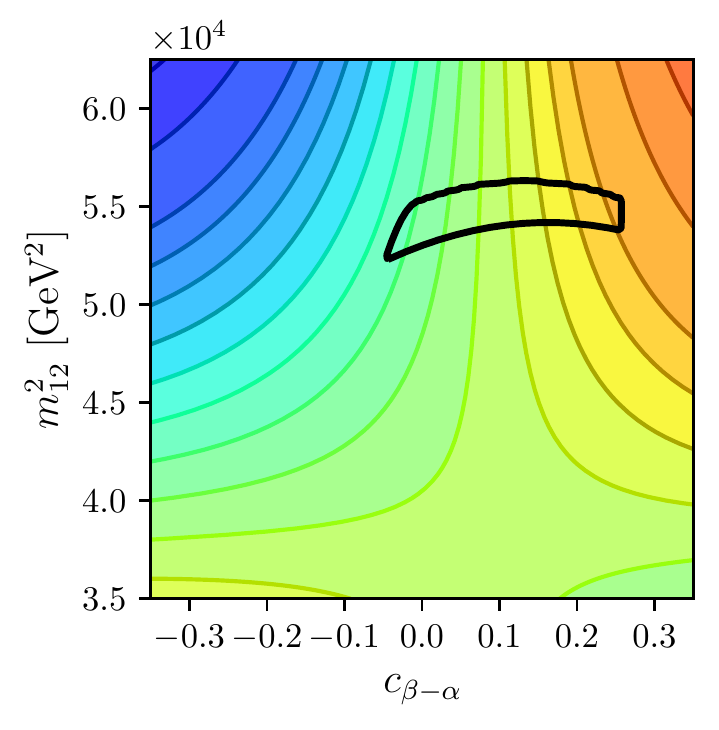}
\includegraphics[width=0.24\textwidth]{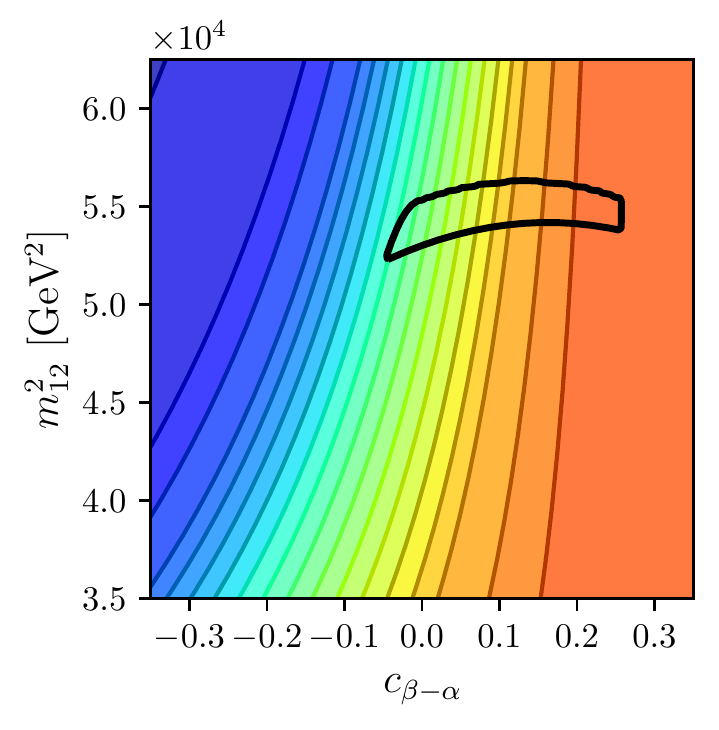}

\includegraphics[width=0.24\textwidth]{h1h1h1_colorbar}
\includegraphics[width=0.24\textwidth]{h1h1h2_colorbar}
\includegraphics[width=0.24\textwidth]{h1h2h2_colorbar}
\includegraphics[width=0.24\textwidth]{h1HpHm_colorbar}\\[1em]
\caption{Allowed areas (dotted regions) from the various constraints (upper row) and triple Higgs couplings (lower row)
  for the benchmark scenario I-2 in the $\CBA$--$\msq$ plane
  with $\tb=7.5$ and $m = 650 \gev$.}
\label{fig:I-2}
\end{figure}

\begin{figure}[!htb]
\centering
\includegraphics[width=0.234\textwidth]{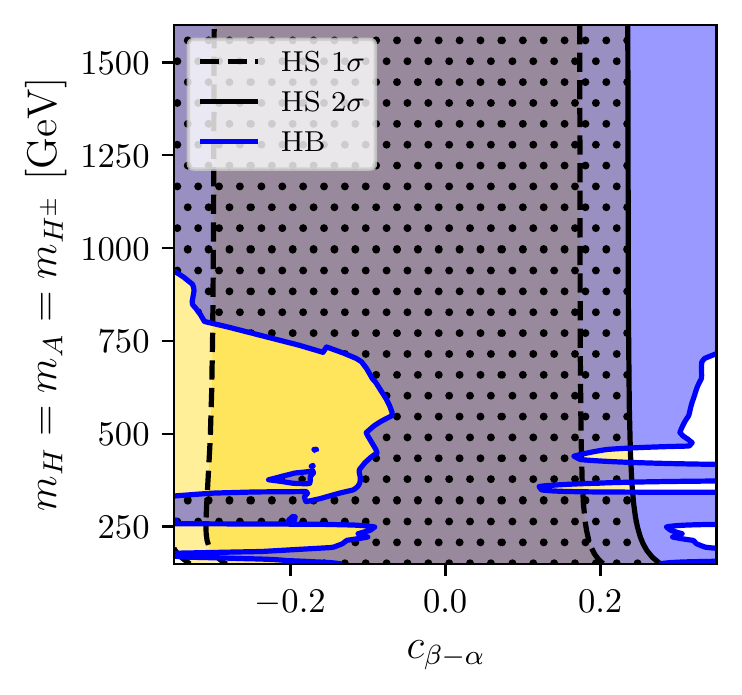}
\includegraphics[width=0.234\textwidth]{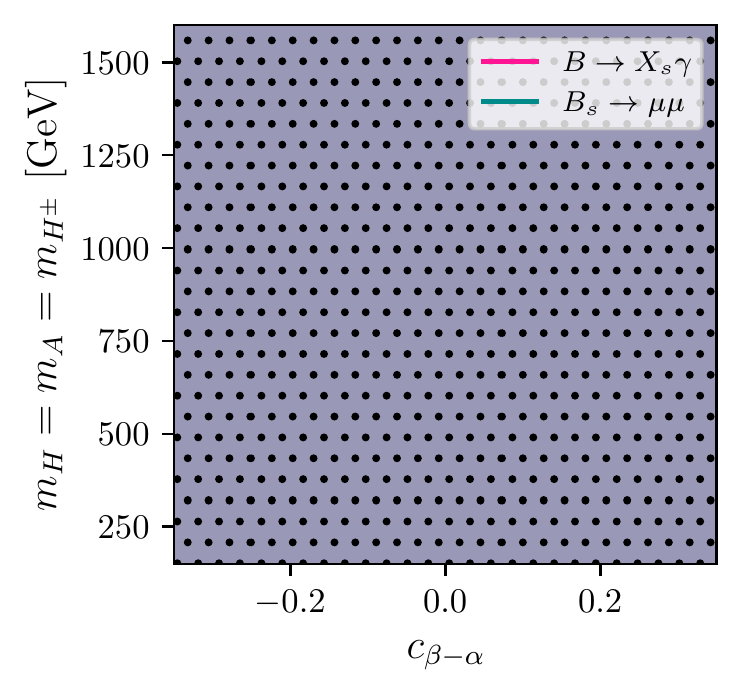}
\includegraphics[width=0.234\textwidth]{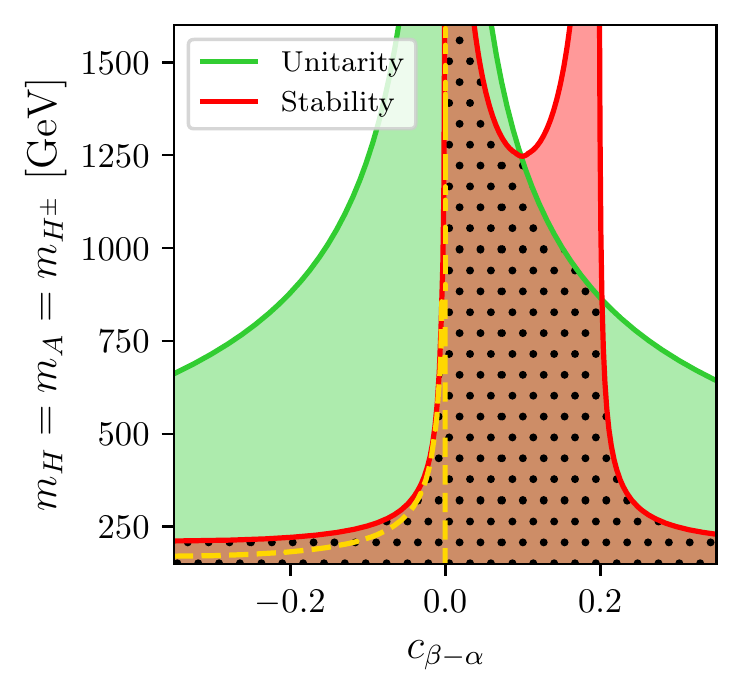}
\includegraphics[width=0.234\textwidth]{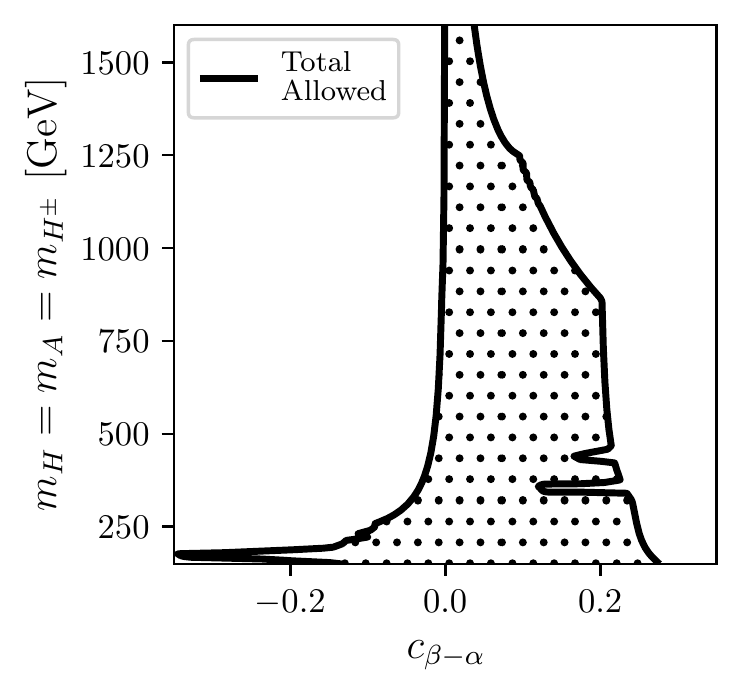} \\[1em]

\includegraphics[width=0.234\textwidth]{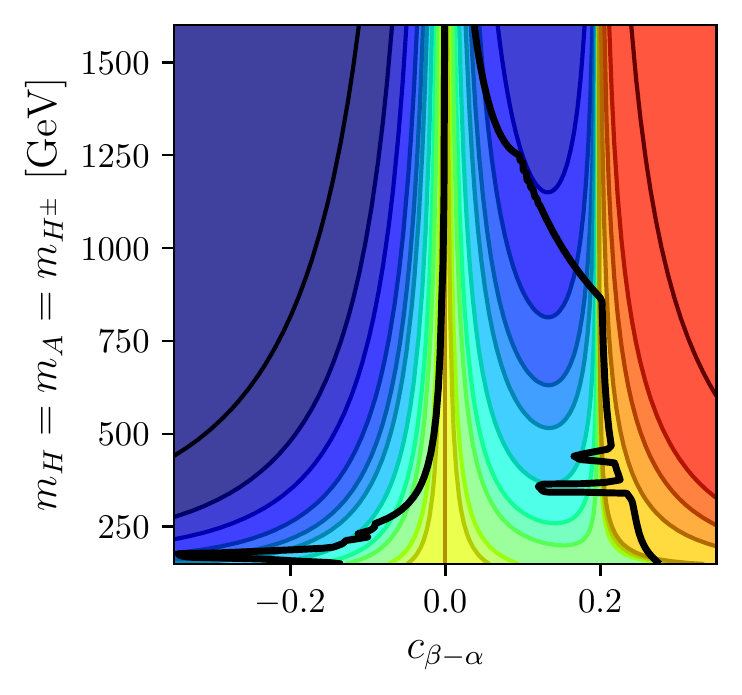}
\includegraphics[width=0.234\textwidth]{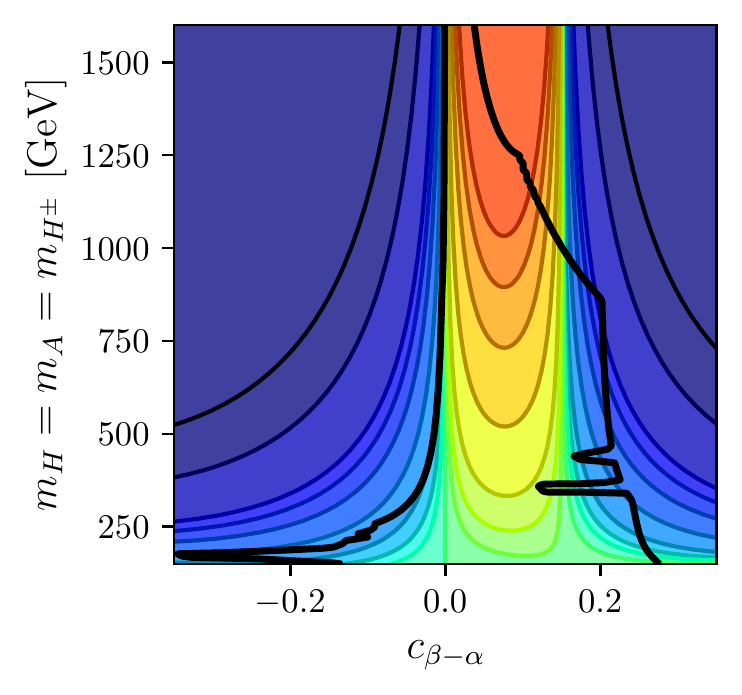}
\includegraphics[width=0.234\textwidth]{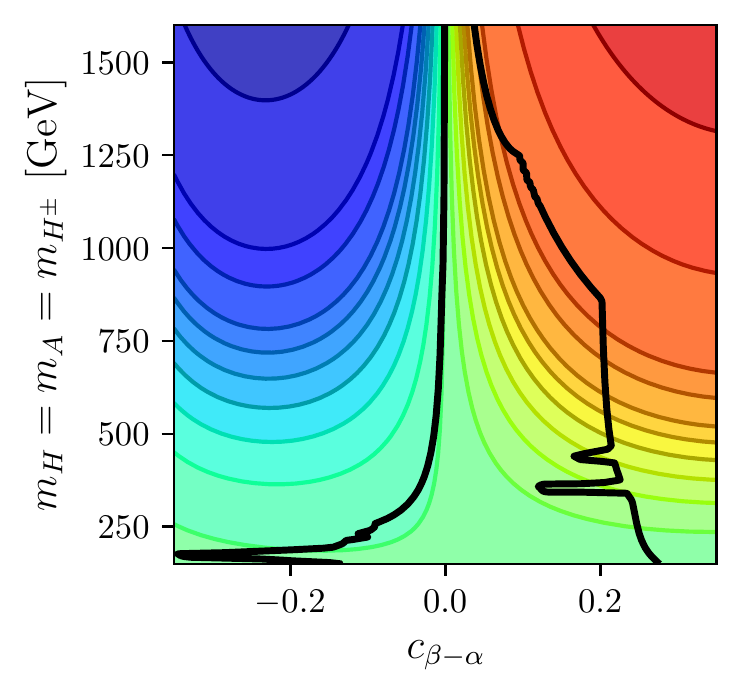}
\includegraphics[width=0.234\textwidth]{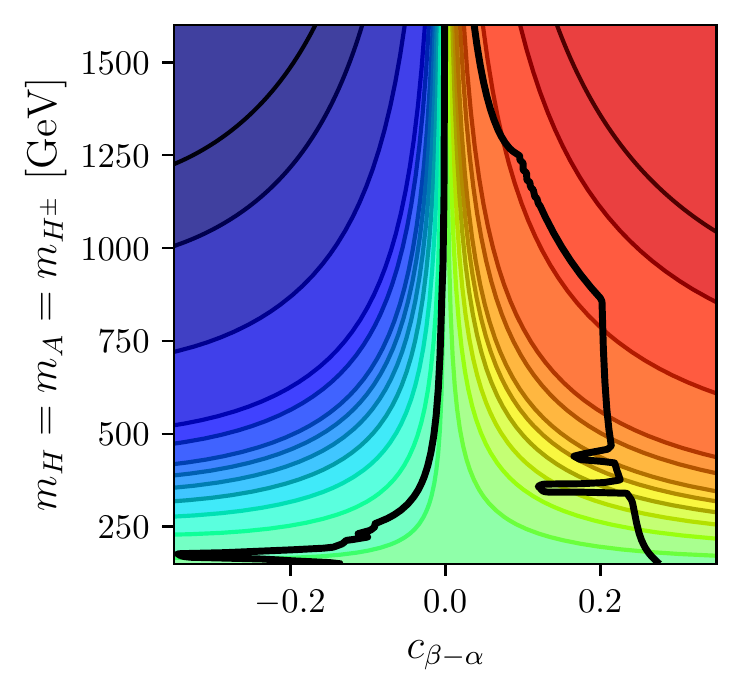}

\includegraphics[width=0.234\textwidth]{h1h1h1_colorbar}
\includegraphics[width=0.234\textwidth]{h1h1h2_colorbar}
\includegraphics[width=0.234\textwidth]{h1h2h2_colorbar}
\includegraphics[width=0.234\textwidth]{h1HpHm_colorbar} 
\caption{Allowed areas (dotted regions) from the various constraints  (upper row) and triple Higgs couplings (lower row)
  for the benchmark scenario I-3 in the $\CBA$--$m$ plane
 with $\tb=10$ and $\msq$ fixed via \refeq{eq:m12special}.}
\label{fig:I-3}
\end{figure}

\begin{figure}[!htb]
\centering
\includegraphics[width=0.234\textwidth]{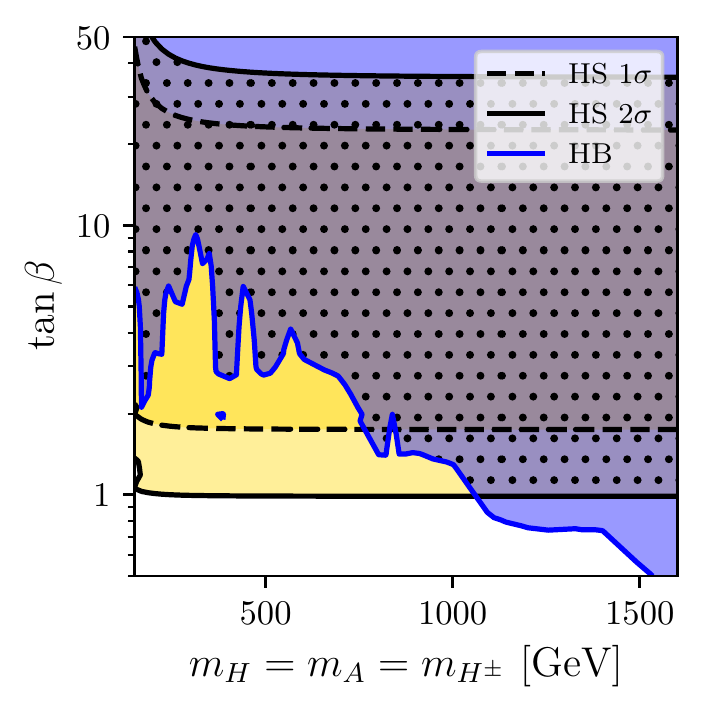}
\includegraphics[width=0.234\textwidth]{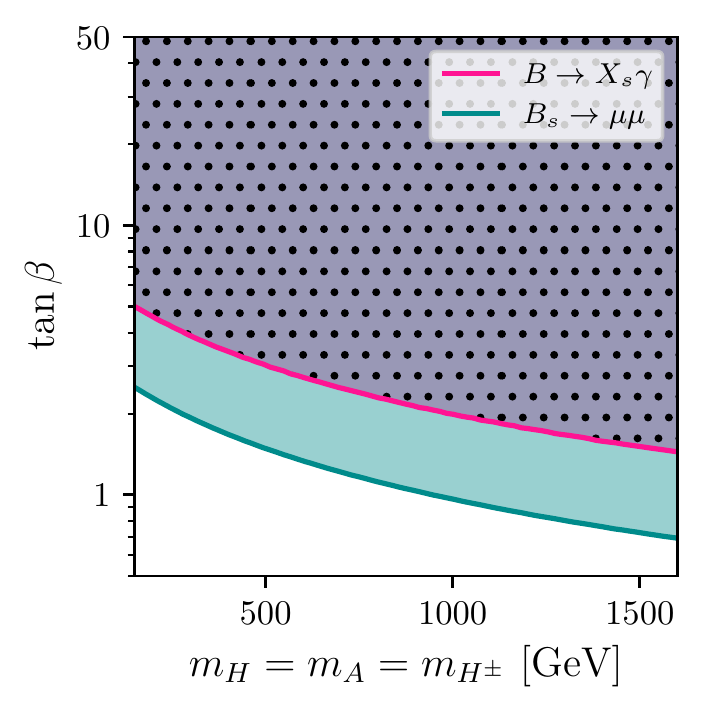}
\includegraphics[width=0.234\textwidth]{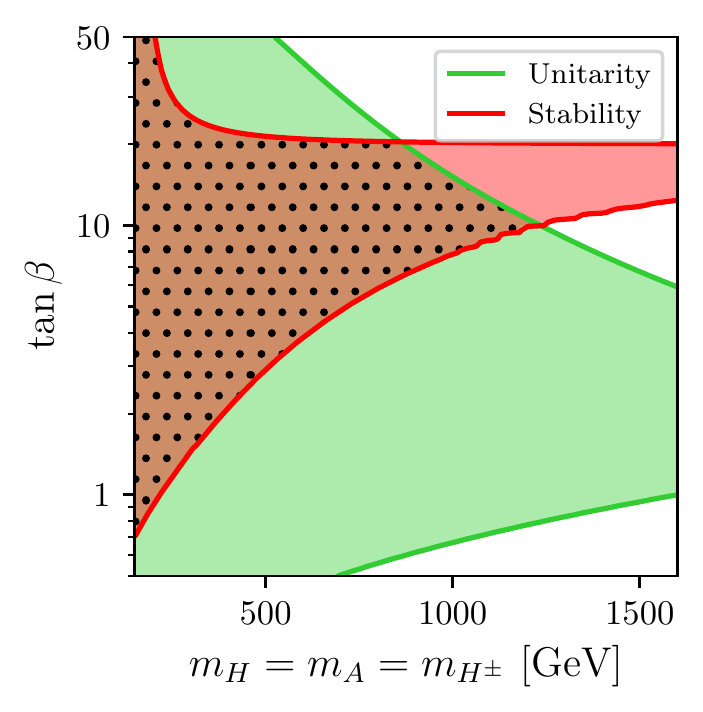}
\includegraphics[width=0.234\textwidth]{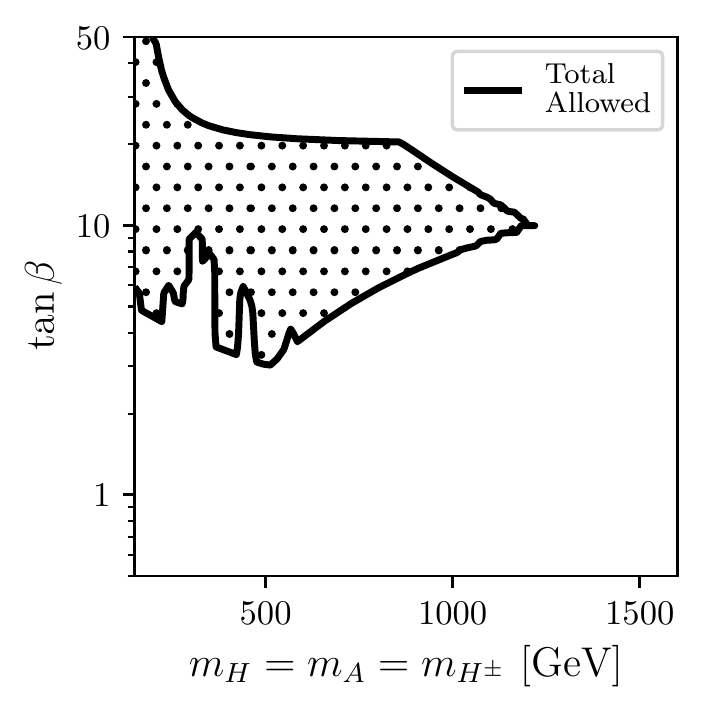} \\[1em]

\includegraphics[width=0.234\textwidth]{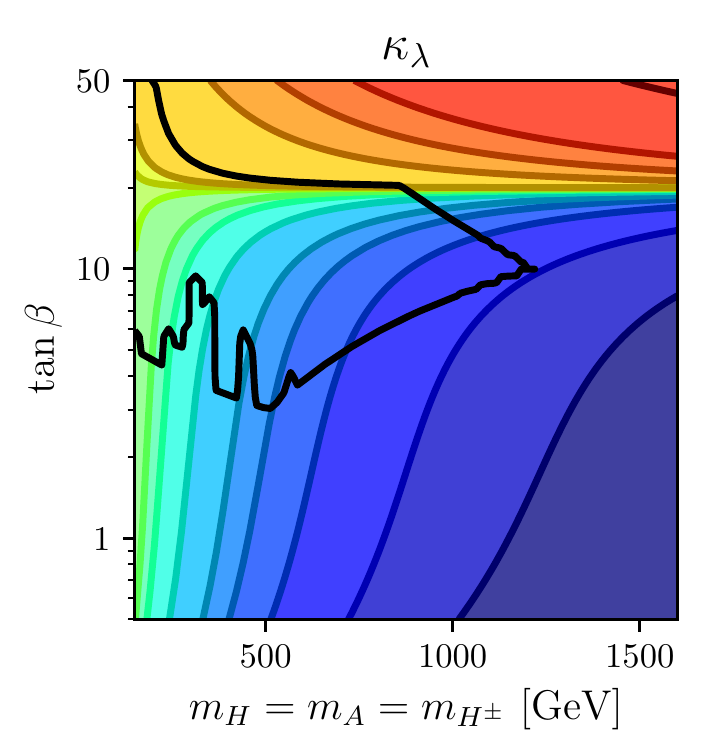}
\includegraphics[width=0.234\textwidth]{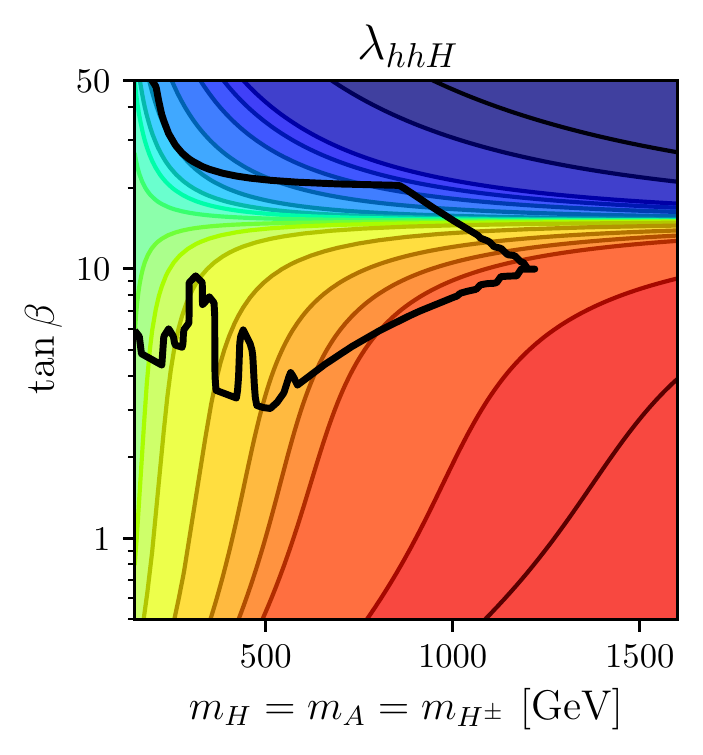}
\includegraphics[width=0.234\textwidth]{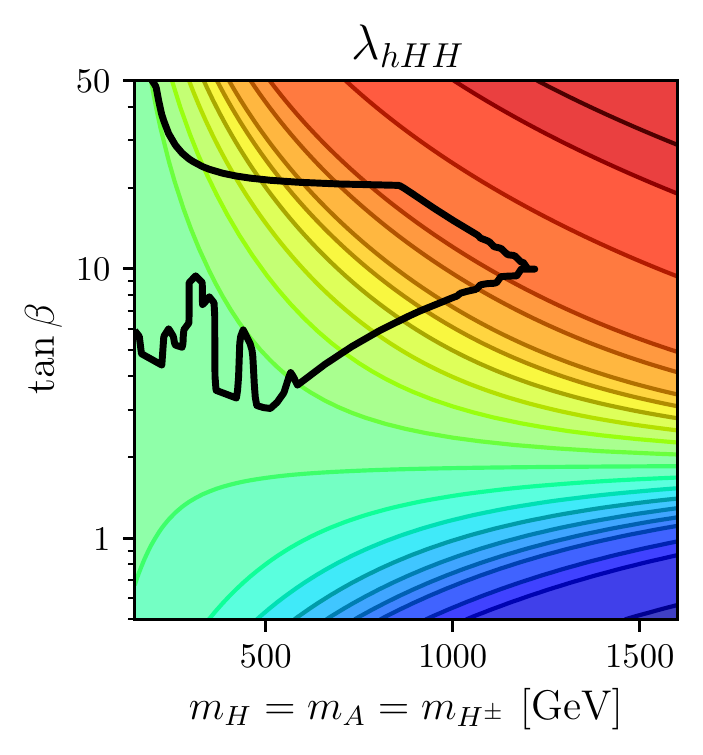}
\includegraphics[width=0.234\textwidth]{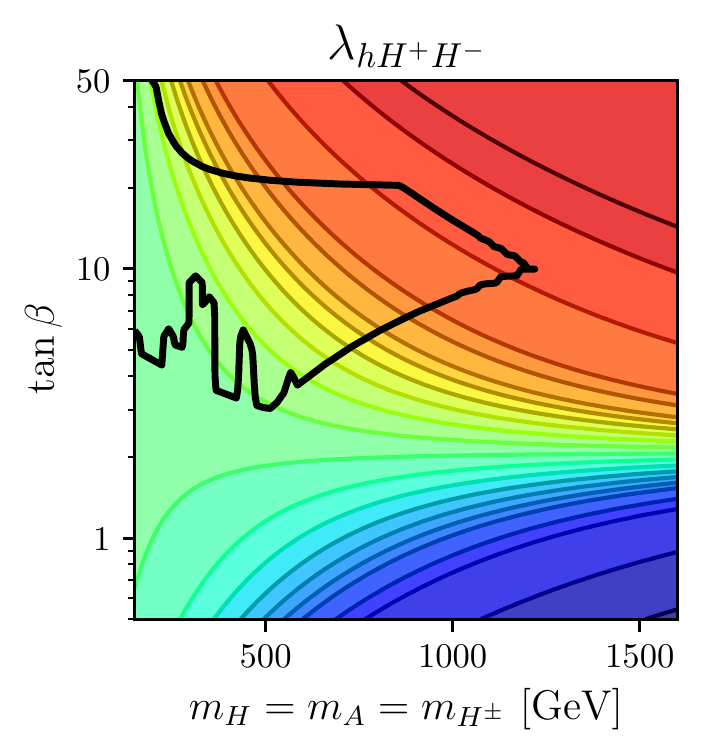}

\includegraphics[width=0.234\textwidth]{h1h1h1_colorbar}
\includegraphics[width=0.234\textwidth]{h1h1h2_colorbar}
\includegraphics[width=0.234\textwidth]{h1h2h2_colorbar}
\includegraphics[width=0.234\textwidth]{h1HpHm_colorbar} 
\caption{Allowed areas (dotted regions) from the various constraints  (upper row) and triple Higgs couplings (lower row)
  for the benchmark scenario I-4 in the $m$--$\tb$ plane
  with $\CBA=0.2$ and $\msq$ fixed via \refeq{eq:m12special}.}
\label{fig:I-4}
\end{figure}

The allowed parameter region in scenario~I-1, as shown in
\reffi{fig:I-1}, is found mainly for positive $\CBA$ with
$\tb \ge 2$. The largest allowed $\CBA$ values of $\sim 0.2$ are found for
$\tb \sim 6$. 
In the first scenario we found $\kala \sim \inter{-0.4}{1}$, where the
smallest values are reached for these largest $\CBA$ points. 
For $\lahhH$ the largest values were found for $\CBA \sim 0.08$ and
$\tb \sim 7.5$, reaching up to $\lahhH \sim 1.2$. 
The other triple Higgs couplings reach their maximum values around
$\CBA \sim 0.06$ and $\tb \sim 27$ with
$\lahHH \approx \lahAA = \lahHpHm/2 \sim 12.5$.

In the second scenario, I-2, shown in \reffi{fig:I-2},
only a very restricted region for $\msq$ is
allowed by the constraints, $\msq \sim \inter{52000 \gev^2}{56000 \gev^2}$.
One finds $\kala = 1$ for $\CBA = 0$, i.e.\ in
the alignment limit, as required. The same value is also found for
$\CBA \sim 0.26$ due to cancellations in $\lahhh$. Overall, we found
$\kala \sim \inter{0.5}{1.2}$, where the largest values are reached for
the largest allowed $\CBA \sim 0.28$.
The values of $\lahhH$ are quite small in this scenario, only reaching
up to $\lahhH \sim 0.5$. The other triple Higgs couplings reach their
maximum values around $\CBA \sim 0.26$ and $\msq \sim 55000 \gev^2$ with
$\lahHH \approx \lahAA = \lahHpHm/2 \sim 6.5$.

The third scenario, I-3, depicted in \reffi{fig:I-3},
exhibits a rather ``large'' allowed parameter space,
where, depending on $m$ we found allowed $\CBA$ values between
$\sim -0.3$ to $\sim +0.3$. As in the second scenario one finds $\kala = 1$ 
not only for $\CBA = 0$, but also for a second branch with $\CBA \ge 0.2$,
partially in the ``allowed'' parameter space.
The values that can be reached by $\kala$ range from $\kala \sim 0.07$
for $\CBA \sim 0.1$ and large $m$ close to $1200 \gev$ to about
$\kala \sim 1.2$ for the largest allowed $\CBA$ values and $m \sim 300 \gev$.
$\lahhH$ reaches its maximum value of $\sim 1.7$ for $\CBA \sim 0.05$
and $m \sim 1500 \gev$. The other triple Higgs couplings reach their
maximum allowed values around $\CBA \sim 0.11$ and $m \sim 1200 \gev$ with
$\lahHH \approx \lahAA = \lahHpHm/2 \sim 12.5$.

The final scenario for type~I, I-4, is shown in \reffi{fig:I-4}. It is
given in the $m$--$\tb$ plane, where for low values of $m$ the largest
values of $\tb \sim 50$ are reached. Direct searches and
stability/unitarity constraints yield bounds of $m \lsim 1200 \gev$ with
$\tb$ ranging between $\sim 3$ and $\sim 20$ (except for the lowest
values of $m$.  As in the previous planes, we find $\kala \sim
\inter{0.05}{1.05}$, where the largest 
(smallest) values are reached for the smallest (largest) values
of~$m$. Similarly, we find $\lahhH \sim \inter{-0.5}{1.3}$,
with the negative values reached for the higher $\tb$ value, and the
largest value around the highest allowed values of~$m$. As in the other
benchmark planes the smallest (largest) values of
$\lahHH \sim \lahAA = \lahHpHm/2$ are found for the smallest (largest)
values of $m$. Their allowed values are found in the interval
$\lahHH \sim \lahAA = \lahHpHm/2\sim \inter{0.2}{12}$.


\subsection{Triple Higgs couplings in the 2HDM types II and III}
\label{sec:typeII}

\begin{figure}[t!]
\centering
\includegraphics[height=0.17\textheight]{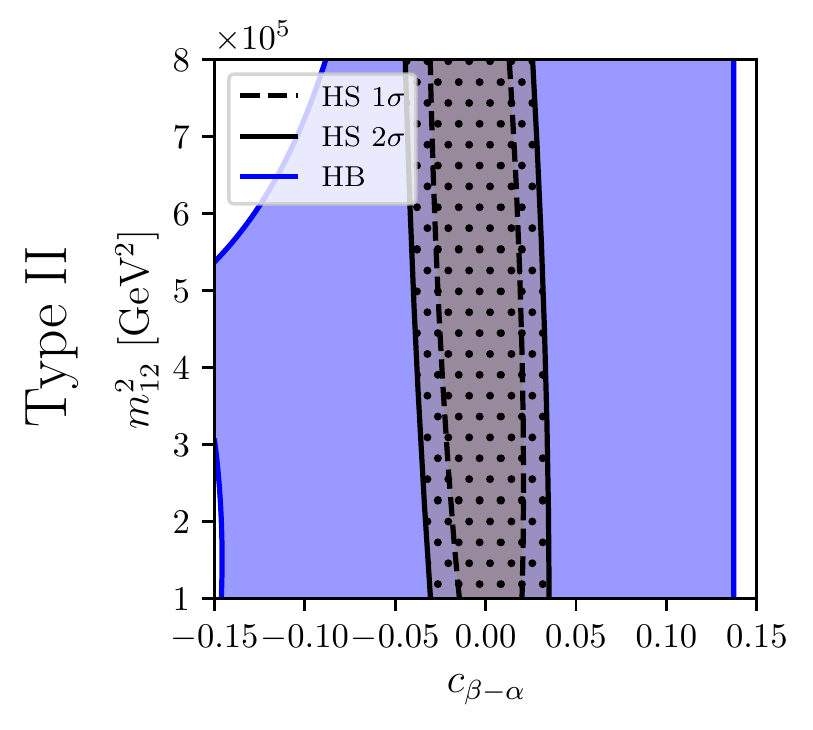}
\includegraphics[height=0.17\textheight]{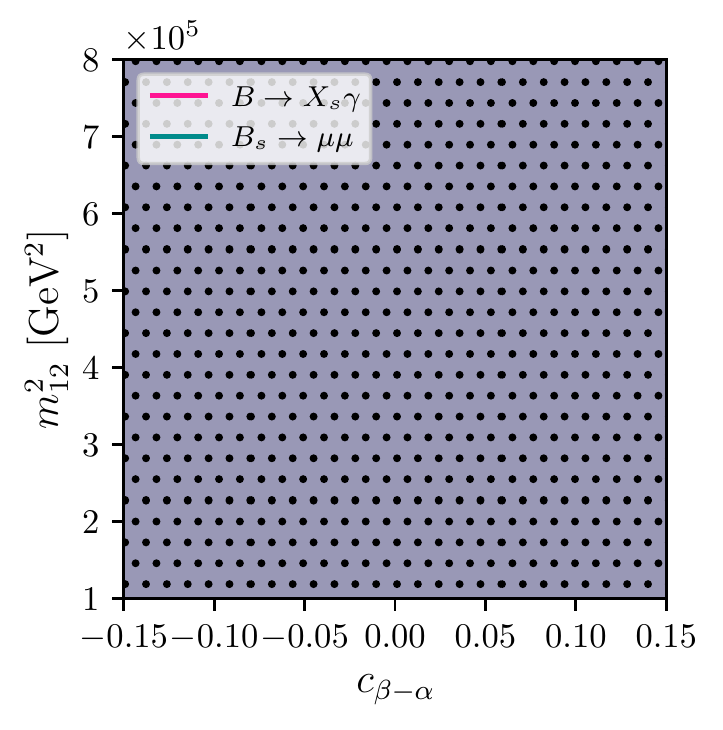}
\includegraphics[height=0.17\textheight]{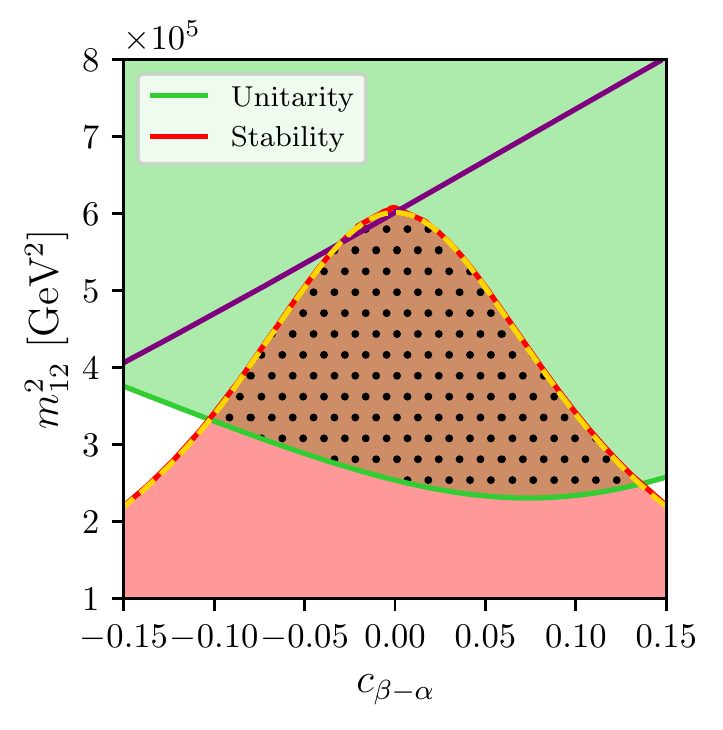}
\includegraphics[height=0.17\textheight]{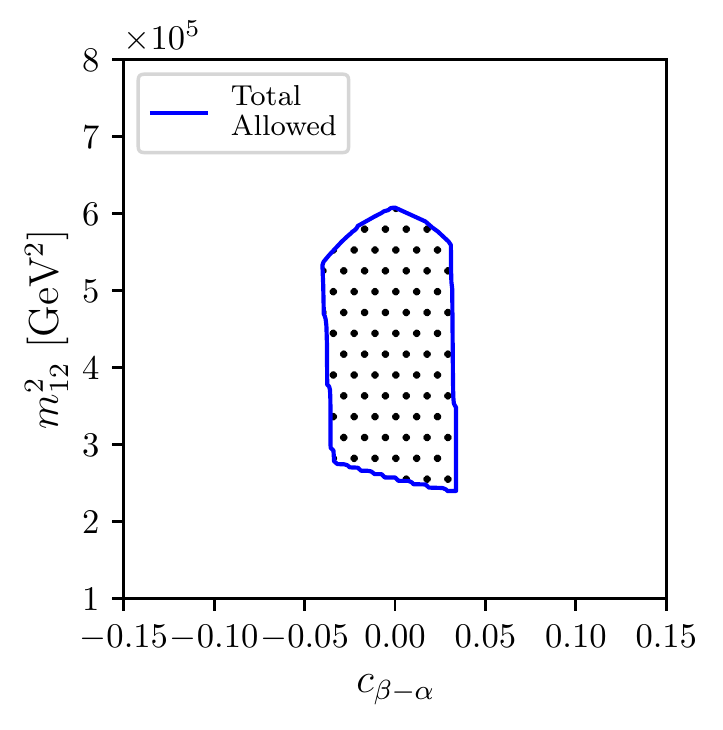}

\includegraphics[height=0.17\textheight]{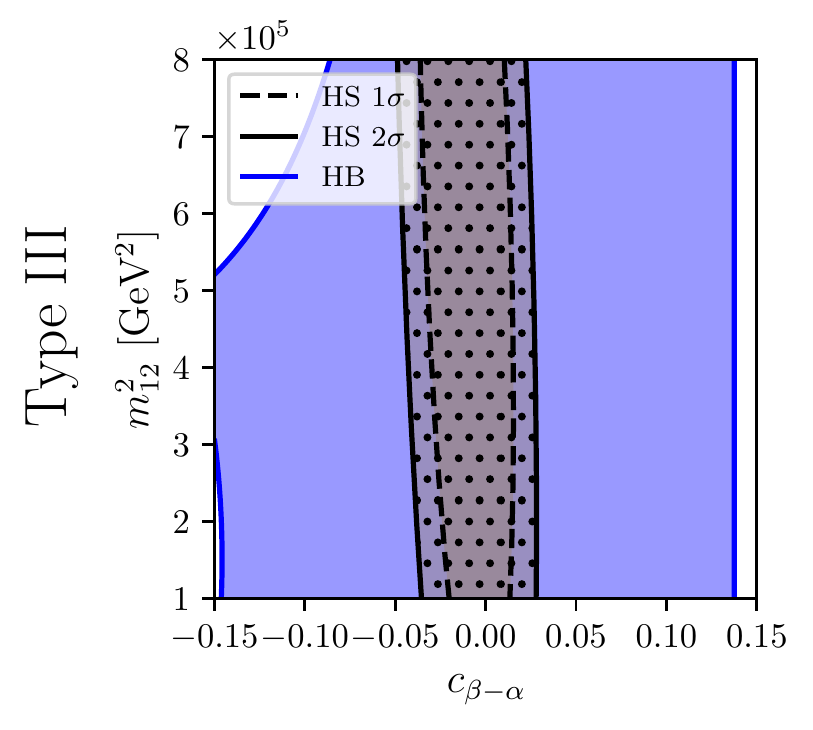}
\includegraphics[height=0.17\textheight]{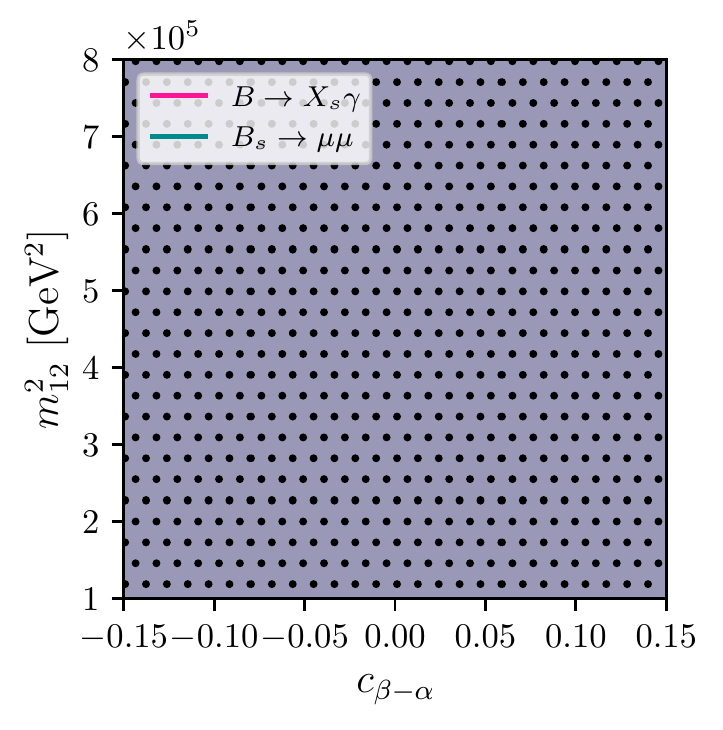}
\includegraphics[height=0.17\textheight]{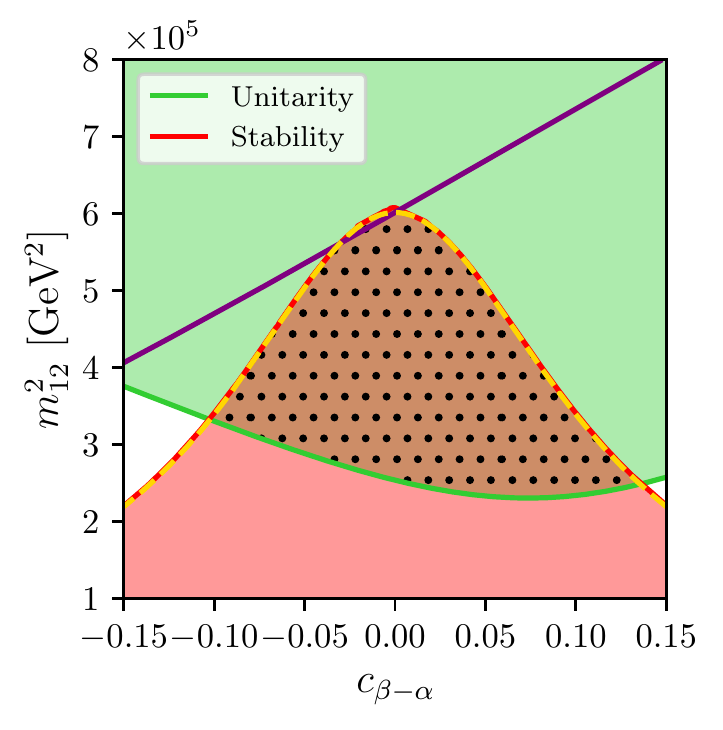}
\includegraphics[height=0.17\textheight]{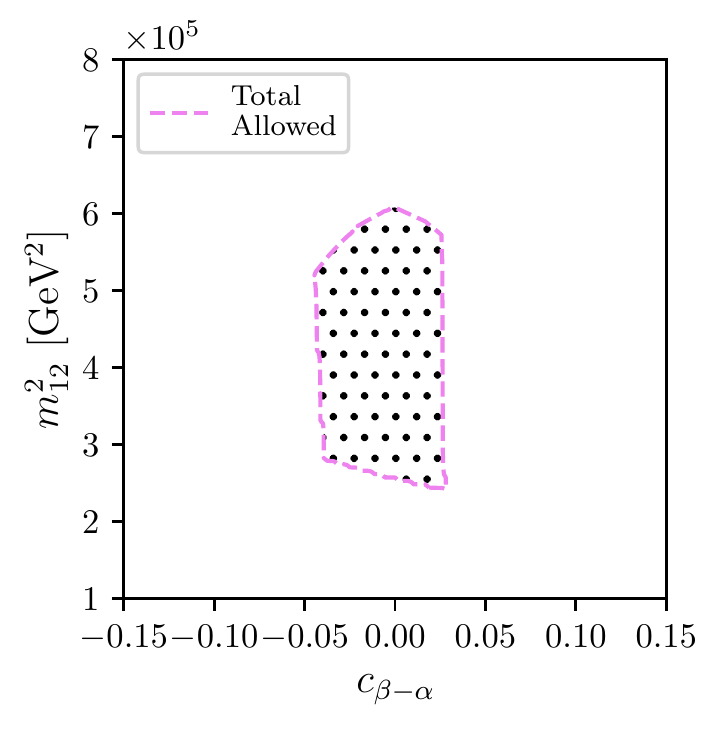}

\includegraphics[width=0.24\textwidth]{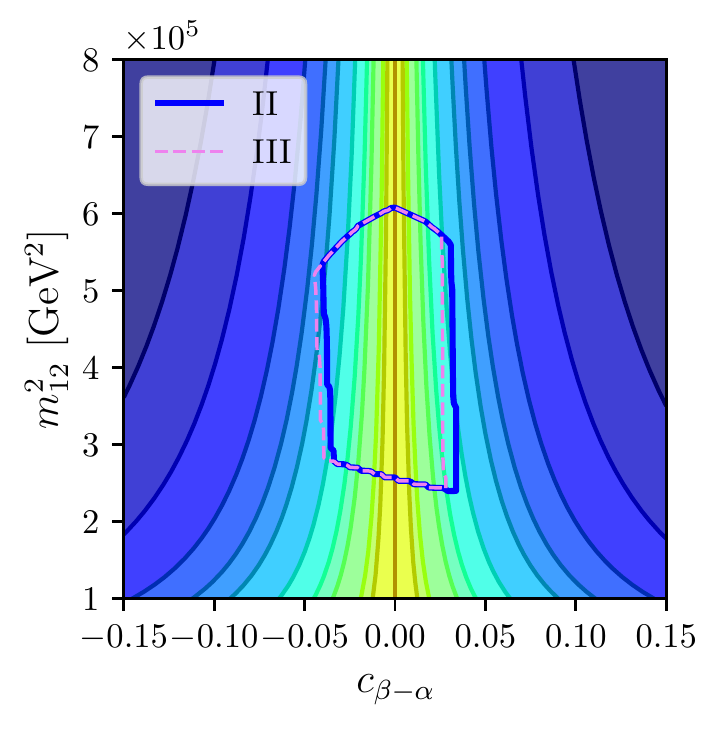}
\includegraphics[width=0.24\textwidth]{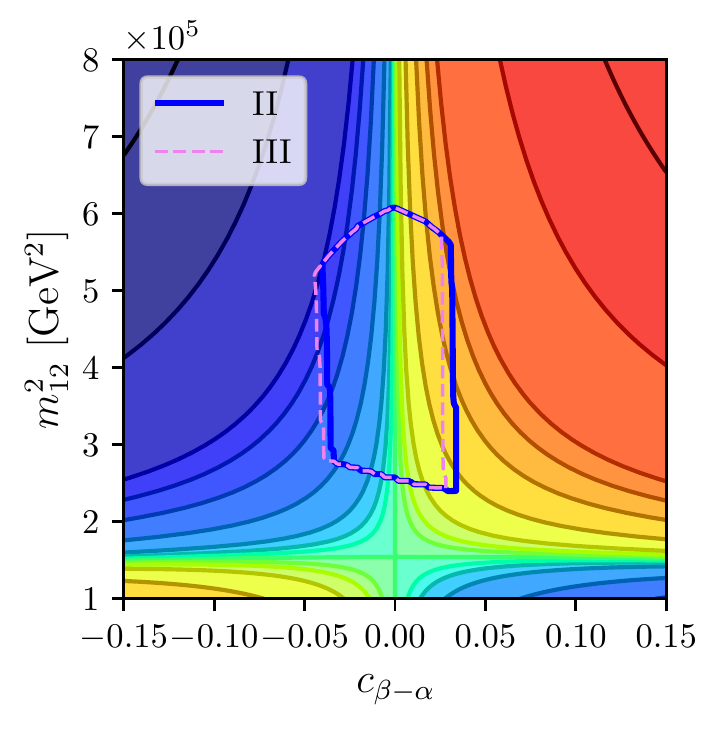}
\includegraphics[width=0.24\textwidth]{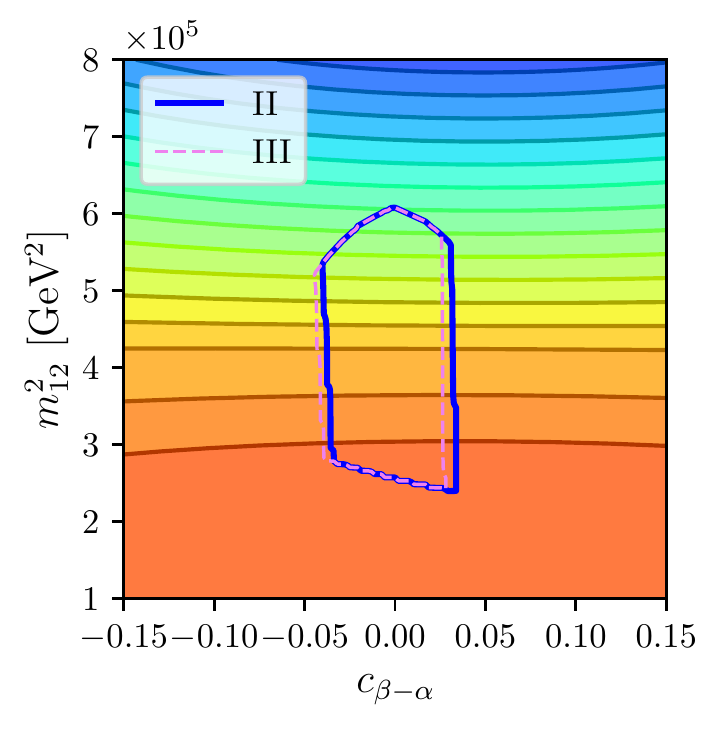}
\includegraphics[width=0.24\textwidth]{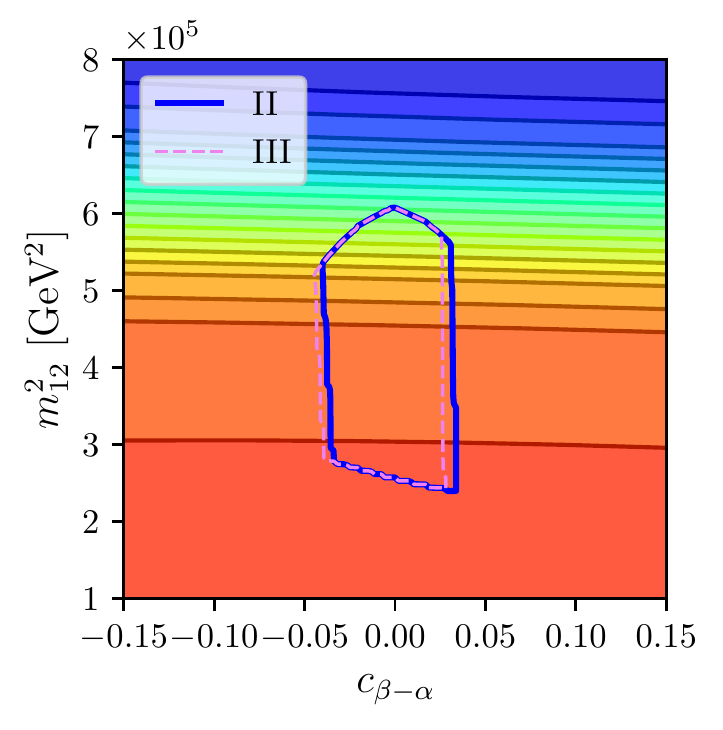}

\includegraphics[width=0.24\textwidth]{h1h1h1_colorbar}
\includegraphics[width=0.24\textwidth]{h1h1h2_colorbar}
\includegraphics[width=0.24\textwidth]{h1h2h2_colorbar}
\includegraphics[width=0.24\textwidth]{h1HpHm_colorbar}
\caption{Allowed areas (dotted regions) from the various constraints  in type~II (upper row) and type~III (middle row)
  and triple Higgs couplings (lower row) 
  for the benchmark scenario II/III-1 in the $\CBA$--$m$ plane
  with $\tb=0.9$ and $m = 1100 \gev$.}
\label{fig:II-III-1}
\end{figure}

The benchmark planes for the 2HDM types~II and~III are defined as (with the
first plane taken over from \citere{Arco:2020ucn}): 

\begin{itemize}

\item[II/III-1:]
  $m \equiv \MHp = \MH = \MA = 1100 \gev$, $\tb = 0.9$,\\
  free parameters: $\CBA$, $\msq$
  
\item[II/III-2:]
  $\CBA=-0.035$,
  $\tb=1.2$,\\
  free parameters: $\msq$, $m\equiv \MH = \MA = \MHp$.
  
\item[II/III-3:]
  $m\equiv \MH = \MA = \MHp = 1300 \gev$,
  $\msq=700000\gev^2$,\\
  free parameters: $\CBA$, $\tb$.
  

\item[II/III-4:]
 $m\equiv \MH = \MA = \MHp = 1000 \gev$,
 $\CBA=-0.035$,\\
 free parameters: $\msq$, $\tb$.

\end{itemize}

\begin{figure}[t!]
\centering
\includegraphics[height=0.16\textheight]{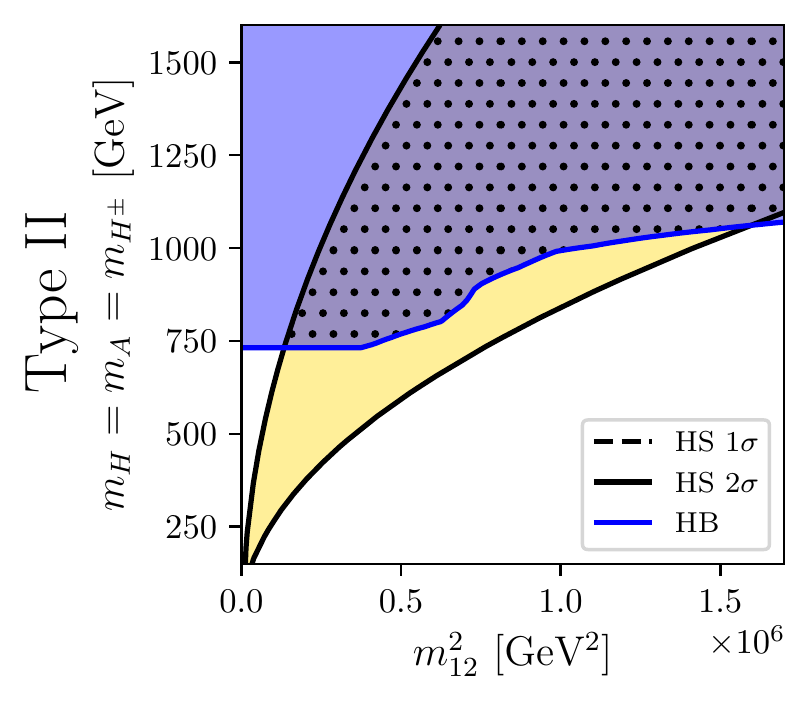}
\includegraphics[height=0.16\textheight]{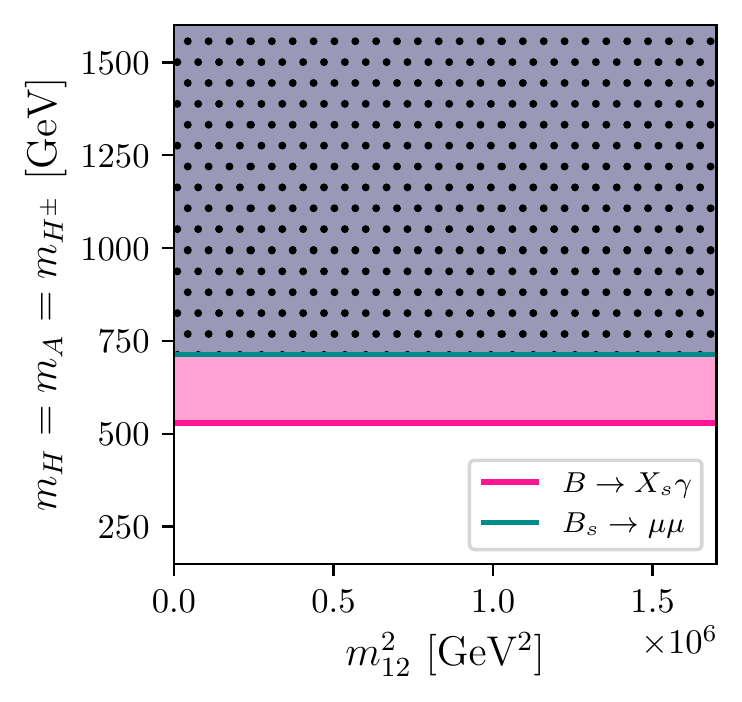}
\includegraphics[height=0.16\textheight]{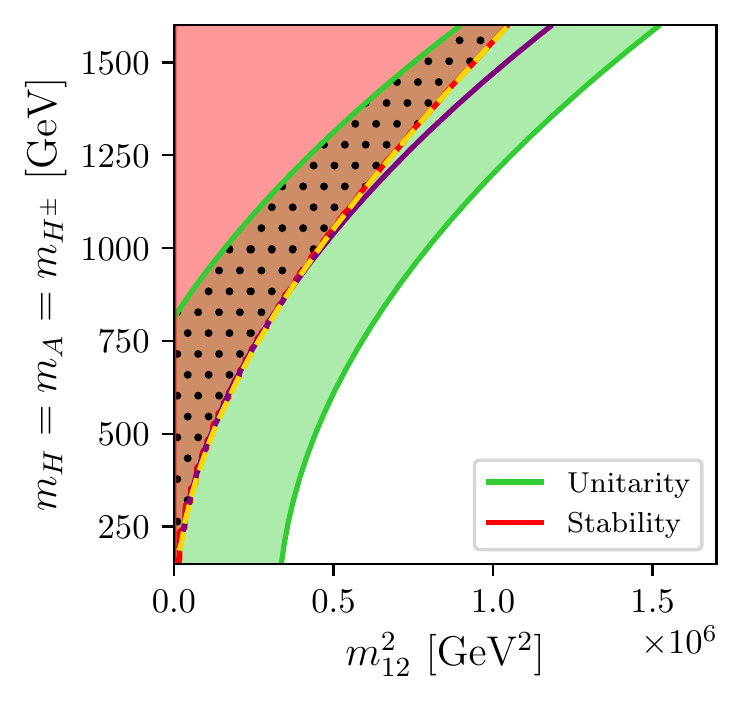}
\includegraphics[height=0.16\textheight]{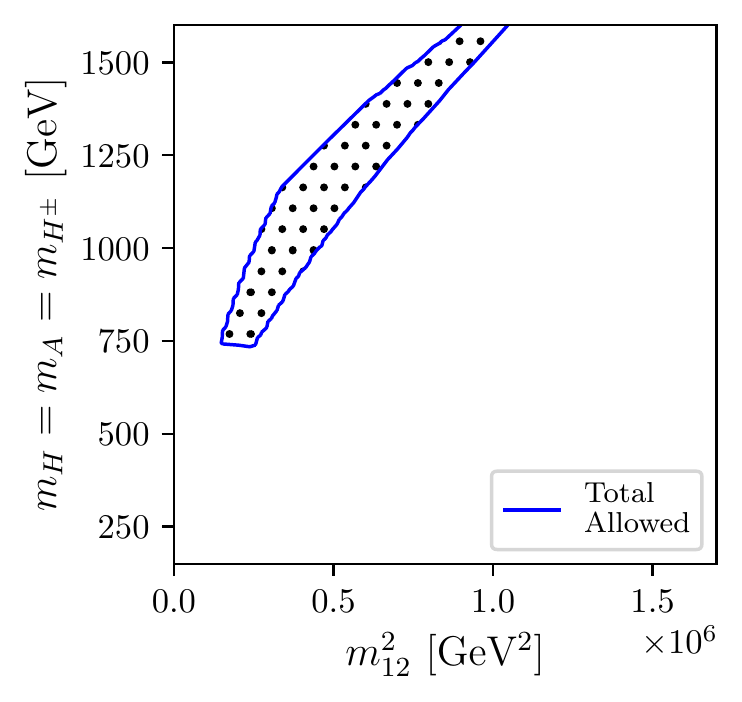}

\includegraphics[height=0.16\textheight]{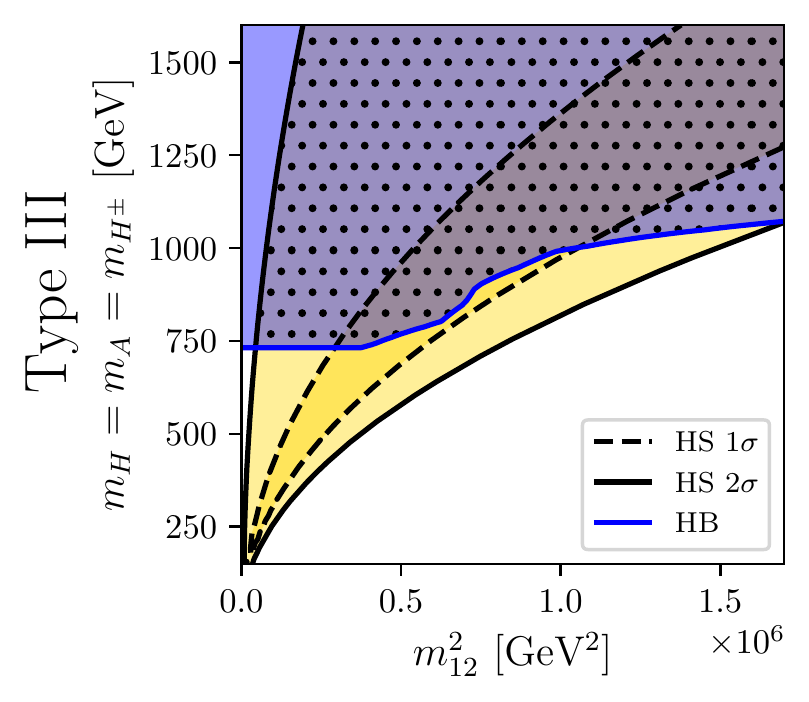}
\includegraphics[height=0.16\textheight]{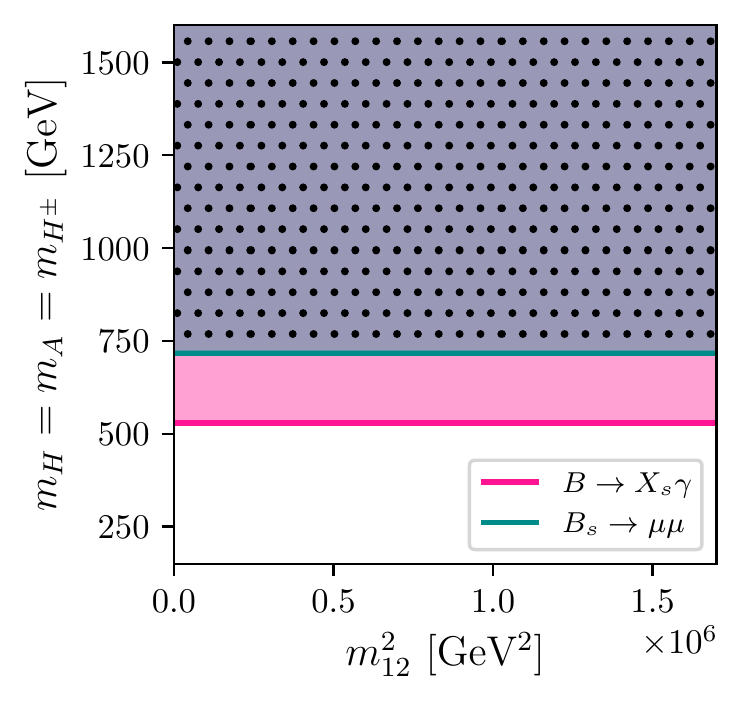}
\includegraphics[height=0.16\textheight]{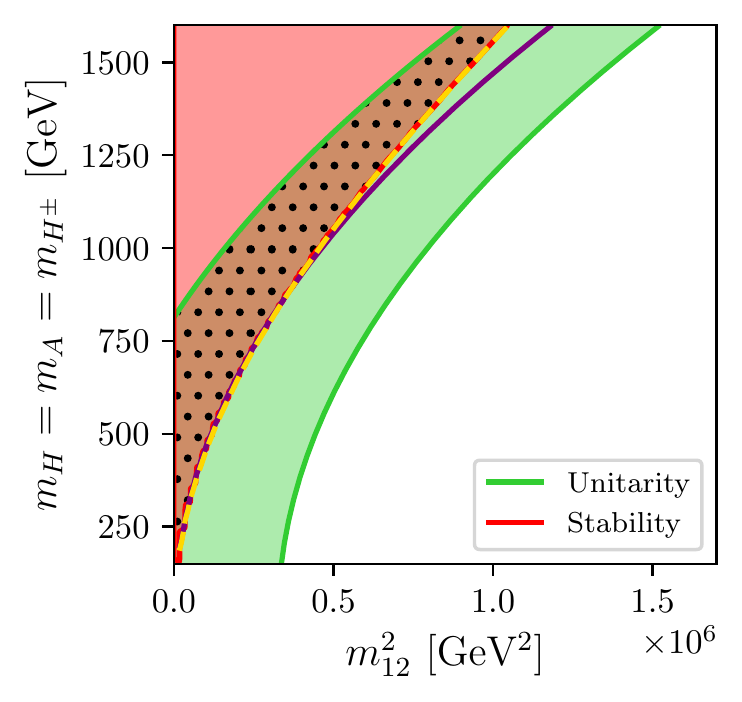}
\includegraphics[height=0.16\textheight]{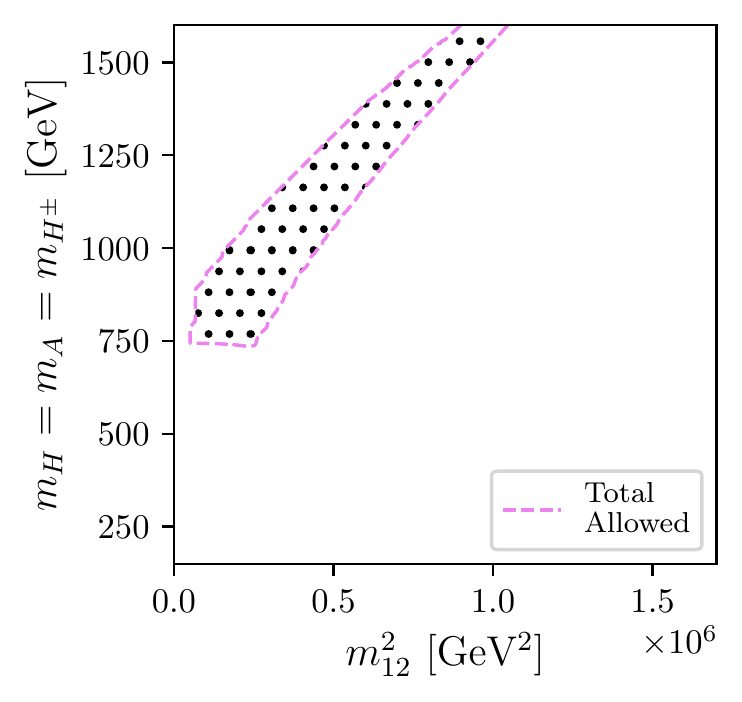}

\includegraphics[width=0.24\textwidth]{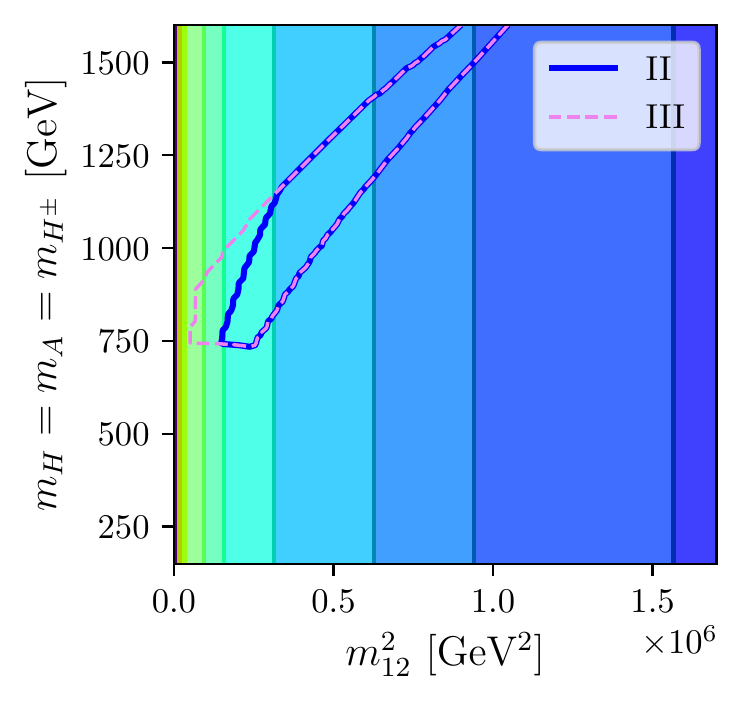}
\includegraphics[width=0.24\textwidth]{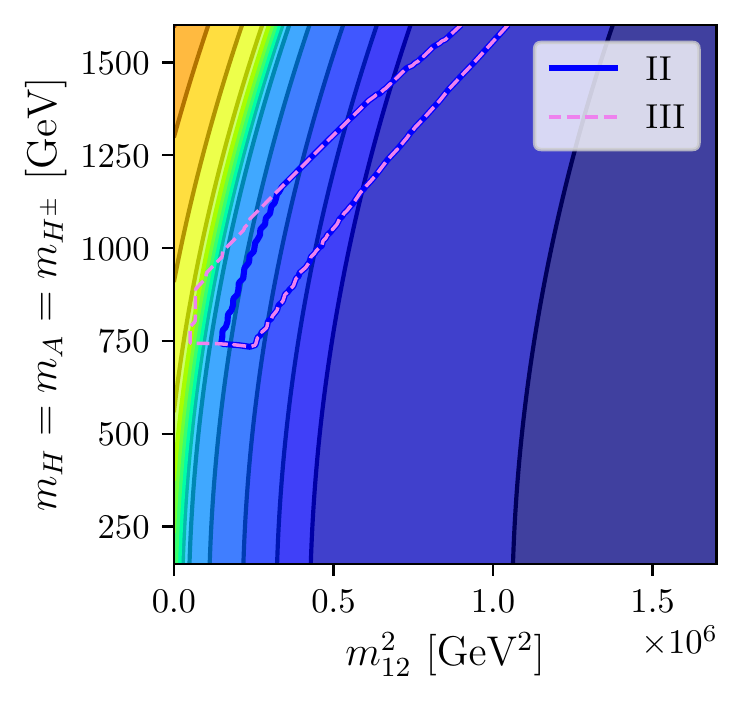}
\includegraphics[width=0.24\textwidth]{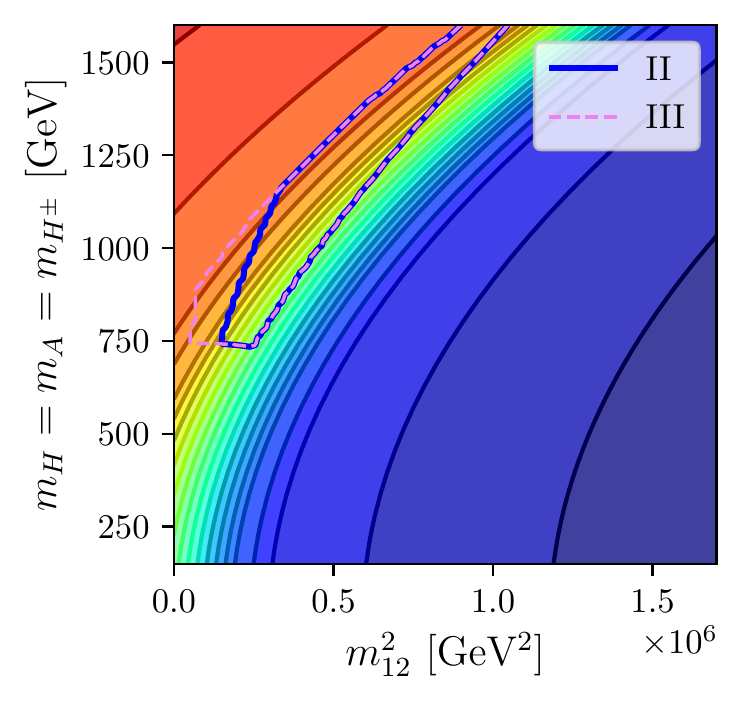}
\includegraphics[width=0.24\textwidth]{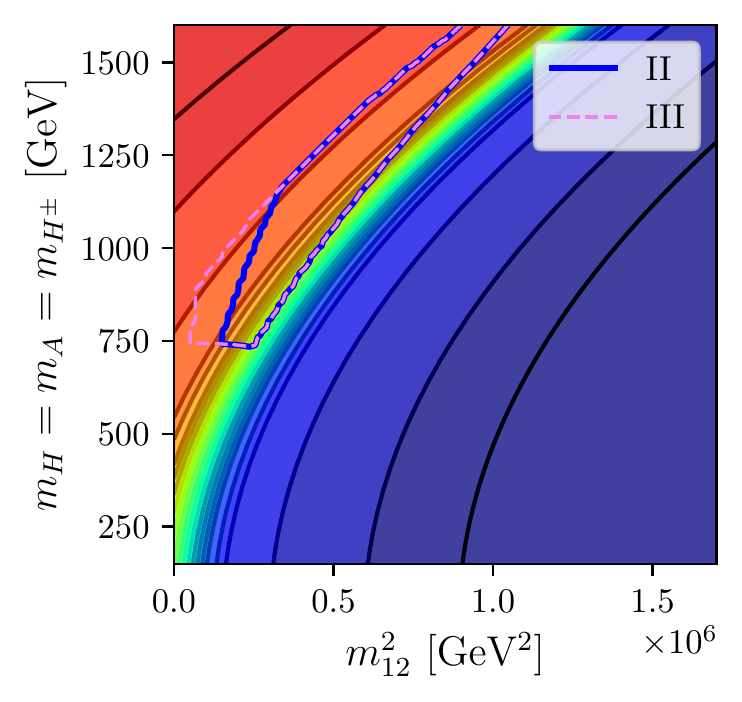}

\includegraphics[width=0.24\textwidth]{h1h1h1_colorbar}
\includegraphics[width=0.24\textwidth]{h1h1h2_colorbar}
\includegraphics[width=0.24\textwidth]{h1h2h2_colorbar}
\includegraphics[width=0.24\textwidth]{h1HpHm_colorbar}
\caption{Allowed areas (dotted regions) from the various constraints in type II (upper row) and type III (middle row)
  and triple Higgs couplings (lower row) 
  for the benchmark scenario II/III-2 in the $\msq$--$m$ plane with
  $\tb=1.2$ and $\CBA=-0.035$.}
\label{fig:II-III-2}
\end{figure}

\begin{figure}[t!]
\centering
\includegraphics[height=0.16\textheight]{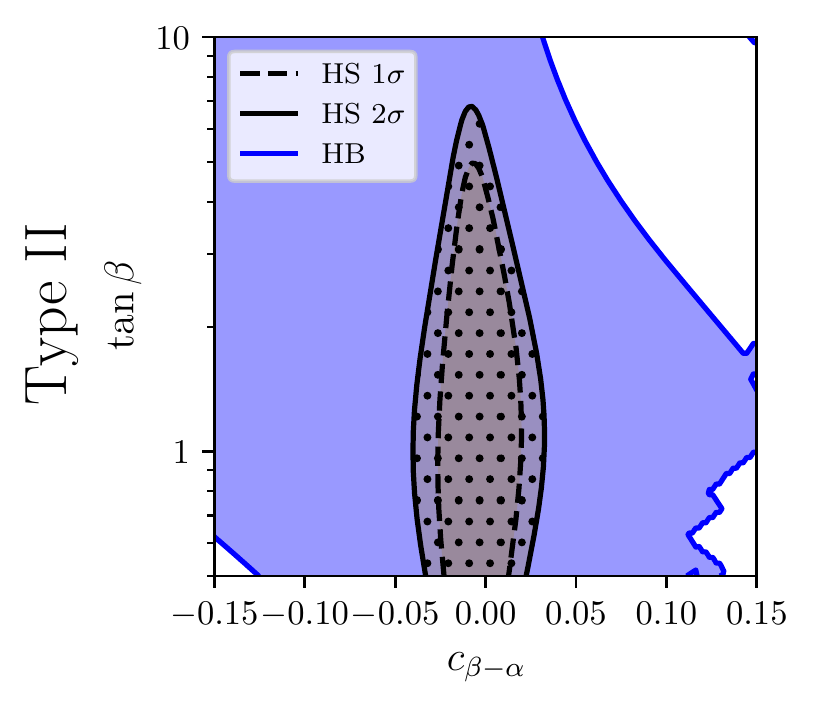}
\includegraphics[height=0.16\textheight]{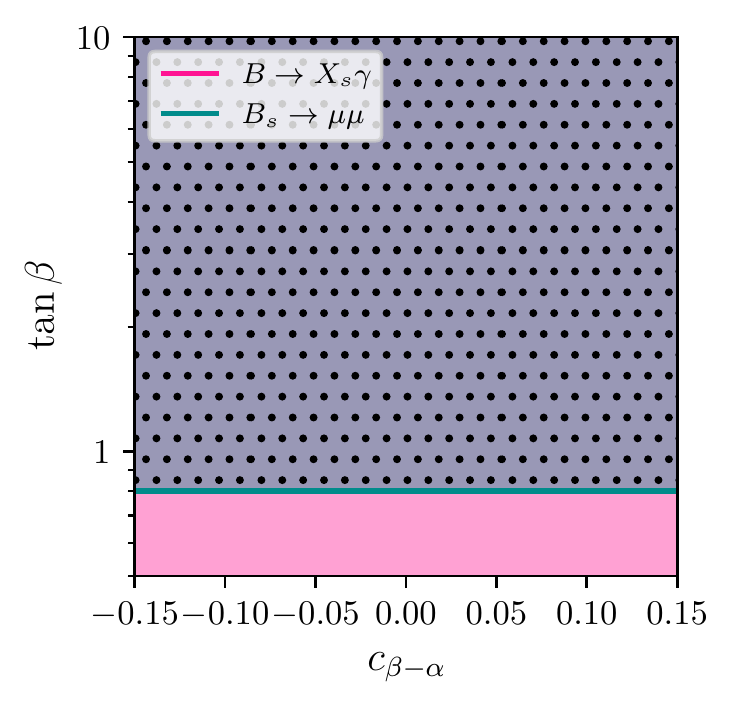}
\includegraphics[height=0.16\textheight]{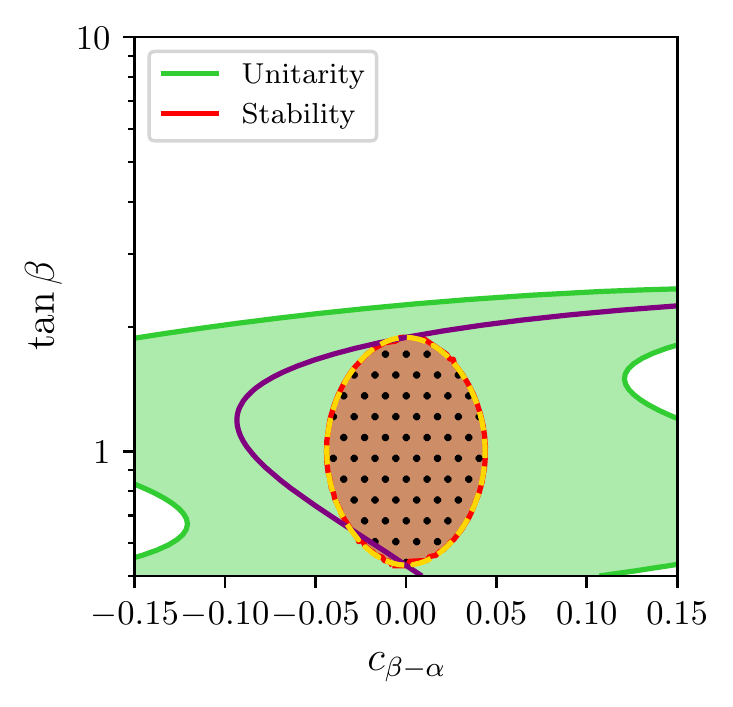}
\includegraphics[height=0.16\textheight]{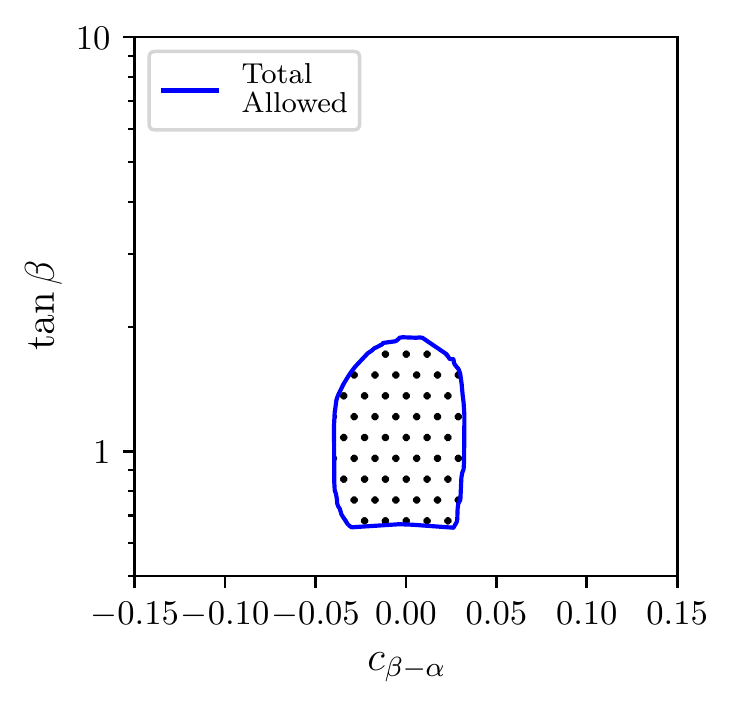}

\includegraphics[height=0.16\textheight]{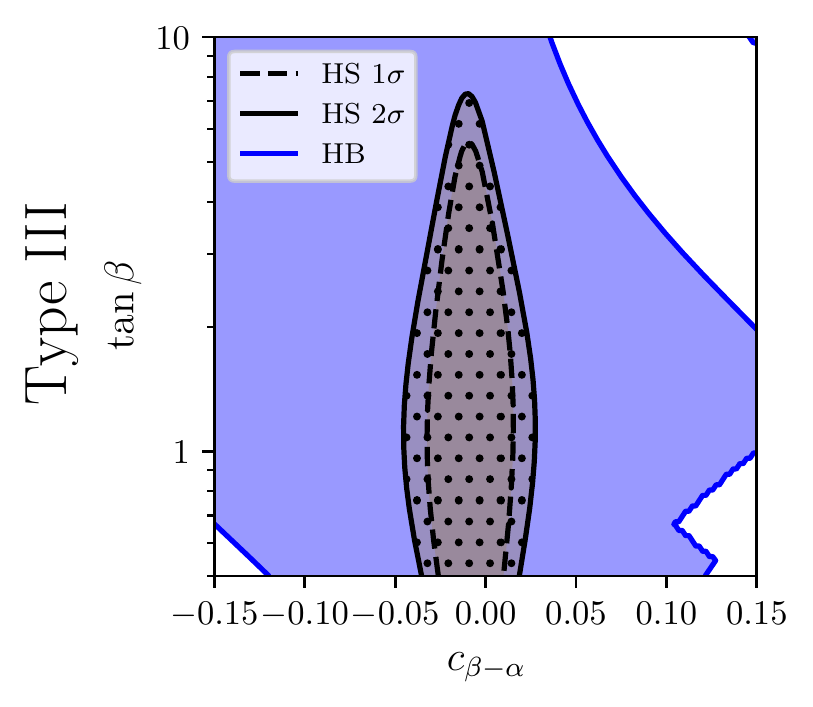}
\includegraphics[height=0.16\textheight]{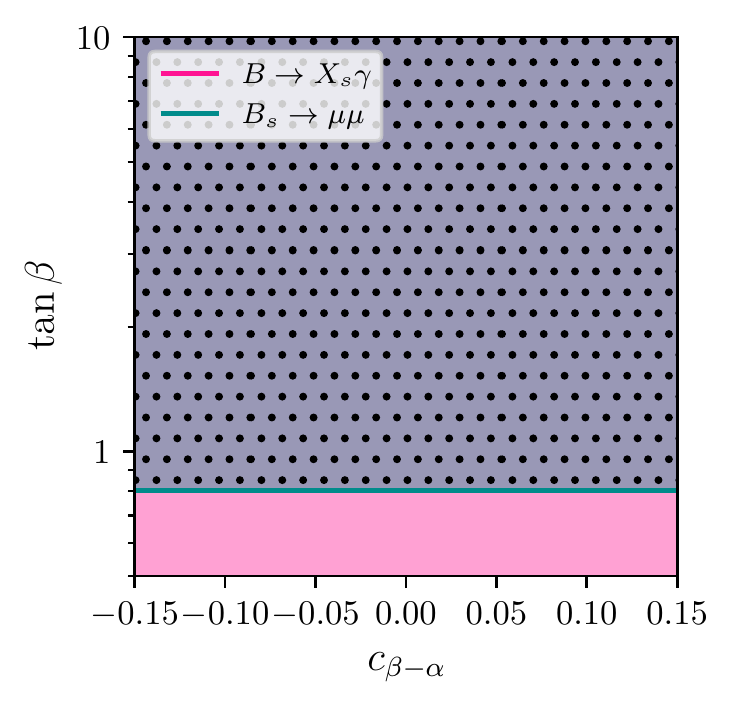}
\includegraphics[height=0.16\textheight]{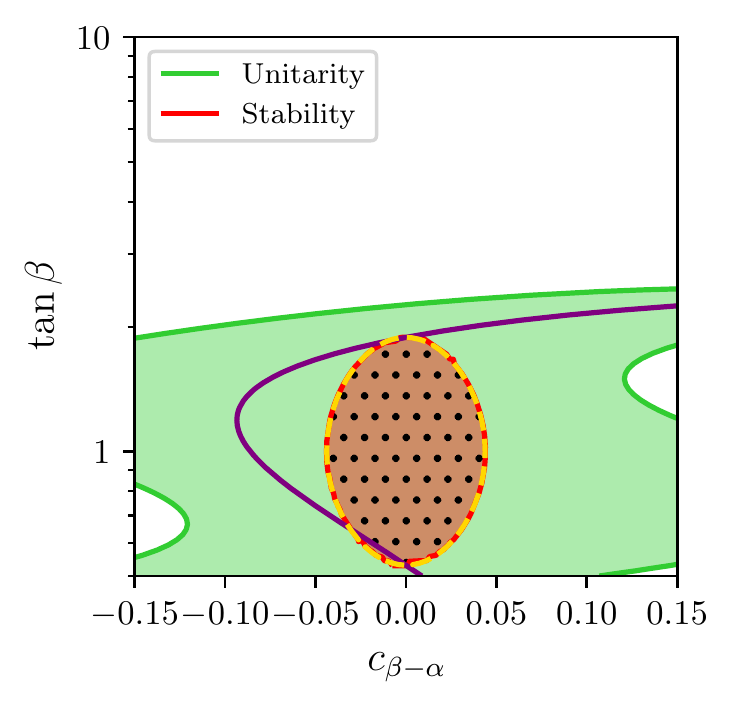}
\includegraphics[height=0.16\textheight]{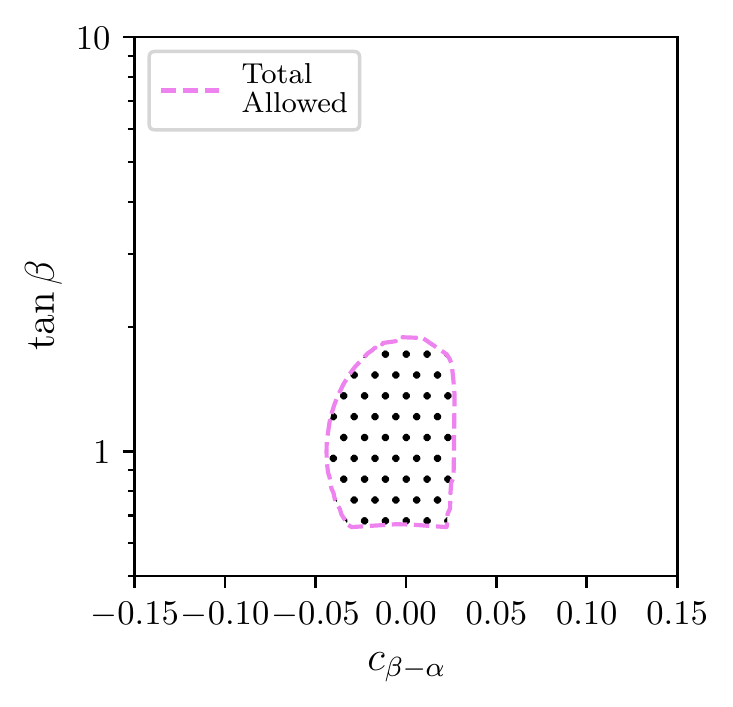}

\includegraphics[width=0.24\textwidth]{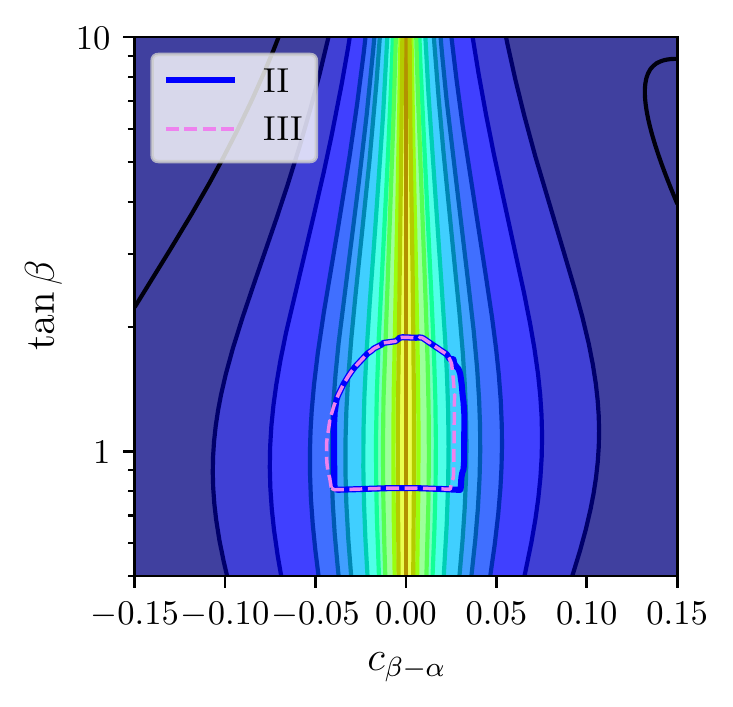}
\includegraphics[width=0.24\textwidth]{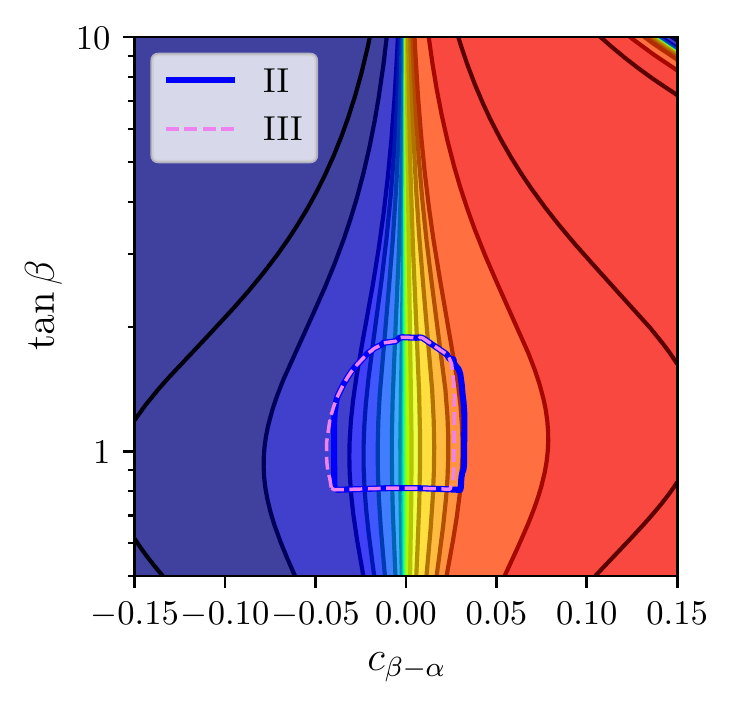}
\includegraphics[width=0.24\textwidth]{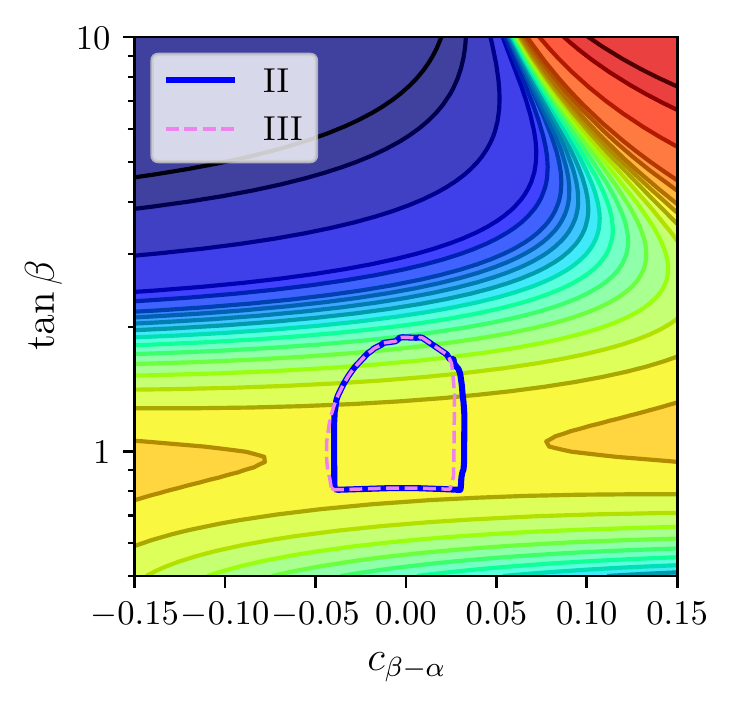}
\includegraphics[width=0.24\textwidth]{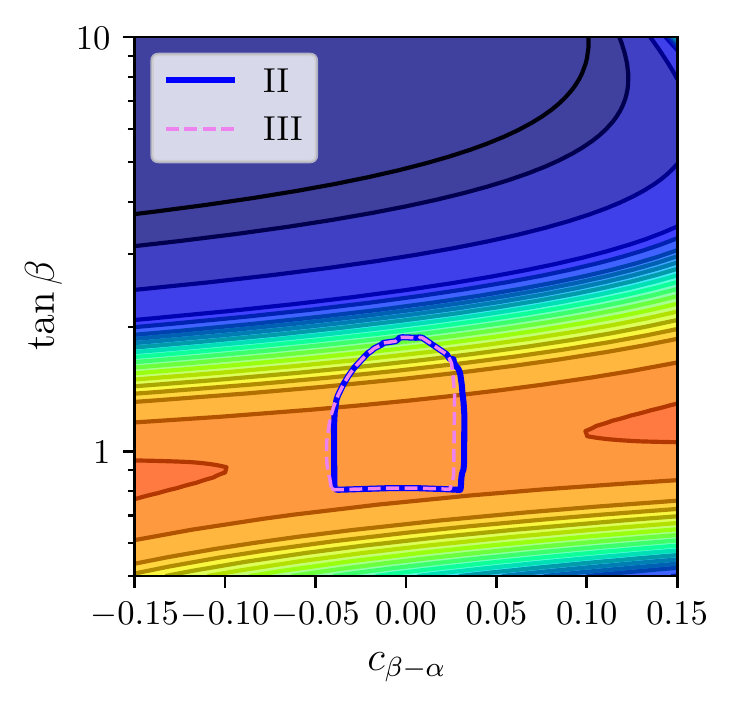}

\includegraphics[width=0.24\textwidth]{h1h1h1_colorbar}
\includegraphics[width=0.24\textwidth]{h1h1h2_colorbar}
\includegraphics[width=0.24\textwidth]{h1h2h2_colorbar}
\includegraphics[width=0.24\textwidth]{h1HpHm_colorbar}
\caption{Allowed areas (dotted regions) from the various constraints in type II (upper row) and type III (middle row)
  and triple Higgs couplings (lower row) 
  for the benchmark scenario II/III-3 in the $\CBA$--$\tb$ plane with
  $m\equiv \MH = \MA = \MHp = 1300 \gev$ and $\msq=700000\gev^2$.}
\label{fig:II-III-3}
\end{figure}

\begin{figure}[h]
\centering
\includegraphics[height=0.17\textheight]{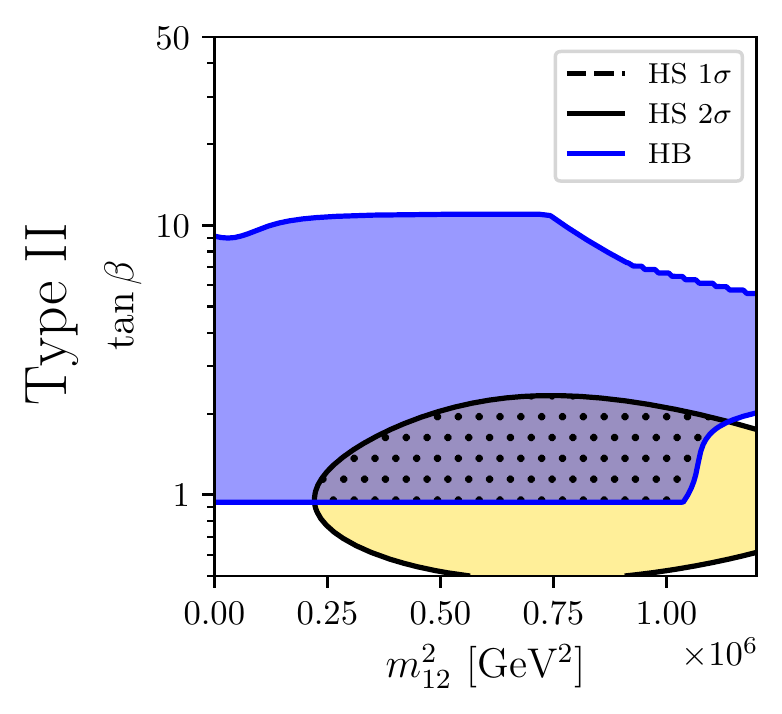}
\includegraphics[height=0.17\textheight]{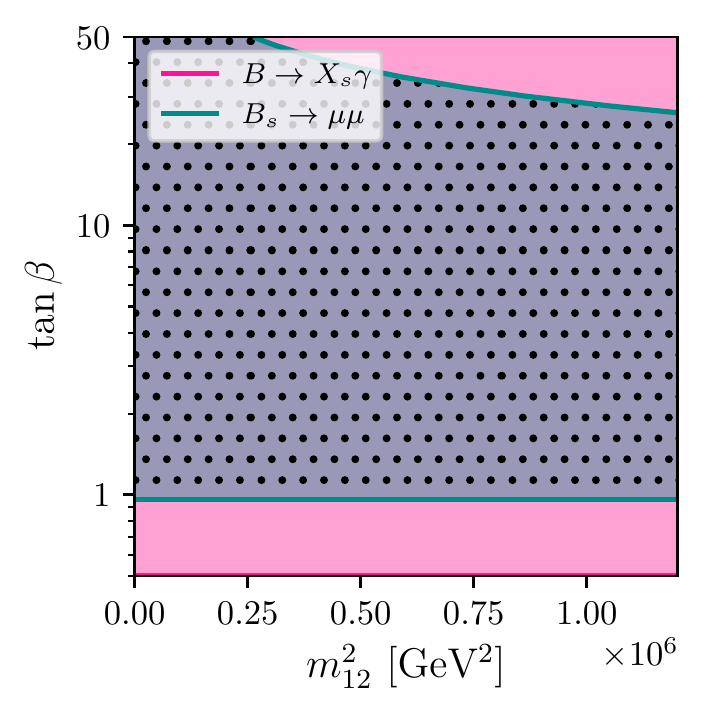}
\includegraphics[height=0.17\textheight]{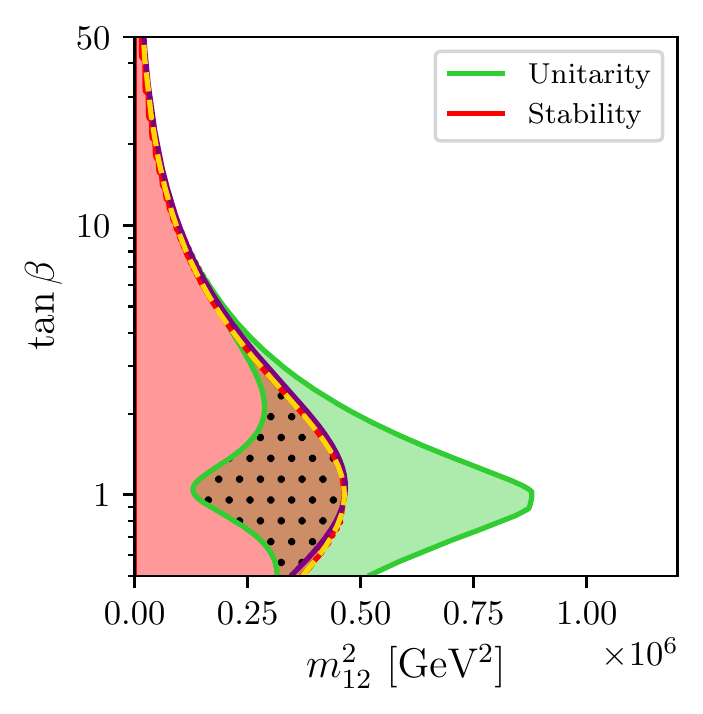}
\includegraphics[height=0.17\textheight]{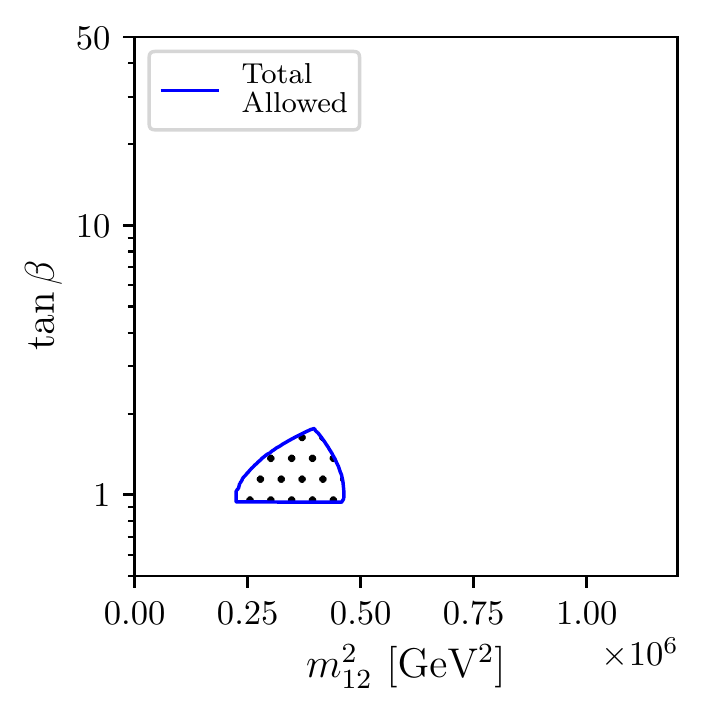}

\includegraphics[height=0.17\textheight]{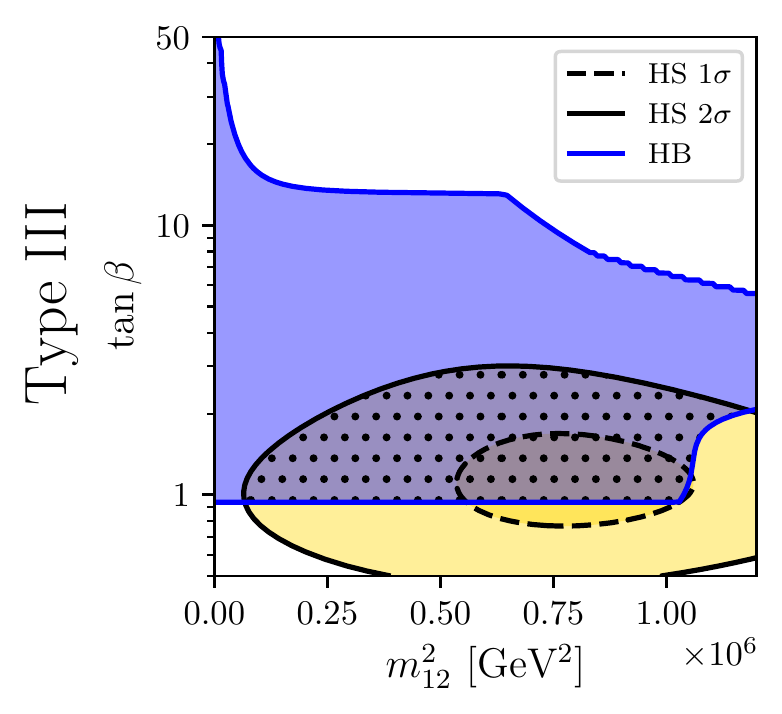}
\includegraphics[height=0.17\textheight]{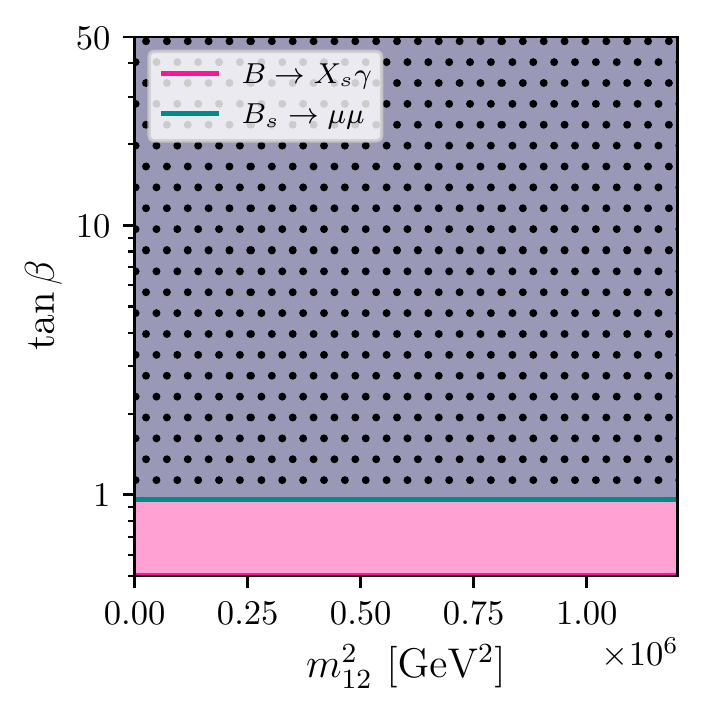}
\includegraphics[height=0.17\textheight]{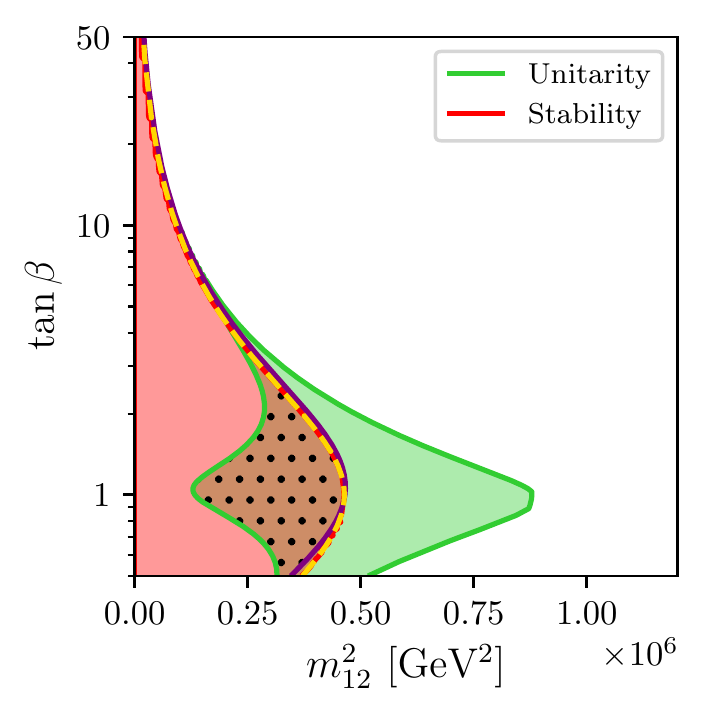}
\includegraphics[height=0.17\textheight]{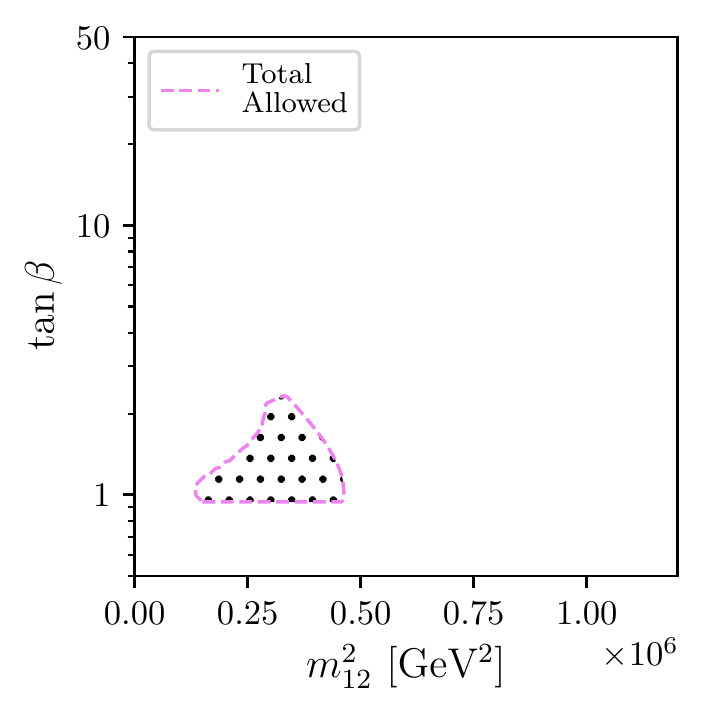}

\includegraphics[width=0.24\textwidth]{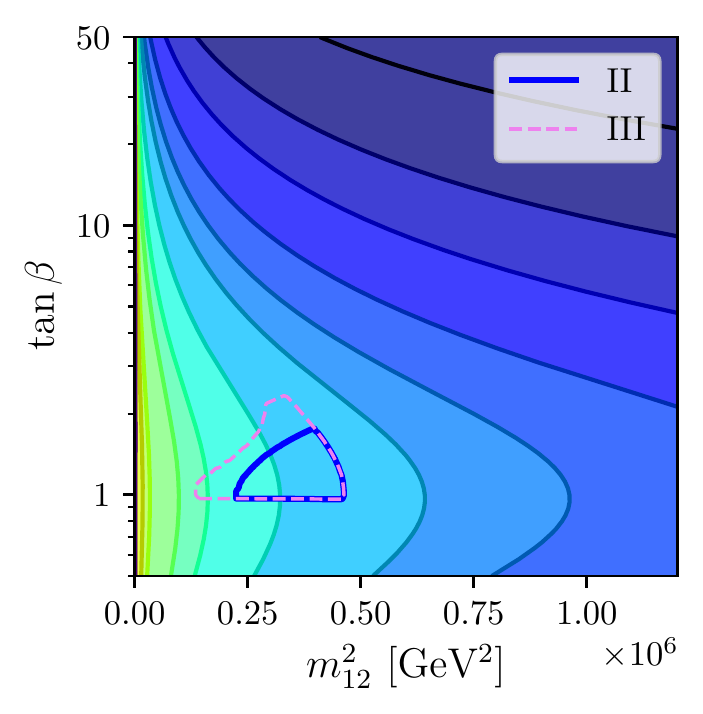}
\includegraphics[width=0.24\textwidth]{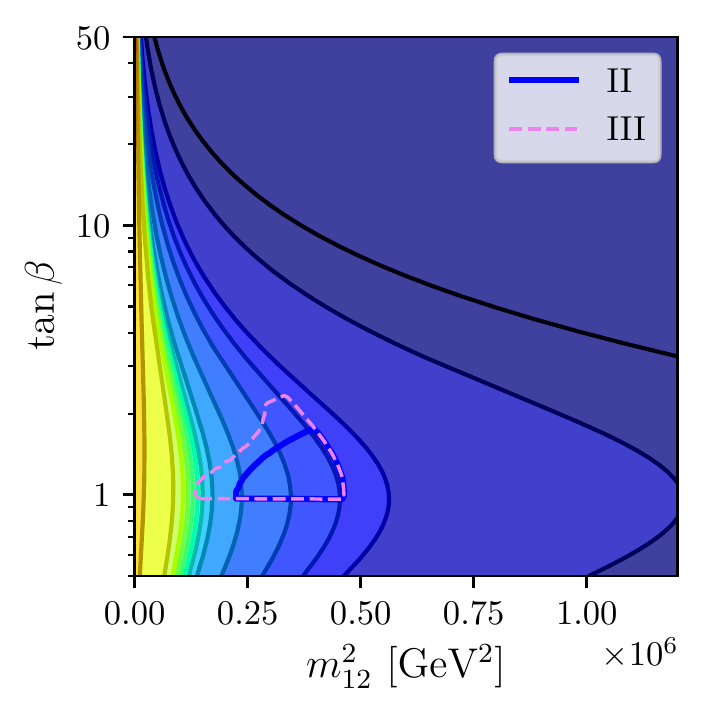}
\includegraphics[width=0.24\textwidth]{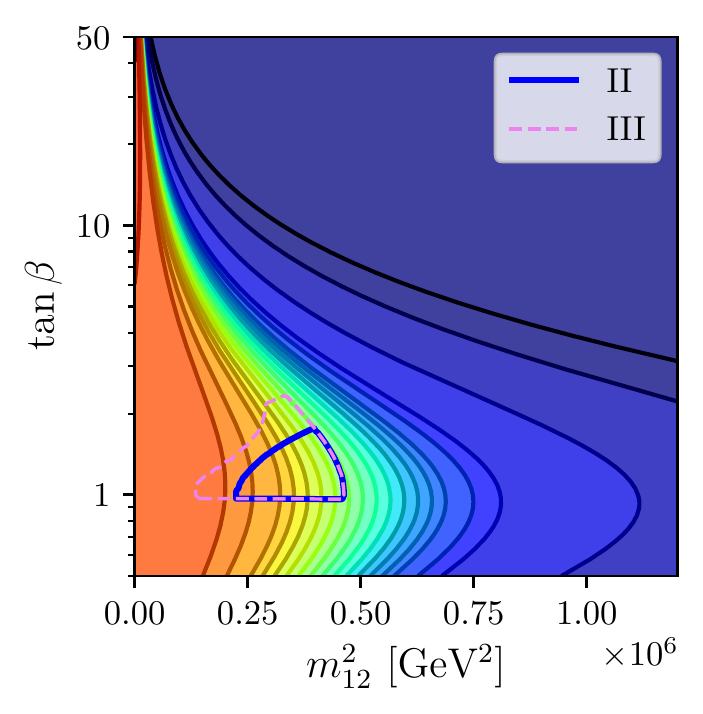}
\includegraphics[width=0.24\textwidth]{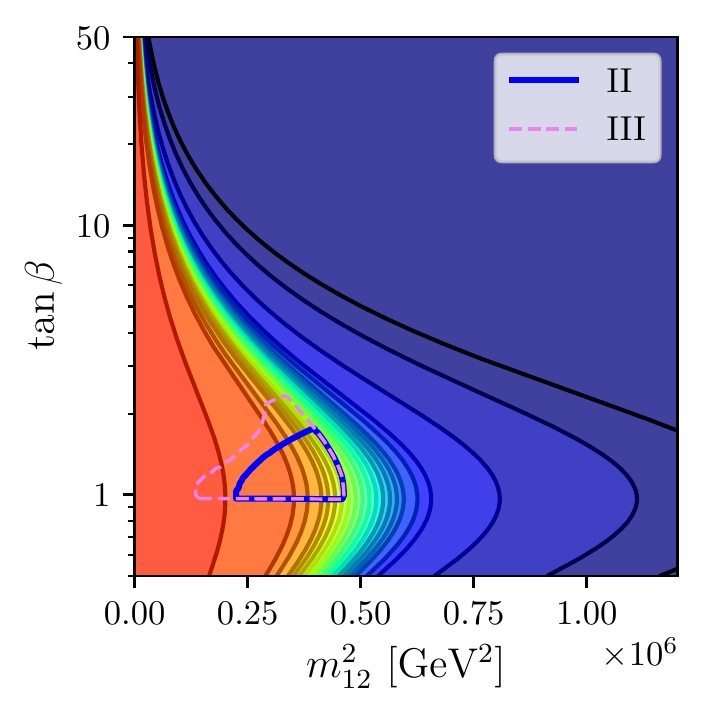}

\includegraphics[width=0.24\textwidth]{h1h1h1_colorbar}
\includegraphics[width=0.24\textwidth]{h1h1h2_colorbar}
\includegraphics[width=0.24\textwidth]{h1h2h2_colorbar}
\includegraphics[width=0.24\textwidth]{h1HpHm_colorbar}
\caption{Allowed areas (dotted regions) from the various constraints (upper row) and triple Higgs couplings (lower row)
  for the benchmark scenario II/III-4 in the $\msq$--$\tb$ plane with
  $m\equiv \MH = \MA = \MHp = 1000 \gev$ and $\CBA=-0.035$.}
\label{fig:II-III-4}
\end{figure}

The results for the first scenario II/III-1, shown in
\reffi{fig:II-III-1}, is an update of the same scenario as presented in 
\citere{Arco:2020ucn}, but now analyzed for the two Yukawa types~II
and~III. It is shown in the $\CBA$--$\msq$ plane with $m = 1100 \gev$
and $\tb = 0.9$.
The main difference for type~II w.r.t.\ the previous analysis
consists in the stronger  
bounds from the Higgs-boson signal-rate measurements (as included by
\HS). This results in particular in a tighter bound on $\CBA$, as can be
seen in the upper left plot of \reffi{fig:II-III-1}, where we find
$\CBA \sim \inter{-0.04}{0.03}$. Flavor constraints
allow the whole plane, whereas unitarity/stability selects a nearly
triangular region, as can be observed in the upper row, middle-right
plot. Together with the tighter bounds from the Higgs-boson rate
measurements the dotted area shown in the upper right plot remains
allowed in this scenario. Nearly identical results are found in the
Yukawa type~III, as can be seen in the middle row of
\reffi{fig:II-III-1}. 
The corresponding allowed regions for the various triple Higgs
couplings are shown for both Yukawa types in the lower row of
\reffi{fig:II-III-1}. Since the results are so similar for type~II
and~III here and for \reffis{fig:II-III-2} - \ref{fig:II-III-4}
we only quote a common set of allowed intervals. 
With the stronger bounds on $\CBA$ we find $\kala \sim \inter{0.8}{1}$,
where the largest deviations from unity are found for the largest
deviations of $\CBA$ from zero, i.e.\ the alignment limit.
Similarly, also $\lahhH$ is more restricted in type~II than in
\citere{Arco:2020ucn},  $\lahhH \sim \inter{-1}{0.8}$.  
The situation is different for the triple Higgs couplings involving two
heavy Higgs bosons. These depend only mildly on $\CBA$, but strongly on
$\msq$. For $\lahHH \sim \lahAA = \lahHpHm/2$ the largest values reached
in the allowed area are $\sim 12$, with the largest values found for the
smallest $\msq$.

The second scenario for Yukawa types~II and~III, denoted as II/III-2, is
shown in \reffi{fig:II-III-2}. The overall allowed parameter space, shown
as dotted area in the upper and middle right plots is found for
$m \sim \inter{750 \gev}{1600 \gev}$ 
(where the upper limit is the end of our scan range) and
$\msq \sim \inter{1.5(0.5)\times10^5\gev^2}{10^6 \gev^2}$ in type~II (III)
(where the upper limit is given by
the upper limit on $m$). It should be noted that in this scenario
the lowest  allowed value of $m \sim 750 \gev$ is set mainly by the
flavor constraints on $\MHp$. The values of $\kala$ in this scenario are all
smaller than~1, where the smallest value of $\kala \sim 0.67$
are found for the largest $m$ and $\msq$ region.
$\lahhH$ is found to be negative, with the smallest values
$\lahhH \sim -1.7$ again for the large $m$, $\msq$
region. The largest values of $\lahHH$ and $\lahHpHm = 2 \lahAA$ are
found for the largest values of $m$ and $\msq$ that are reached in
the upper right corner of the allowed region,  reaching values of 
$\sim 12$ and $\sim 24$, respectively. 

The third scenario of Yukawa types~II and~III, denoted as II/III-3, is
presented in \reffi{fig:II-III-3} in the $\CBA$--$\tb$ plane. The
overall allowed 
parameter space is given as a combination of all types of constraints
and found for $\CBA \sim \inter{-0.04}{0.03}$ and
$\tb \sim \inter{0.7}{1.8}$. 
The deviations in $\lahhh$ from the SM value in this scenario are
relatively small, $\kala \sim \inter{0.66}{1}$, where the smallest values are
found for the lowest allowed $\CBA$. The values $\lahhH$ range in
$\lahhH \sim \inter{-1.4}{0.9}$, depending mainly on $\CBA$.
The values of the triple Higgs couplings involving two heavy Higgs
bosons, on the other hand, depends mainly on $\tb$ with the largest
values, $\lahHH \sim \lahAA = \lahHpHm/2 \sim 5$, are
found around $\tb \sim 1$. The smallest values of $\sim 0.1$ are reached at
$\tb \sim 1.8$. 

The final scenario chosen for Yukawa types~II and~III, denoted as
II-III-4, is shown in \reffi{fig:II-III-4} in the $\msq$--$\tb$ plane. The
strongest constraints, particularly in $\msq$ are given by a
combination of the unitarity/stability limits and the Higgs-boson
rate measurements, where the latter yields a reduction of the allowed
parameter space in type~II w.r.t.\ type~III. Consequently, we will quote
different (particularly upper) limits for the tripe Higgs couplings for the two
Yukawa types in this scenario. We find for type~II~(III)
$\msq \sim \inter{2.2 (1.3) \times 10^5 \gev^2}{4.6 \times 10^5 \gev^2}$.
The lower $\tb$ limit of $\tb \gsim 0.95$ is due to the BSM Higgs
searches, where the charged Higgs-boson searches yield the exclusion.
The dependences of all triple Higgs couplings on the two free parameters
are similar in this plane. The smallest (largest) values are found for
the largest (smallest) values of $\msq$. 
The ranges found in this scenario are
$\kala \sim \inter{0.85}{0.93}$,
$\lahhH \sim \inter{-0.79}{-0.22}$ and
$\lahHH \sim \lahAA = \lahHpHm/2 \sim \inter{1.1}{9.2}$
for type II and 
$\kala \sim \inter{0.85}{0.96}$,
$\lahhH \sim \inter{-0.8}{-0.01}$ and
$\lahHH \sim \lahAA = \lahHpHm/2 \sim \inter{1.0}{12.2}$
for type III.



\subsection{Triple Higgs couplings in the 2HDM type IV}
\label{sec:typeIV}

We finish our overview of the four Yukawa types of the 2HDM with three
benchmark planes in type~IV, which are defined as:

\begin{itemize}
\item[IV-1:]
  $m\equiv \MH = \MA = \MHp = 1300 \gev$,
  $\CBA=-0.02$,\\
  free parameters: $\msq$, $\tb$
  
\item[IV-2:]
  $m\equiv \MH = \MA = \MHp= 1300 \gev$,
  $\msq$ fixed by \refeq{eq:m12special2}, \\
  free parameters: $\CBA$, $\tb$

\item[IV-3:]
  $\CBA = 0.02$, 
  $\msq$ fixed by \refeq{eq:m12special}, \\
  free parameters $m\equiv \MH = \MA = \MHp$,$\tb$,
  
\item[IV-4:]
  $\msq$ fixed by \refeq{eq:m12special},
  $\tb$ fixed via \refeq{eq:wrongsign} (wrong sign Yukawa limit),\\
  free parameters: $\CBA$, $m\equiv \MH = \MA = \MHp$
\end{itemize}

\begin{figure}[t!]
\centering
\includegraphics[width=0.239\textwidth]{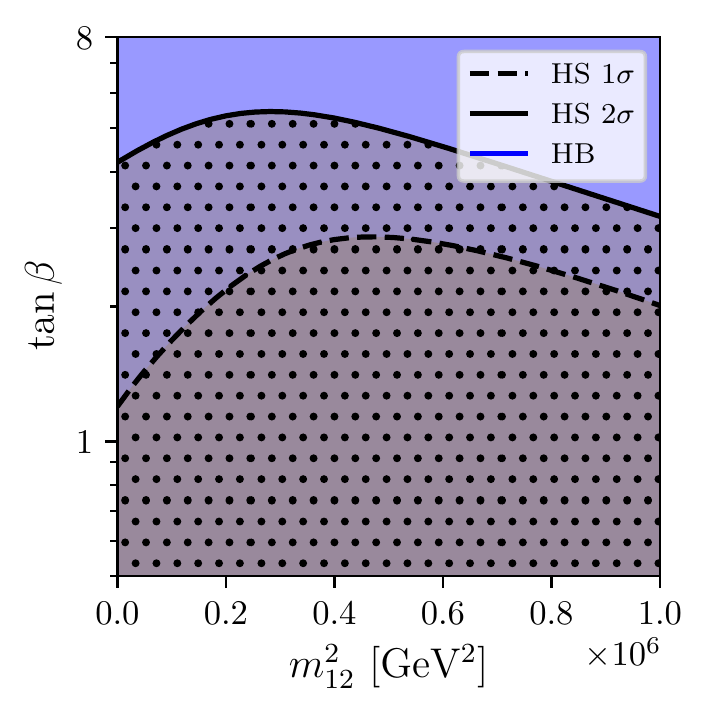}
\includegraphics[width=0.239\textwidth]{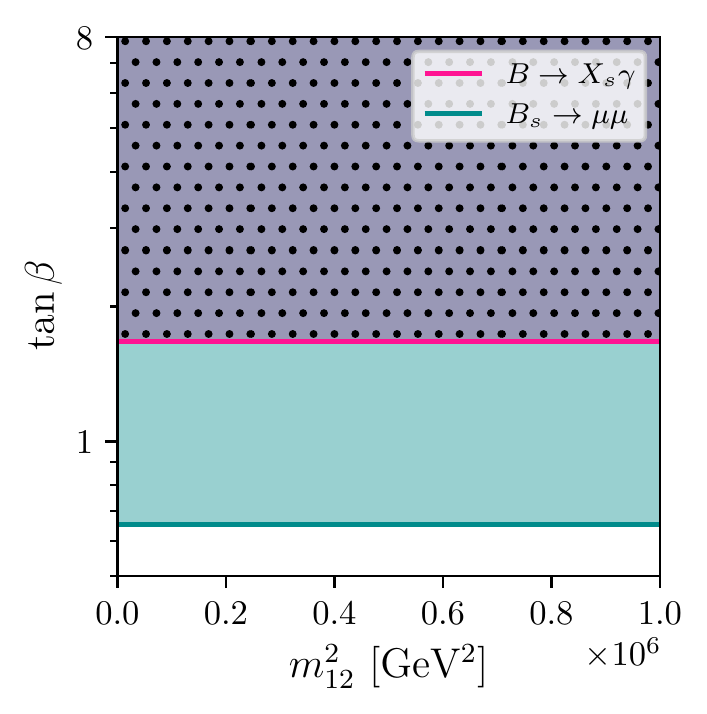}
\includegraphics[width=0.239\textwidth]{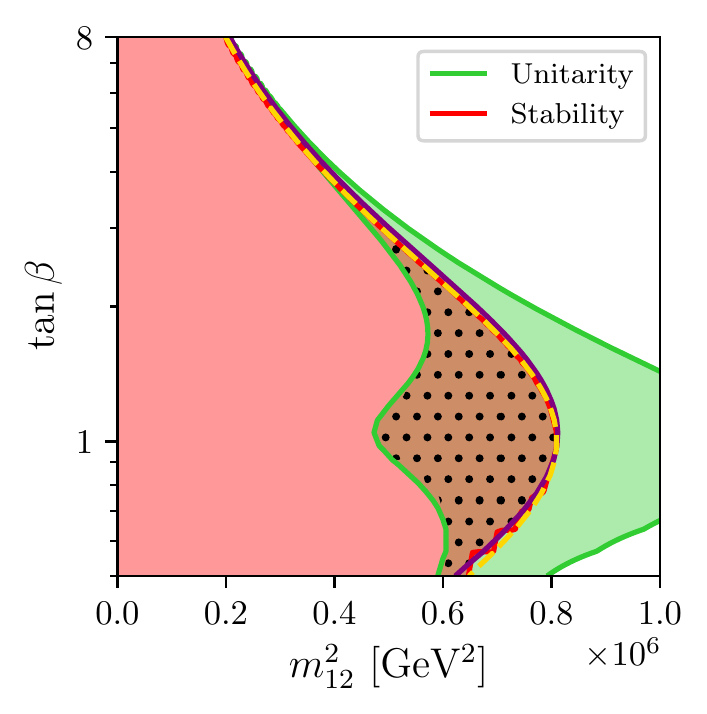}
\includegraphics[width=0.239\textwidth]{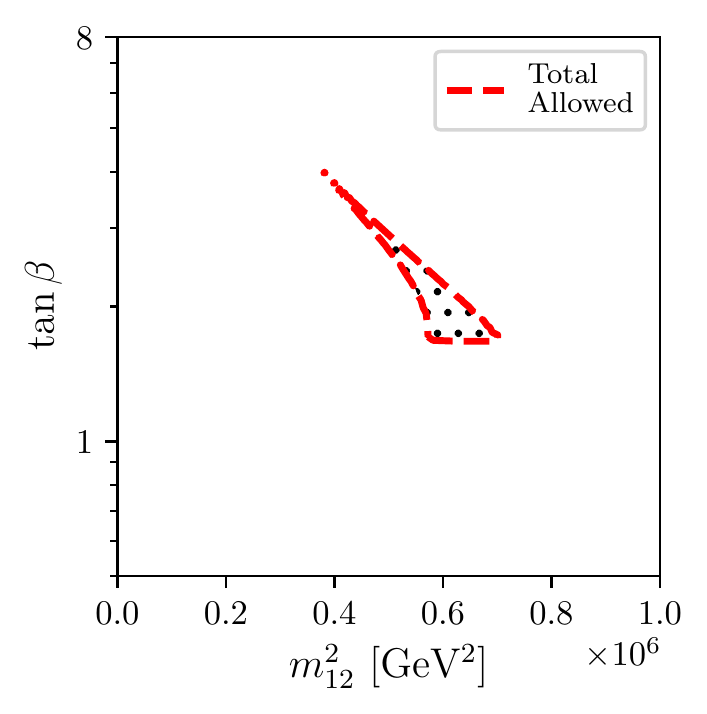}

\includegraphics[width=0.239\textwidth]{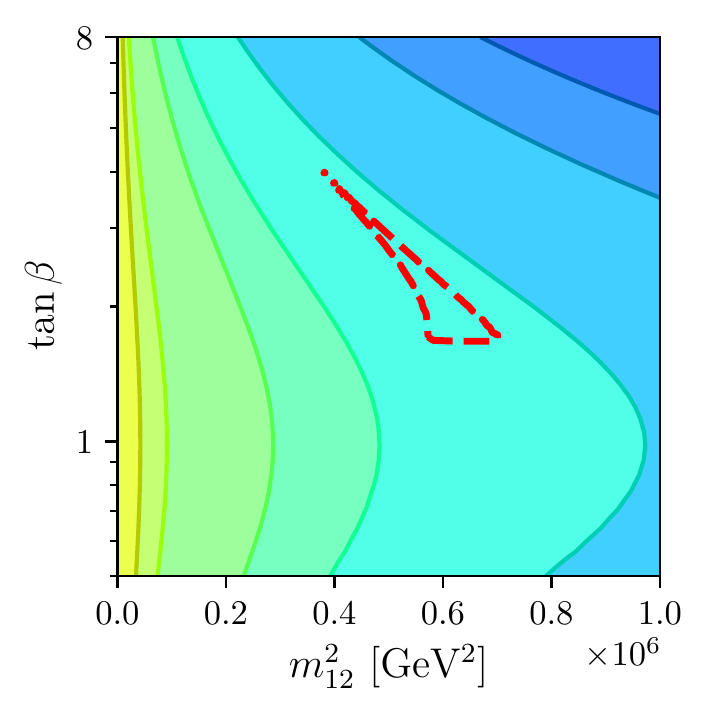}
\includegraphics[width=0.239\textwidth]{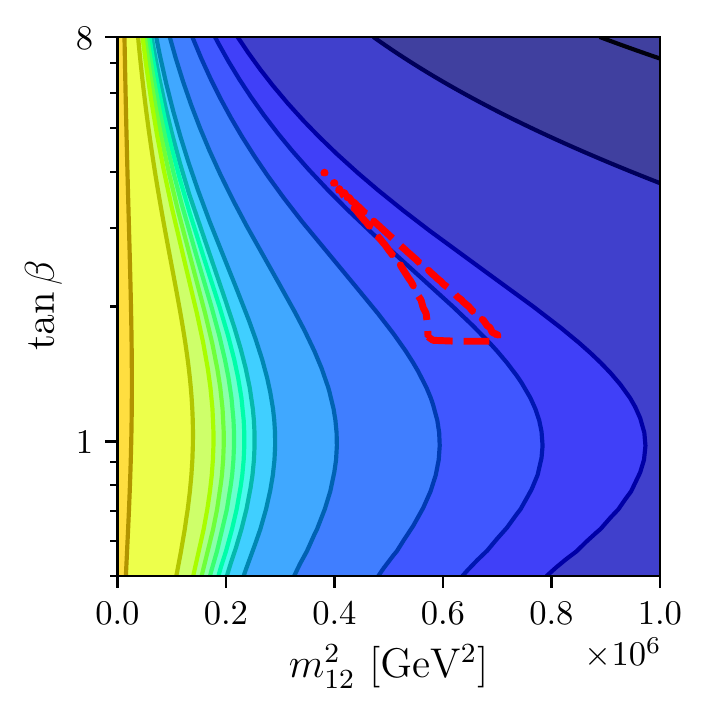}
\includegraphics[width=0.239\textwidth]{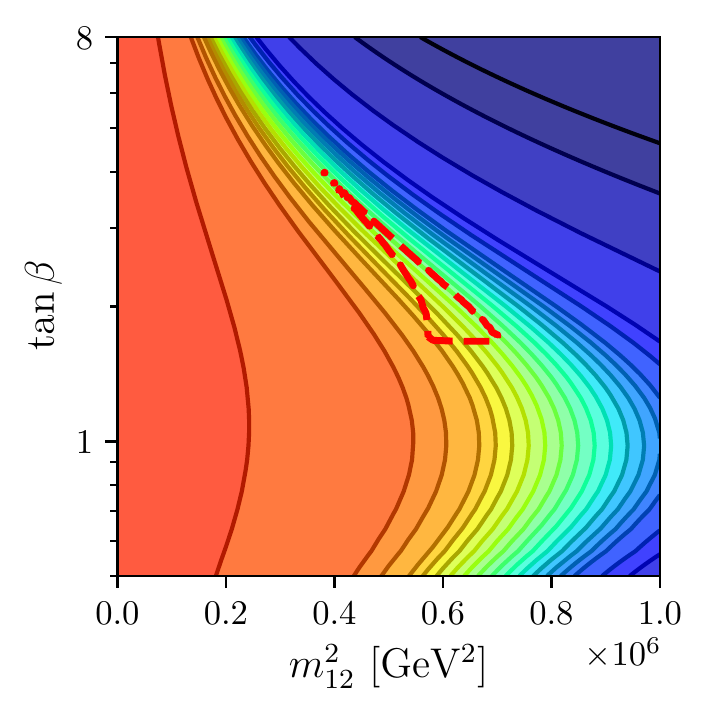}
\includegraphics[width=0.239\textwidth]{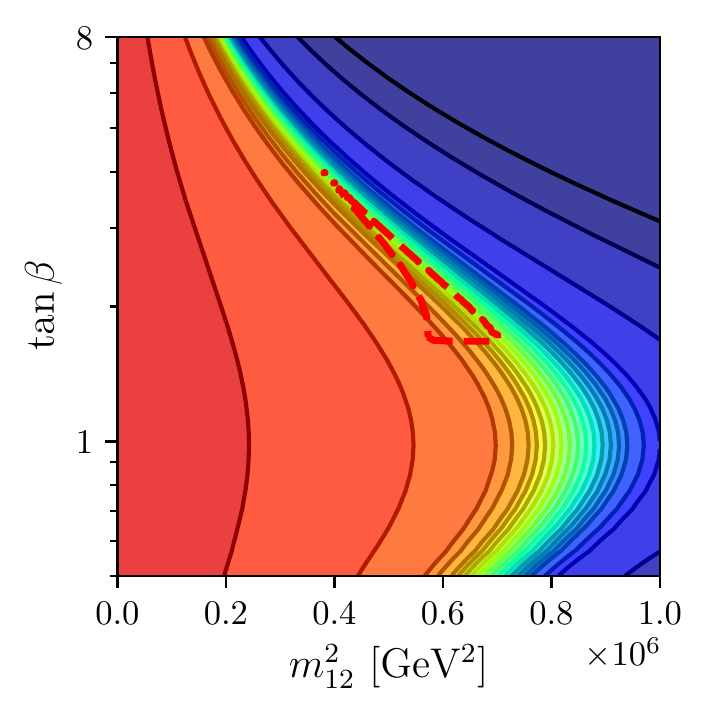}

\includegraphics[width=0.239\textwidth]{h1h1h1_colorbar}
\includegraphics[width=0.239\textwidth]{h1h1h2_colorbar}
\includegraphics[width=0.239\textwidth]{h1h2h2_colorbar}
\includegraphics[width=0.239\textwidth]{h1HpHm_colorbar}
\caption{Allowed areas (dotted regions) from the various constraints (upper row) and triple Higgs couplings (lower row)
  for the benchmark scenario IV-1 in the $\msq$--$\tb$ plane with
  $m\equiv \MH = \MA = \MHp = 1300 \gev$ and $\CBA=-0.02$.}
\label{fig:IV-1}
\end{figure}

The first scenario of type~IV, denoted as IV-1, is presented in
\reffi{fig:IV-1} in the $\msq$--$\tb$ plane with $m = 1300 \gev$ and
$\CBA = -0.02$. The lower bound on $\tb$ is given by $\br(B \to X_s \ga)$
at around $\tb \sim 1.7$. The unitarity/stability
constraints then restrict the allowed area to a triangular shape
reaching up to $\tb \sim 4$. The variations of $\kala$ and $\lahhH$ are
very small in this small allowed parameter space, with values of
$\kala \sim 0.92$ and $\lahhH \sim -0.7$. 
The largest values of the other triple Higgs couplings are found for the
lowest $\tb$ and at the same time smallest $\msq$. They are given by
$\lahHH \sim \lahAA = \lahHpHm/2 \sim 6$. 

\begin{figure}[ht!]
\centering
\includegraphics[width=0.239\textwidth]{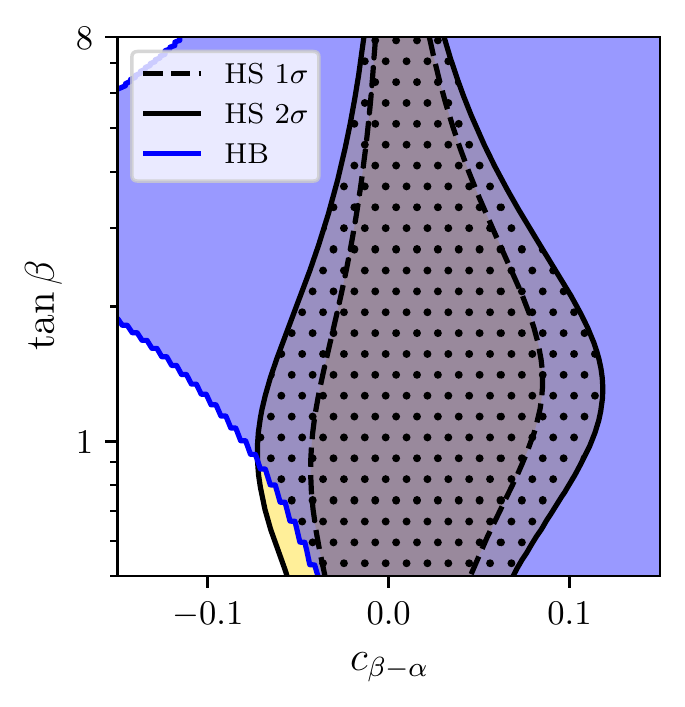}
\includegraphics[width=0.239\textwidth]{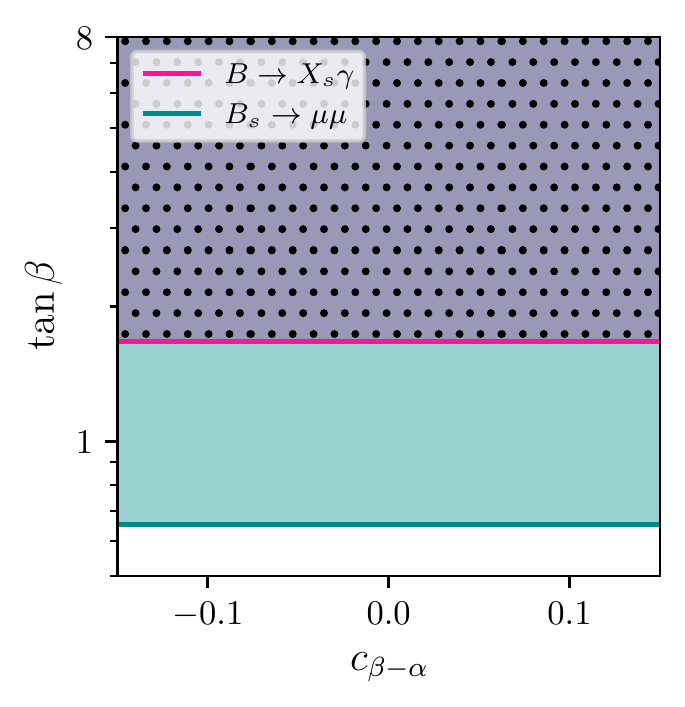}
\includegraphics[width=0.239\textwidth]{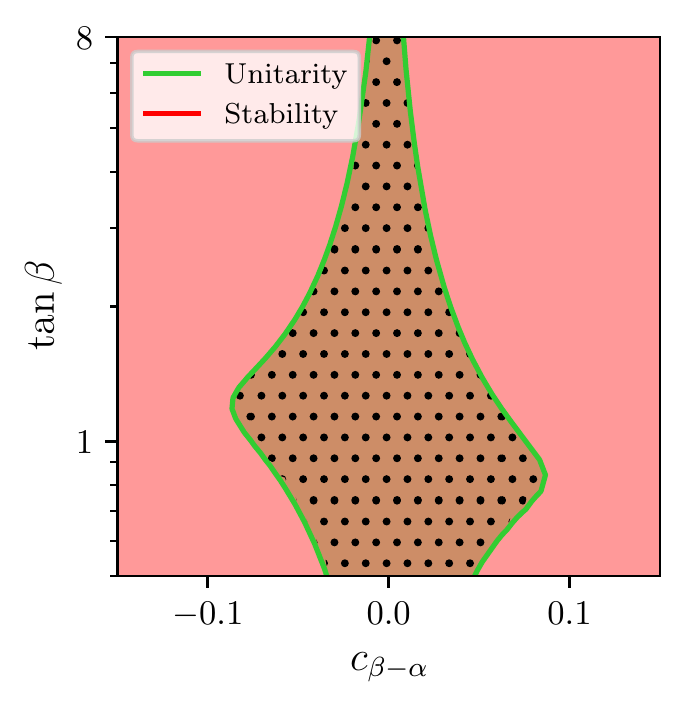}
\includegraphics[width=0.239\textwidth]{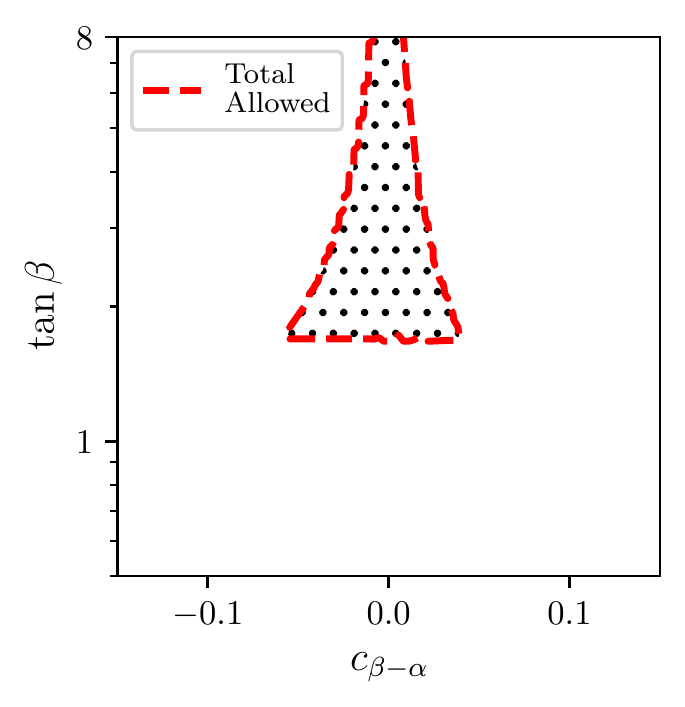}

\includegraphics[width=0.239\textwidth]{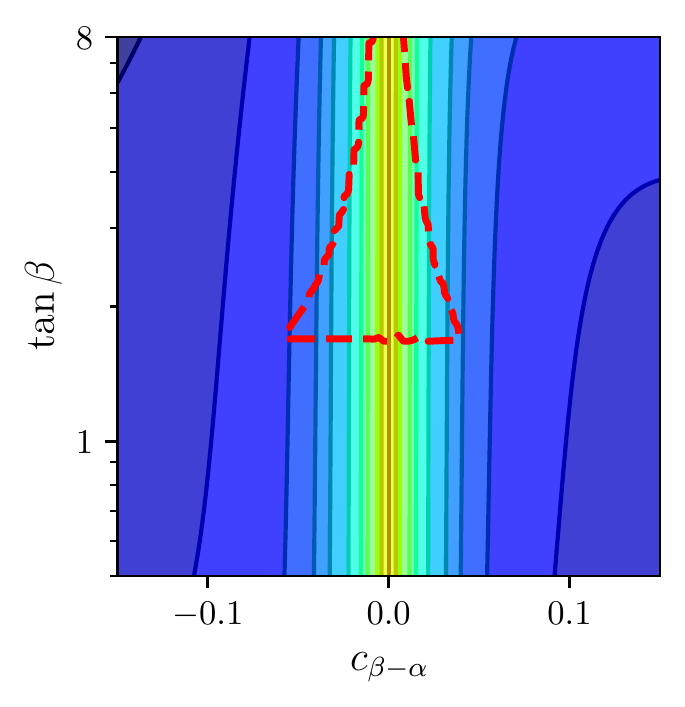}
\includegraphics[width=0.239\textwidth]{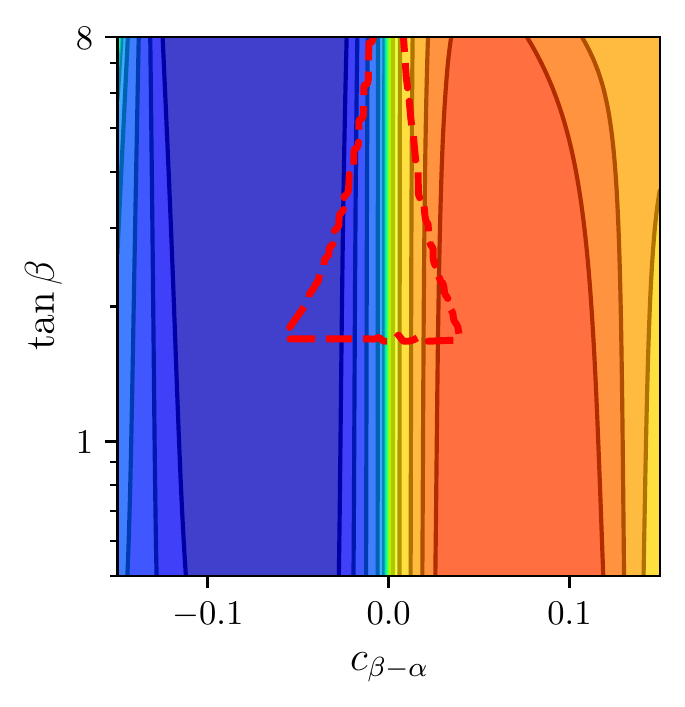}
\includegraphics[width=0.239\textwidth]{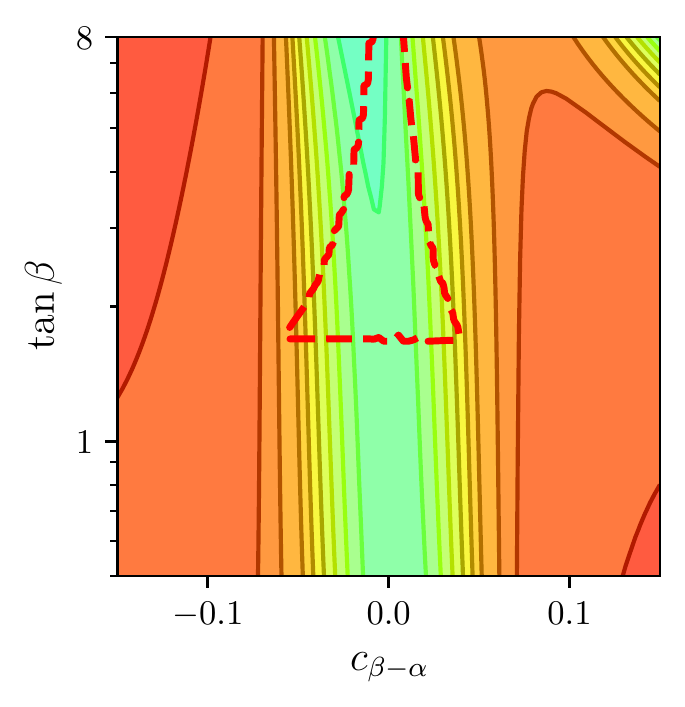}
\includegraphics[width=0.239\textwidth]{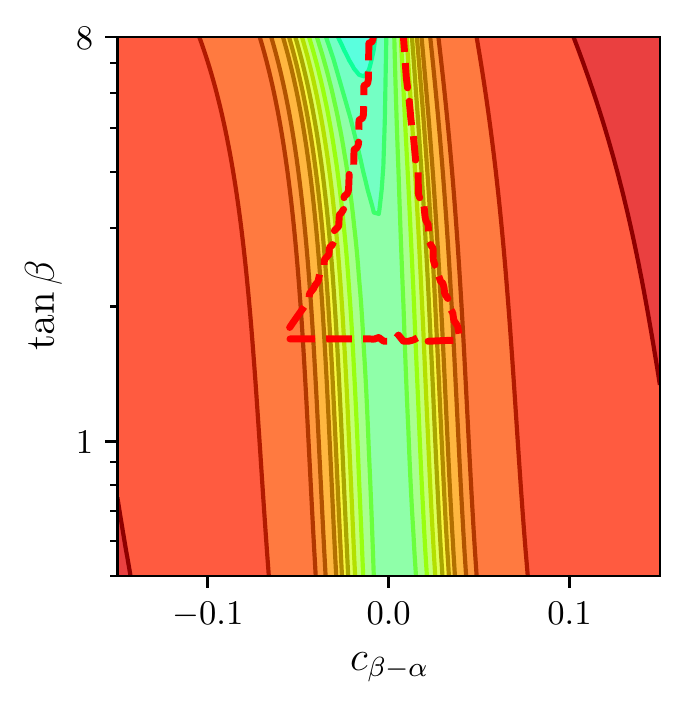}

\includegraphics[width=0.239\textwidth]{h1h1h1_colorbar}
\includegraphics[width=0.239\textwidth]{h1h1h2_colorbar}
\includegraphics[width=0.239\textwidth]{h1h2h2_colorbar}
\includegraphics[width=0.239\textwidth]{h1HpHm_colorbar}
\caption{Allowed areas (dotted regions) from the various constraints (upper row) and triple Higgs couplings (lower row)
  for the benchmark scenario IV-2 in the $\CBA$--$\tb$ plane with
  $m\equiv \MH = \MA = \MHp= 1300 \gev$ and $\msq$ fixed by
  \refeq{eq:m12special2}.}
\label{fig:IV-2}
\end{figure}

\begin{figure}[th!]
\centering
\includegraphics[width=0.24\textwidth]{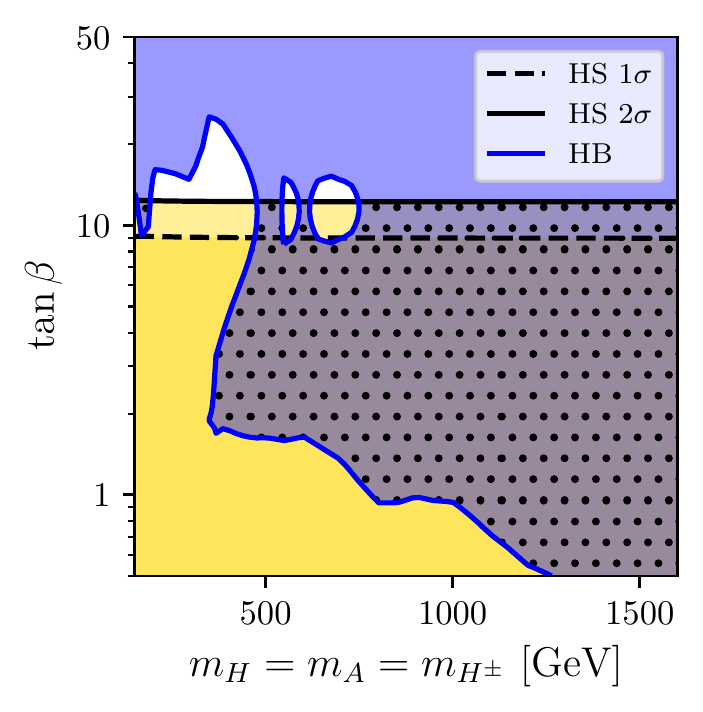}
\includegraphics[width=0.24\textwidth]{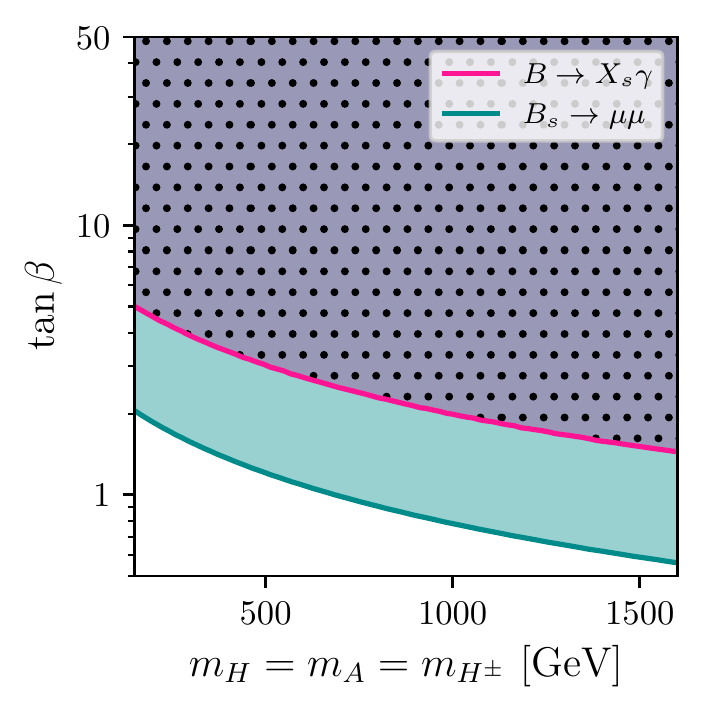}
\includegraphics[width=0.24\textwidth]{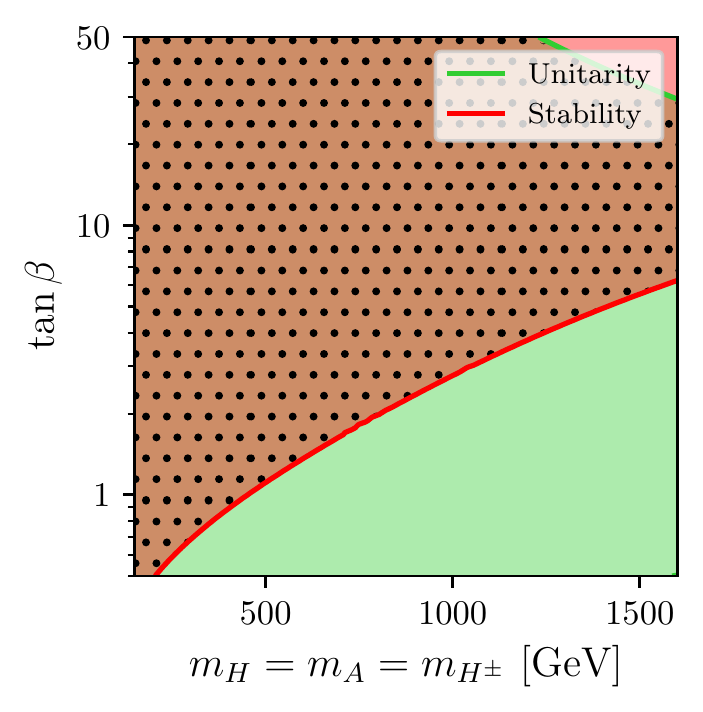}
\includegraphics[width=0.24\textwidth]{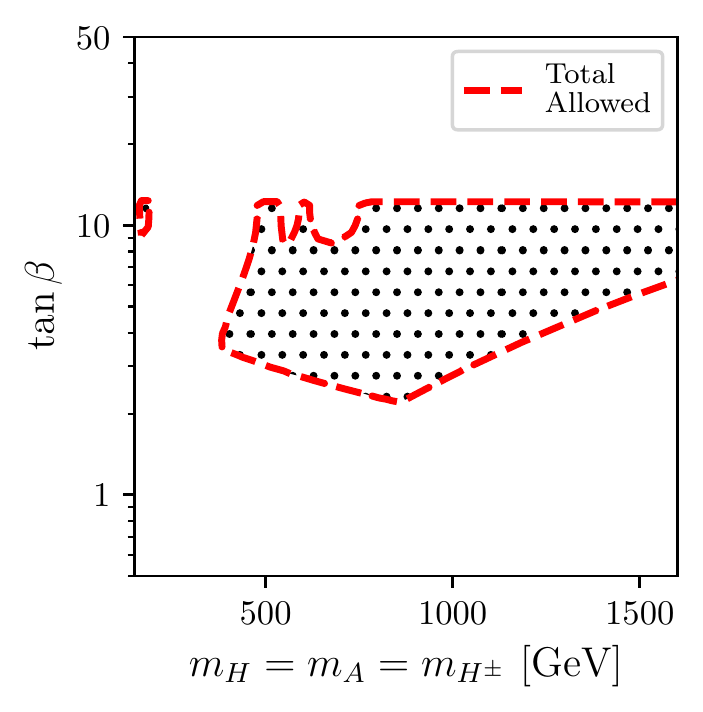}

\includegraphics[width=0.24\textwidth]{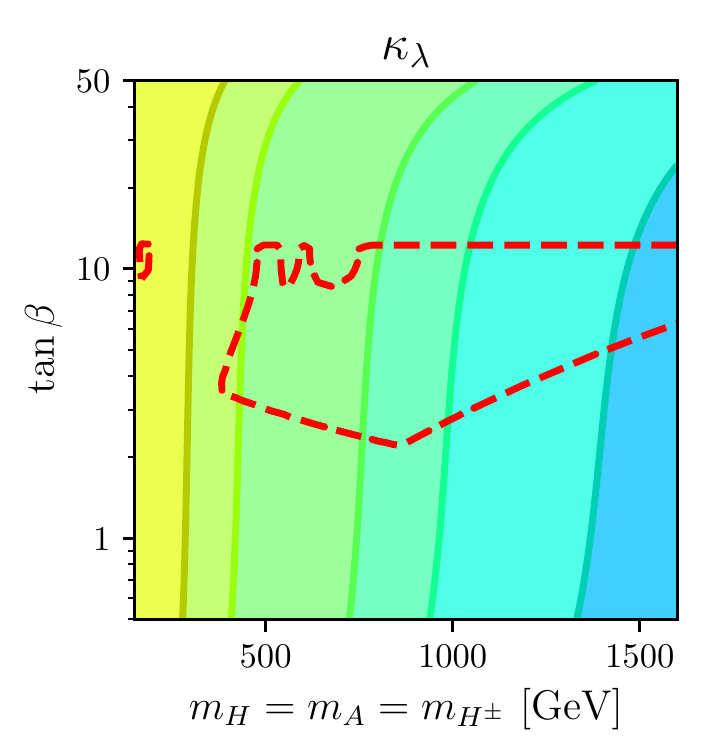}
\includegraphics[width=0.24\textwidth]{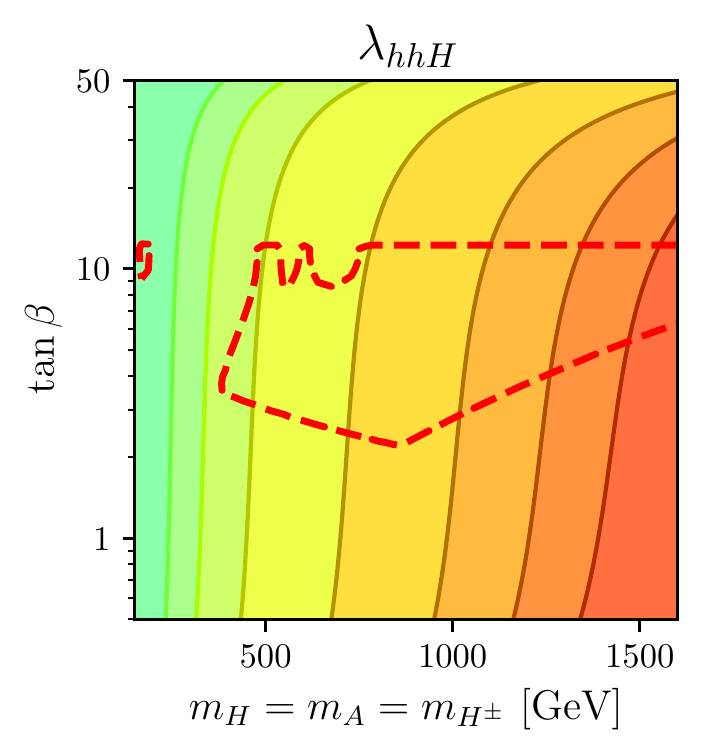}
\includegraphics[width=0.24\textwidth]{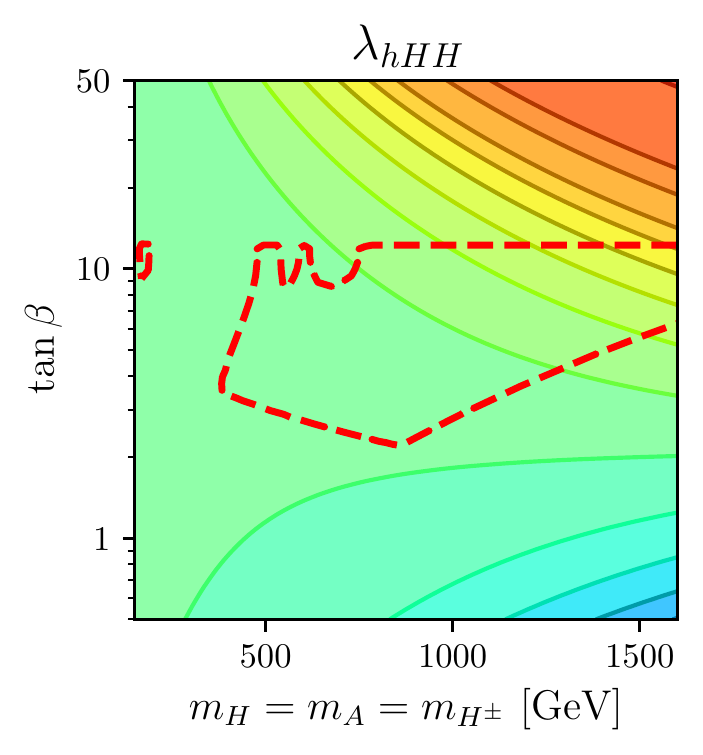}
\includegraphics[width=0.24\textwidth]{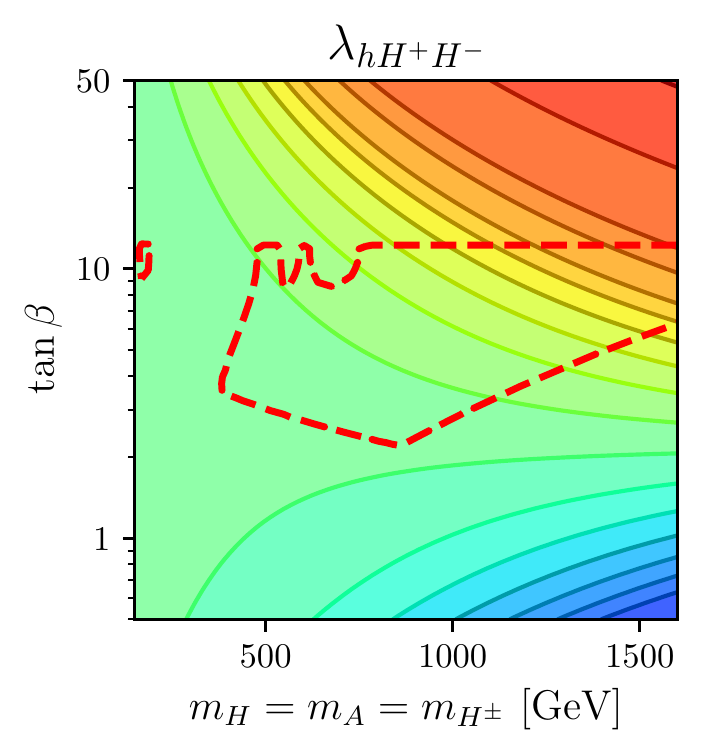}

\includegraphics[width=0.24\textwidth]{h1h1h1_colorbar}
\includegraphics[width=0.24\textwidth]{h1h1h2_colorbar}
\includegraphics[width=0.24\textwidth]{h1h2h2_colorbar}
\includegraphics[width=0.24\textwidth]{h1HpHm_colorbar}
\caption{Allowed areas (dotted regions) from the various constraints (upper row) and triple Higgs couplings (lower row)
  for the benchmark scenario IV-3 in the $m\equiv \MH = \MA =
  \MHp$--$\tb$ plane with 
  $\CBA=0.02$ and $\msq$ fixed by
  \refeq{eq:m12special}.}
\label{fig:IV-3}
\end{figure}

The second scenario, IV-2, is shown in \reffi{fig:IV-2} in the
$\CBA$--$\tb$ plane with 
$m = 1300 \gev$ and $\msq$ fixed by \refeq{eq:m12special2}. $\tb$ is
restricted by $\br(B \to X_s \ga)$ to $\tb \gsim 1.7$. The remaining
parameter space is constrained by unitarity/stability, going up to 
$\tb = 8$, where the scan range ends. $\CBA$ is found in the interval
$\inter{-0.05}{0.04}$, reached for the smallest allowed $\tb$.
$\kala = 1$ is found for $\CBA = 0$ in the alignment limit, going down
to $\kala \sim 0.5$ for the smallest allowed $\CBA$.
$\lahhH \sim \inter{-1.59}{1.26}$ is found going from the
smallest to the largest allowed $\CBA$ values. The other three triple
Higgs couplings take values around $0$ for $\CBA \sim 0$ and large
$\tb$. They reach their largest value of
$\lahHH \sim \lahAA = \lahHpHm/2 \sim 6$ at $\CBA \sim -0.05$.

The third type~IV scenario, IV-3, is shown in \reffi{fig:IV-3} in the
$m$--$\tb$ plane with $\CBA=0.02$ and $\msq$ fixed by
\refeq{eq:m12special}. Upper and lower limits on $\tb$ are given by the Higgs
rate measurements at $\tb = 10$ and by the stability bound,
respectively. Low values of 
$m$ are excluded by the BSM Higgs-boson searches at the LHC, leaving a
range of about $400 \gev$ to $1600 \gev$ (where the scan stopped). 
$\kala$ is found in the interval $\kala \sim \inter{0.88}{1.00}$,
with the smallest (largest) values for large
(small)~$m$, and nearly independent of $\tb$. Also $\lahhH$ is nearly
independent of $\tb$ in the allowed parameter range,
$\lahhH \sim \inter{0.01}{1.2}$, where now the largest
values are found for large~$m$.
$\lahHH \sim \lahAA = \lahHpHm/2$ is found around~1 for lower values
of~$m$ and~$\tb$, reaching the highest values of $\sim 5$ for the
largest allowed~$m$ and $\tb \sim 10$.

\begin{figure}[h]
\centering
\includegraphics[width=0.24\textwidth]{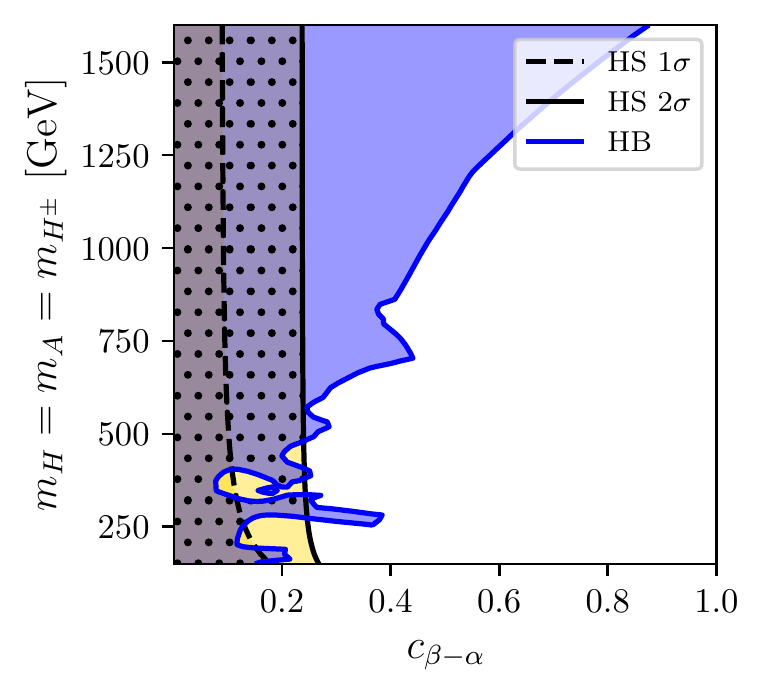}
\includegraphics[width=0.24\textwidth]{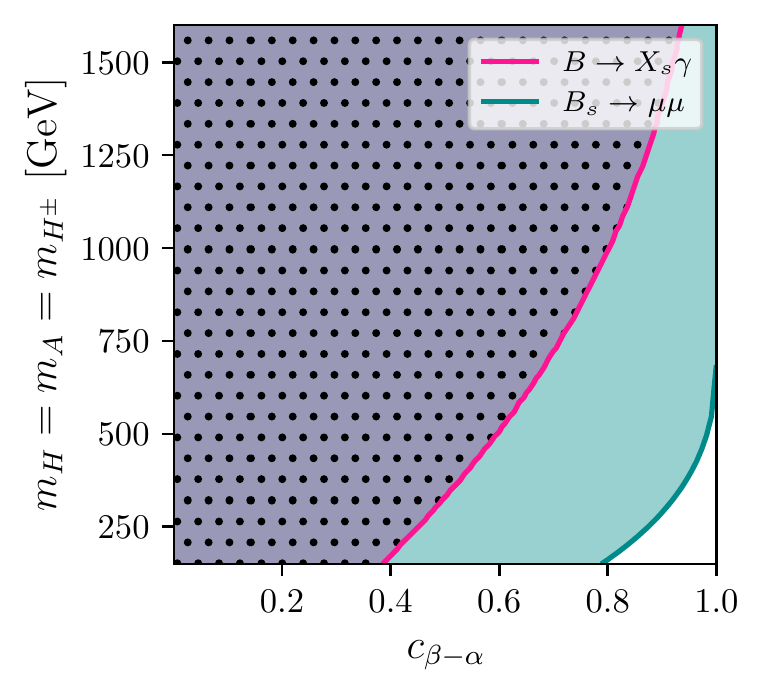}
\includegraphics[width=0.24\textwidth]{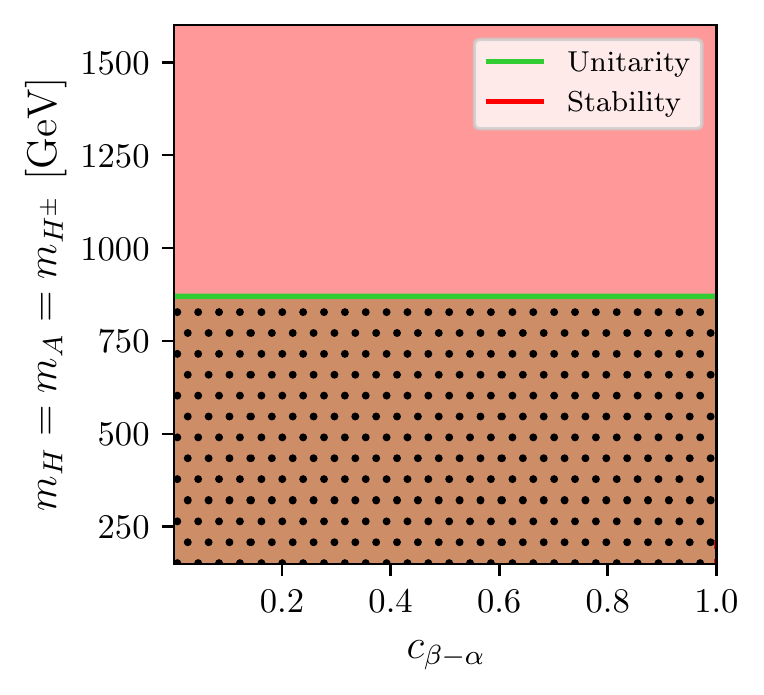}
\includegraphics[width=0.24\textwidth]{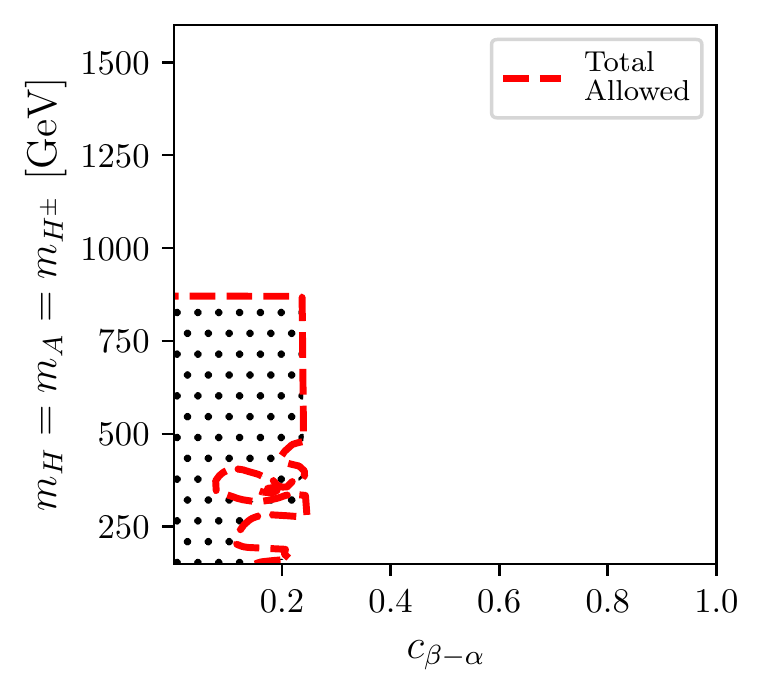}

\includegraphics[width=0.24\textwidth]{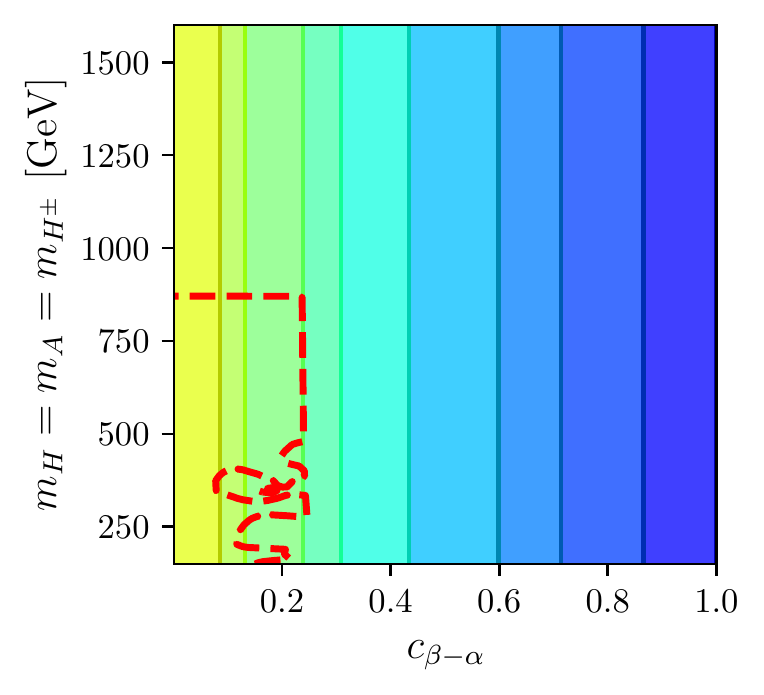}
\includegraphics[width=0.24\textwidth]{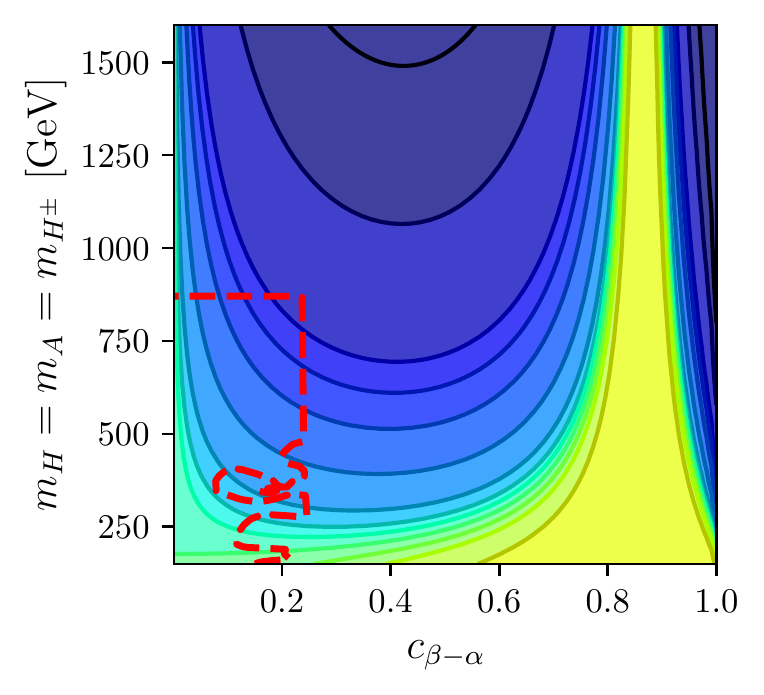}
\includegraphics[width=0.24\textwidth]{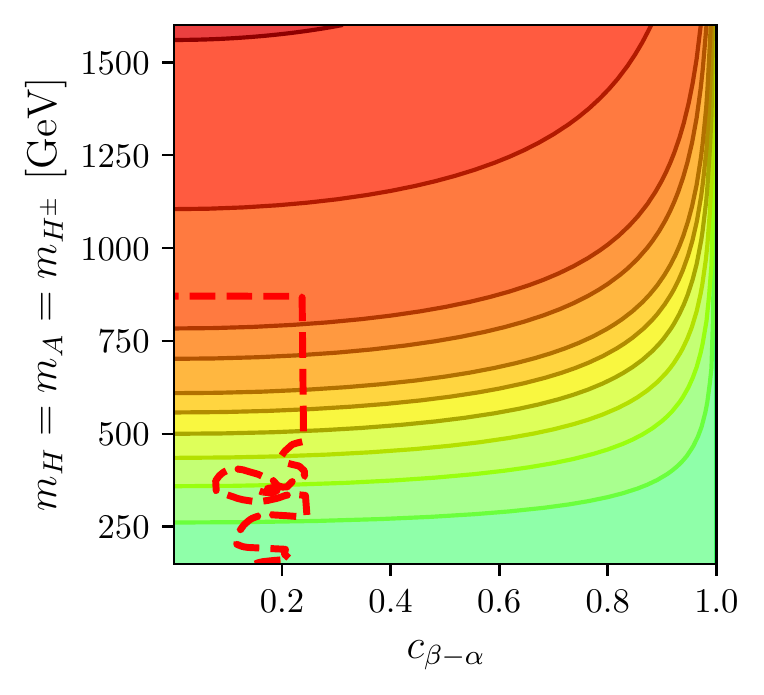}
\includegraphics[width=0.24\textwidth]{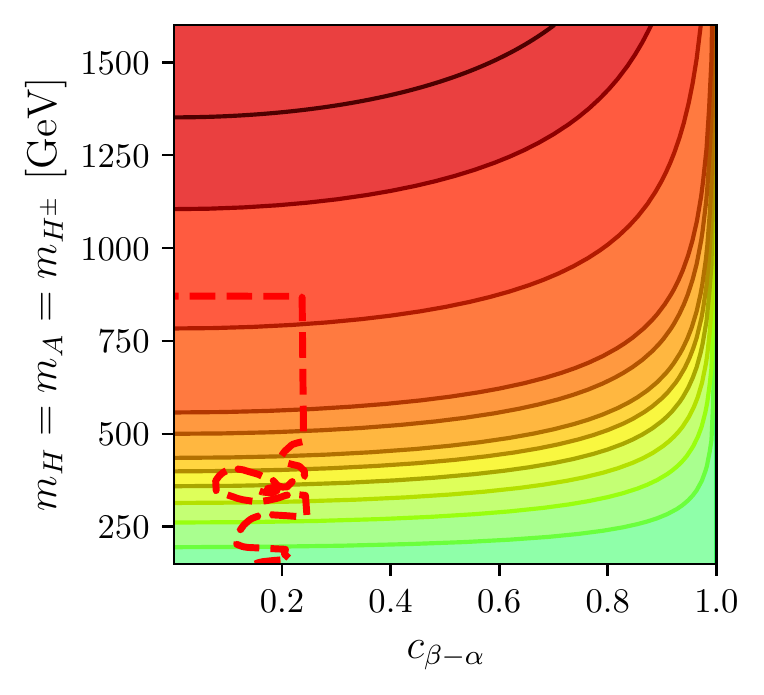}

\includegraphics[width=0.24\textwidth]{h1h1h1_colorbar}
\includegraphics[width=0.24\textwidth]{h1h1h2_colorbar}
\includegraphics[width=0.24\textwidth]{h1h2h2_colorbar}
\includegraphics[width=0.24\textwidth]{h1HpHm_colorbar}
\caption{Allowed areas (dotted regions) from the various constraints (upper row) and triple Higgs couplings (lower row)
  for the benchmark scenario IV-4 in the $\CBA$--$m$ plane with
  $\msq$ fixed by \refeq{eq:m12special} and
  $\tb$ fixed via \refeq{eq:wrongsign} (wrong sign Yukawa limit).}
\label{fig:IV-4}
\end{figure}

The last scenario for Yukawa type~IV, IV-4, is presented in
\reffi{fig:IV-4} in the $\CBA$--$m$ plane with $\msq$ fixed by
\refeq{eq:m12special}, and 
$\tb$ is given by \refeq{eq:wrongsign}, i.e.\ such that the wrong
  sign Yukawa limit is reached.
The main restrictions for low
$m$ are given by the LHC Higgs rate measurements and the BSM Higgs
searches, restricting $\CBA \sim 0.25$. The upper limit on $m$ is given
by the unitarity constraint, yielding $m \lsim 850 \gev$. 
$\kala$ is smaller than~1, but reaching only deviations of
$\kala \sim 0.97$. $\lahhH$ is found in the interval
$\inter{-1.2}{0.01}$ with the smallest values reached at large $\CBA$ and
large $m$. The triple Higgs couplings involving two heavy Higgs bosons
depend mainly on $m$, reaching their largest values of
$\lahHH \sim \lahAA = \lahHpHm/2 \sim 12$ at the highest allowed values
of $m$ in this scenario, $m \sim 850 \gev$.

\newcommand{\mheavy}{m_{\rm heavy}}
\subsection{Complete picture of allowed  triple Higgs  couplings}
\label{sec:intervals}

In order to find the overall allowed ranges of the various triple Higgs
couplings in the four Yukawa types we have performed a parameter
scan. The free parameters were randomly varied in the ranges given in
\refta{tab:scan}.%
\footnote{
  A similar strategy for $\kala$ in type~I and~II was followed in
  \citere{Abouabid:2021yvw}.}%
~Following \citere{Arco:2020ucn}, we here also investigate the
possibility of a non-fully degenerate scenario with $\MA = \MHp$ and
$\MH$ as independent mass parameter 
(scenario~A). For scenario~C, with degenerate Higgs bosons masses,
$\MA = \MH = \MHp$, 10000 valid points were
generated.  For scenario~A,  with $\MA = \MHp$ and $\MH$ as additional free
parameter, 30000 valid points were generated.  From now on, we will
refer generically to the heavy mass $\mheavy$ in this section as the
degenerate mass $m=\MH=\MA=\MHp$ in scenario~C, and to both independent
masses $\MH$ and  $\MA = \MHp$ in scenario~A. 
Naturally, in scenario~A slightly larger intervals for the triple Higgs
couplings are expected. We consider only these two possibilities,
C~and~A, because in the alternative non-fully degenerate scenarios with 
$\MA=\MH$  and $\MHp$ as independent masses,
(named scenario~B in \citere{Arco:2020ucn}), 
sizable contributions to the $T$ parameter 
can appear at two-loop level that may be in conflict with data
\cite{Hessenberger:2016atw}.  
Under these assumptions, we always have $2 \lahAA = \lahHpHm$, and 
in this section we will only refer to $\lahHpHm$.

The final allowed intervals for the various triple Higgs couplings 
are summarized in \refta{tab:intervals}. 
One can see that in all four types, 
$\kala$ and $\lahhH$ can reach their maximum allowed ranges already in 
the fully degenerate scenario (with slightly larger possible
values of $\kala$ in type~I). On the other hand, the couplings
of the light Higgs with two heavy Higgs bosons, 
$\lahHH$, and $\lahHpHm$ can have larger values if some non-degeneracy
between  $\MH$ and $\MA=\MHp$ is allowed (scenario~A).
In the following we discuss the intervals displayed in \refta{tab:intervals}, based on our analyses of the benchmark planes in \refses{sec:typeI} - \ref{sec:typeIV}.

\begin{table}[t!]
\centering
\begin{tabular}{c||c|c|c|c}
 & Type I & Type II & Type III & Type IV\tabularnewline
\hline 
$\mheavy$ & [150, 1600] & [450, 1600] & [450, 1600] & [150, 1600]\tabularnewline
$\tb$ & [1, 50] & [0.5, 50] & [0.5, 50] & [1, 50]\tabularnewline
$\CBA$ & [-0.35, 0.35] & [-0.06, 0.06] & [-0.06, 0.06] & [-0.08, 0.15]\tabularnewline
$\msq$ & [0, $M_{\rm heavy}^{2}$] & [0, $M_{\rm heavy}^{2}$] & [0, $M_{\rm heavy}^{2}$] &
    [0, $M_{\rm heavy}^{2}$]\tabularnewline
\end{tabular}
\caption{Ranges for the input parameters of the 2HDM in our numerical
  scan. $\mheavy$ (given in GeV) refers to
  $m \equiv \MH = \MA = \MHp$ in scenario~C and to both
    independent masses $\MH$ and
  $\MA = \MHp$ in scenario~A.  The maximum value for $\msq$ taken in our
    scans is $M_{\rm heavy}^{2}$,  where $M_{\rm heavy}$  is the largest
    of the heavy Higgs boson masses (coinciding with  $\mheavy$ only in
    scenario~C).}  
\label{tab:scan}
\end{table}

\begin{table}[t!]
\centering
\begin{tabular}{c||cc|cc}
 & \multicolumn{2}{c|}{Type I} & \multicolumn{2}{c}{Type II}\tabularnewline
\cline{2-5} 
& scenario C & scenario A & scenario C & scenario A
\tabularnewline
\hline 
\kala & [-0.48, 1.23] & [-0.48, 1.28] & [0.62, 1.00] & [0.62, 1.00]
\tabularnewline
\lahhH & [-1.69, 1.62] & [-1.69, 1.62] & [-1.80, 1.46] & [-1.80, 1.46]
\tabularnewline
\lahHH & [-0.7, 11.5] &  [-0.7, 14.5] & [-0.2, 12.3] & [-0.5, 16.2]
\tabularnewline
\lahHpHm = 2\lahAA & [-1.8, 22.6] & [-1.8, 32.8]  & [-0.5, 24.6] & [-1.4, 32.7]
\tabularnewline
\end{tabular}

\mbox{}\vspace{1em}

\begin{tabular}{c||cc|cc}
& \multicolumn{2}{c|}{Type III} &
\multicolumn{2}{c}{Type IV$^\dagger$ }\tabularnewline
\cline{2-5} 
& scenario C & scenario A & scenario C & scenario A
\tabularnewline
\hline 
\kala & [0.55, 1.00] & [0.55, 1.00] & [0.53, 1.00] & [0.53, 1.01]
\tabularnewline
\lahhH & [-1.81, 1.34] & [-1.81, 1.34] & [-1.75, 1.36] & [-1.75, 1.36]
\tabularnewline
\lahHH & [-0.3, 12.3] & [-0.2, 15.7]  & [-0.6, 8.6] & [-0.6, 9.2]  
\tabularnewline
\lahHpHm = 2\lahAA & [-0.7, 24.7] & [-1.3, 32.6] & [-1.2, 16.4] & [-1.7, 32.7]
\tabularnewline
\end{tabular}
\caption{Final allowed ranges for the couplings $\la_{hh_ih_j}$
  (for details of the scan, see text).
  ``scenario~C'' refers to the fully degenerate case with
  $m=\MH = \MA = \MHp$, ``scenario~A'' to the non-fully degenerate case
  with $\MA = \MHp$ and $\MH$ being independent mass parameters 
  (see text).
$^\dagger$The ranges of type~IV do not include the wrong sign Yukawa region.
}
\label{tab:intervals}
\end{table}

Focusing first on $\kala$, the 2HDM type~I
is the only type that can accommodate $\kala > 1$,  which can be
understood as follows. In type~I
large values of $\tb$ together with large values of $\CBA$ up to 
$\sim \pm 0.3$ are allowed, as it can be seen in \refses{sec:comparison}
and \ref{sec:thc-anal}. Specifically, those $ \kala > 1$ values can  
be reached in type I when the heavy Higgs boson masses are 
$\mheavy \lesssim 500 \gev$,  $\tb \gtrsim 5$ and
$\CBA \gtrsim 0.2$.
Type I is also found to be the unique one allowing for negative  $\kala$
values.  The minimum allowed value is $\kala \sim -0.5$,  which is
found for $\mheavy \sim 800 \gev$, $\tb \sim 7$ and 
$\CBA$ is at its maximum allowed value around~0.25. 
In these parts of the parameter space of type~I with such large
values for $\tb$, close to 10, $\msq$ has to be close the value
given by \refeq{eq:m12special} to satisfy the theoretical constraints.
In contrast to type~I, in the other three Yukawa types, the lower values
of $\kala$ that can be reached are around~0.5, corresponding to
deviations of around 50\%  below the SM prediction. 
They are found for the largest value of the
heavy Higgs boson masses $\mheavy$ considered in the scan, the lowest allowed 
value for $\tb$ and the largest allowed 
value of  $|\CBA|$,  especially for the case of negative
$\CBA$. 
In these cases, setting $\msq$ close to the value given by
\refeq{eq:m12special2} can help 
to maximize the deviation on $\kala$ from~1 while respecting the
theoretical constraints. 

Regarding the other types, we see that in type~IV the minimum allowed
values of $\tb$ around~1 are larger than in types~II and III,
which are closer to~0.5, due to the $B\to X_s\gamma$ constraint, and the
effect on $\kala$  is expected to be smaller. 
However, this milder effect at low $\tb$ on $\kala$ is compensated by
the fact that type~IV can accommodate larger values of $|\CBA|$ than in
types~II and~III. It is also worth mentioning that the negative deviation from
$\kala = 1$ could be larger with larger heavy Higgs boson masses than
those considered in our scans. 

In the case of $\lahhH$, we find that for all four types the largest
values reached for this coupling are roughly $\sim \pm 1.5$. In all four
types, the minimum (maximum) value is reached for the mass range close
to the maximum scanned value for the heavy Higgs mass
$\mheavy$,   
$\tb \sim 1$ and $\CBA \sim - 0.03$ ($+0.03$).  
In type~I values of $\lahhH \sim 1.5$ can also be reached for
$\tb \sim 10$.  Again, larger values of $\mheavy$
could lead to a larger absolute values for this coupling.

Now we turn to the maximum allowed value for $\lahHH$. In types~I,~II
and~III one can achieve large values up to $\sim 12$ in the fully
degenerate scenario~C and up to $\sim 16$ in scenario~A with non
degenerate masses, $\MH\neq\MA=\MHp$.  
However, the region of the parameter space in which those extreme
values are achieved are different depending on the 2HDM type. 
In type~I with scenario~C, the largest allowed values for $\lahHH$ 
are achieved when all heavy masses are around $1 \tev$ for rather large
values of $\CBA \gtrsim 0.1$ and $\tb \gtrsim 7$, with $\msq$ fixed to
\refeq{eq:m12special}. In scenario~A, 
this coupling can be enhanced for $\MH \sim 1 \tev > \MA = \MHp$.
The situation for types~II and~III is different,  
as they can accommodate extreme values for $\lahHH$ with
$\tb \sim 1$ and being very close to the alignment limit, i.e.\ near
$\CBA \sim 0$, 
for $\mheavy \gtrsim 1 \tev$ in the degenerate scenario and for
$\MH \gtrsim 1 \tev$ and $\MH > \MA = \MHp$ in the non degenerate scenario.
In type~IV,  $\lahHH$ can only acquire values up to $\sim 8$ 
in the fully degenerate scenario~C and up to $\sim 9$ in scenario~A.
These large values of $\lahHH$ close to~10, can only be achieved for
very large values of $\tb > 10$  and being very close to the
alignment limit with $\msq$ set via \refeq{eq:m12special}. 

Turning to the other couplings of the light Higgs to two heavy bosons,
$\lahHpHm = 2\lahAA$, we find that very large values up to $\sim 16$ and
$\sim 32$ are allowed in the four 2HDM types,  
in the fully degenerate and the non-degenerate scenarios, respectively.
In scenario~C with degenerate masses, the maximum allowed values for
these couplings, $\lahHpHm$ and $\lahAA$,  
are found  in the same 2HDM parameter space regions, where we have found
the maximum value for $\lahHH$.  
However,  for the scenario~A the situation is different.
In all four types,  the maximum values are found for
$\MA = \MHp \gtrsim 1 \tev$ and $\MA = \MHp > \MH$, 
for smaller values of $|\CBA|$, close to the alignment limit, and
for values of $\tb \sim 2$.

For the Yukawa type~IV
the wrong sign Yukawa limit is still allowed,
where $\tb$ is given by \refeq{eq:wrongsign}.
In scenario~C
within this particular limit some triple Higgs couplings
can reach larger values than in the above discussed parameter
regions (in which the wrong-sign limit is not reached),
as we have seen in \reffi{fig:IV-4}.
We found that values for $\lahHH$ and $\lahHpHm$ up to $\sim 12$ and
$\sim 24$ are allowed for $\CBA \sim 0.25$ and $\mheavy \sim 800 \gev$. 
We did not consider this limit in scenario~A. 

\smallskip
Finally,  in the last part of this section, we present some concrete
examples of benchmark points within  the 2HDM,  where we find sizeable
effects on the triple Higgs couplings. We have focused both on finding
sizeable departures from $\kala =1$ and on finding large triple
couplings of the light Higgs to the heavy Higgs bosons.  We summarize
our proposed points in  \refta{tab:points}.  We have provided examples
in the four 2HDM-types and, for simplicity, they all have been chosen
within the scenario~C with degenerate heavy masses,  $m = \MH = \MA = \MHp$.
It should be noted,  that type~II and~III are presented together
since they exhibit
practically the same results for the selected benchmark points.   

\begin{table}[t!]
\centering
\begin{tabular}{c|c|c|c|c|c|c|c|c}
Type  & $m$ & $\tan\beta$ & $c_{\beta-\alpha}$ & $m_{12}^{2}$ & $\kala$ &
$\lahhH$ & $\lahHH$ & $\lahHpHm = 2 \lahAA$ \tabularnewline
\hline 
I & 750 & 5.5 & 0.25 & \refeq{eq:m12special} & \textbf{-0.39} & 0.4 & 7 & 12\tabularnewline
 I & 400 & 12 & 0.22 & 12600 & \textbf{1.26} & -0.5 & 3 & 6\tabularnewline
I & 650 & 6 & 0.2 & \refeq{eq:m12special} & 0.13 & 0.5 & 4 & 8\tabularnewline
 I & 1500 & 1.55 & -0.03 & \refeq{eq:m12special2} & 0.62 & \textbf{-1.7} & 7 & 13\tabularnewline
I & 1500 & 2 & -0.025 & 820000  & 0.83 & -1.25 & 3 & 6\tabularnewline
I & 600 & 10 & 0.2 & \refeq{eq:m12special} & 0.99 & -0.5 & 6 & 12\tabularnewline
I & 1000 & 7.5 & 0.2 & \refeq{eq:m12special} & -0.26 & 0.07 & \textbf{13} & \textbf{24}\tabularnewline
\hline 
II/III & 1500 & 1.0 & -0.04 & \refeq{eq:m12special2} & \textbf{0.63} & \textbf{-1.7} & 7 & 14\tabularnewline
II/III & 1000 & 1.2 & -0.035 & 470000 & 0.8 & -0.8 & 3 & 6\tabularnewline
II/III & 1000 & 1.0 & 0.0 & 140000 & 1.0 & 0.0 & \textbf{12} & \textbf{24}\tabularnewline
II/III & 750 & 0.02 & 0.02 & 0 & 0.99 & -0.1 & 9 & 19\tabularnewline
II/III & 550 & 1.8 & 0.01 & 15000 & 0.99 & 0.02 & 5 & 9\tabularnewline
\hline 
IV & 1200 & 2.0 & -0.05 & \refeq{eq:m12special2} & \textbf{0.61} & -1.4 & 4 & 8\tabularnewline
IV & 1200 & 1.8 & -0.055 & \refeq{eq:m12special2} & \textbf{0.55} & -1.4 & 5 & 9\tabularnewline
IV & 1500 & 1.55 & -0.045 & \refeq{eq:m12special2} & \textbf{0.55} & \textbf{-1.8} & \textbf{8} & \textbf{16}\tabularnewline
IV & 700 & 2.5 & 0.09 & \refeq{eq:m12special2} & \textbf{0.65} & 0.7 & 2 & 5\tabularnewline
IV & 400 & 3.8 & 0.06 & 24000 & 0.96 & 1.0 & 1.3 & 2.6\tabularnewline
IV & 550 & 3.0 & 0.045 & 60000 & 0.95 & 0.16 & 2 & 4\tabularnewline
IV & 850 & \refeq{eq:wrongsign} & 0.2 & \refeq{eq:m12special} & 0.97 & -1.05 & \textbf{12} & \textbf{23}\tabularnewline
\end{tabular}
\caption{Example points in the 2HDM types~I II, III and IV that shows
  sizable triple Higgs couplings with at least one 
light $\CP$-even Higgs boson,  still allowed by the actual data.  Bold
values are near the extreme value allowed, shown in
\refta{tab:intervals}.  All points shown are for scenario~C with fully
degenerate heavy Higgs bosons. $m=\MH=\MA=\MHp$ and $\msq$ are expressed
in GeV and GeV$^2$ respectively.} 
\label{tab:points}
\end{table}

As a general remark,  each of the points collected in \refta{tab:points}
exhibits the characteristic phenomenological features
of the particular  type it belongs to,
which have already been described above.  In particular in type~I,
several examples with large triple couplings  of the light Higgs boson
to the heavy Higgs bosons,  or/and large deviations from $\kala=1$ are
shown, with a larger variation in the values of $\tb$,  either
small and close to~1-2, or moderate and close to~10.
This is not the case for the examples found in the other three Yukawa
types, where the largest triple couplings correspond always to a rather small   
value of $\tb \sim 1-2$.

It is interesting to note that
values for $\tb > 10$ are in principle allowed in all four 2HDM
types close to the 
alignment limit, but they do not lead to sizable triple Higgs couplings. 
With such large values for $\tb$, the unitarity and stability conditions
forces $\msq$ to be close to the value given by \refeq{eq:m12AL}.
In the fully degenerate scenario, this would lead to the following
triple Higgs couplings: $\kala=1$, $\lahhH=0$ and 
$\lahHpHm = 2\lahHH= m_h^2/v^2 \simeq 0.26$.
Some BSM boson searches and $B_s\to\mu\mu$ in type II
can pose additional constraints, but heavier Higgs bosons would be
able elude them. 
Regarding the values for  $\CBA$ in this table of points,  they
basically display a variation in the small window allowed,  which is
already quite narrow in the types~II/III.
In type~I the
largest triple couplings appear at the extremes of the allowed interval
$\CBA$, i.e.\  around~0.2. 

The interest of showing these specific
benchmark points is that they can provide interesting scenarios to study
at the future colliders.  In particular, these scenarios could lead to a
remarkable  BSM phenomenology  in the production of two Higgs bosons,
since the triple couplings are involved in a relevant way in those
processes.  The importance of the triple Higgs couplings in the
production of the various (neutral) di-Higgs channels,  $hh$, $hH$, $HH$
and $AA$  have already been studied for the types I and II and for the
future $e^+e^-$  linear colliders in
\citeres{Arco:2021bvf,Arco:2021ecv,Arco:2021zhb}, with
encouraging results.
We leave an extension of these collider studies to the complete picture
of the four 2HDM Yukawa types for future work.


\section{Conclusions}
\label{sec:conclusions}

The measurement of the triple Higgs coupling $\lahhh$ is one of the
important tasks at current and future colliders. Depending on its size
relative to the corresponding SM value, higher (or lower) accuracies can
be expected at certain collider options. Going beyond $\lahhh$, 
large values of triple Higgs couplings involving BSM Higgs
bosons (i.e.\ Higgs bosons in addition to the one at $\sim 125 \gev$)
can play an important role in the di-Higgs production cross sections
at the (HL-)LHC and future $e^+e^-$ colliders. 

In this paper we have investigated triple Higgs couplings in the 
Two Higgs Doublet Models (2HDM), treating equally all four Yukawa types,
focusing on couplings involving at least one light, SM-like Higgs boson.
This is an extension of a previous work~\cite{Arco:2020ucn},
where we focused on the Yukawa types~I and~II. 
We analyze the allowed parameter ranges in the four Yukawa types,
taking into account all relevant theoretical and experimental
constraints. These comprise from the theory side 
unitarity and stability conditions. From the
experimental side we require agreement with measurements of the SM-like
Higgs-boson rates as measured at the LHC, as well as with 
the direct BSM Higgs-boson searches.
Furthermore, we require agreement with flavor observables and
the $T$~parameter, representing the most relevant electroweak precision
observable.
Particularly for type~II we find important differences w.r.t.\ our
previous analysis~\cite{Arco:2020ucn} due to updates in the experimental
LHC constraints, whereas type~I is much less affected. 

It is interesting to note that
for the unitarity/stability constraints $\msq$ plays an important role:
lower (higher) values are favored by the tree-level stability (unitarity)
constraint, where $\msq$ controls the size of the intersection region.
In order to enlarge the allowed parameter region by these constraints
we have employed on several occasions \refeqs{eq:m12special} and
(\ref{eq:m12special2}). 
Concerning the Higgs-boson rate measurements at the LHC, $\msq$ enters
particularly in $\lahHpHm$, and thus in the prediction of
$\Ga(h \to \ga\ga)$. Similarly, but less pronounced, it enters via
$\lahHpHm$ and $\la_{HH^+H^-}$
in the 2HDM prediction for $B_s\to\mu^+\mu^-$
via the $h$ and $H$ 
Higgs penguins contributions with charged Higgs bosons in the loops.

In a first step of our phenomenological analysis we analyze the four
2HDM in three benchmark planes, chosen identical for the four Yukawa
types (and with $\MH = \MA = \MHp$).
This allows us to directly compare the four types to each
other. Overall we find broadly that type~I and type~IV resemble each
other taking all constraints into account, where the allowed parameter
range for type~I is usually somewhat larger than for
type~IV. Conversely, also type~II and~III resemble each other without larger
differences in the allowed parameter ranges. These two types are in
general more restricted at larger values of $\tb$ due to the Higgs-boson
rate measurements and the BSM Higgs-boson searches at the LHC. 
On the other hand, flavor observables in general lead to stronger
restrictions in type~I and~IV at low $\tb$. The parameter associated to
the alignment limit (in which $h$ becomes SM-like), $\CBA$ has larger
allowed ranges particularly in type~I, and somewhat less in type~IV.
These general differences have a clear impact on the allowed sizes of
the various triple Higgs couplings (see below).

In the second step of our analysis we define four benchmark planes
individually for each of the four Yukawa types (and again with
$\MH = \MA = \MHp$), exemplifying where
$\lahhh$ shows larger deviations from $\laSM$, or where larger values of
the other triple Higgs couplings are found. Since type~II and~III show a
very similar phenomenology, we choose the same planes for these two
types. Within these benchmark planes we mark the regions allowed by all
theoretical and experimental constraints. In this way these planes can
be readily used for further phenomenological analyses. As a relevant
example we display the triple Higgs couplings involving at least one
light Higgs in these planes.

In a third step we determine the overall allowed ranges for the various
triple Higgs couplings in the four Yukawa types. These ranges reflect
the overall differences found in the first step of our analysis, see
above. The ranges were determined in a parameter scan, where besides the
``scenario~C'' with $\MH = \MA = \MHp$ we also investigated the case of
``scenario~A'' with $\MH \neq \MA = \MHp$ (which naturally results in
slightly larger 
allowed ranges). Concerning $\kala := \lahhh/\laSM$, in types~II, III
and~IV allowed intervals of $\kala \sim \inter{0.5}{1}$ are found. 
Only in type~I
values below $\sim 0.5$ and above $\sim 1$ are allowed with the overall
interval of $\kala \sim \inter{-0.48}{1.28}$.
The allowed intervals of $\lahhH$ are again similar for types~II, III
and~IV with $\lahhH \sim \inter{-1.8}{1.4}$, whereas for type~I one
finds $\lahhH \sim \inter{-1.7}{1.6}$. 
Concerning the triple Higgs couplings involving two heavy Higgs bosons,
the upper and the lower limits roughly follow
$\lahHH \sim \lahAA \sim \lahHpHm/2$ in agreement with the symmetry
factor in \refeq{eq:lambda}. We roughly find lower allowed limits of
$\lahHH \sim \lahAA \sim -0.8 (-0.4)$ in types~I, II, IV (type III).
For the upper limits,  we find in scenario~C values up to $\lahHH \sim \lahAA \sim  \lahHpHm/2 \sim 12-13$ in all Yukawa types. 
Substantially larger values are found in scenario~A as compared to
scenario~C in all four Yukawa types.  For $\MH \neq \MA = \MHp$ the upper
allowed values in the explored mass range are found at
$\lahHH \sim \lahAA \sim \lahHpHm/2 \sim 16$. 
However, it should be kept in mind that an analysis 
allowing for heavier BSM Higgs bosons could possibly
lead to even larger values for the triple Higgs couplings.

These triple Higgs couplings can have a very strong impact on the heavy
di-Higgs production at $pp$ and $e^+e^-$
colliders~\cite{Arco:2021bvf,Arco:2021ecv,Arco:2021zhb}. 
As was discussed in these references, large coupling values can
possibly facilitate the discovery of heavier 2HDM Higgs bosons.  
However, here it must be kept in mind that the larger values of triple
Higgs couplings involving two heavy Higgs bosons are always realized for
larger values of the respective heavy Higgs-boson mass. 
Therefore, the effects of the large coupling and the heavy mass
always go in opposite directions. 

To facilitate more detailed analyses, see e.g.\ \citere{Arco:2021bvf},
we provide a list of benchmark points that exemplify large deviations
from unity in $\kala$ or large (positive or negative) values of the
other triple Higgs couplings, while being in agreement with the
experimental and theoretical constraints. The benchmark points are given
for the choice $m \equiv \MH = \MA = \MHp$, and they are identical for
Yukawa type~II and~III, reflecting the similarity of these two types.
In order to represent the broad phenomenology that the Higgs-boson
sector of the 2HDM offers, they vary substantially in their choice of
$m$, $\tb$, $\CBA$ and how $\msq$ is determined. We leave a more
detailed analysis of their phenomenology at the LHC and future $e^+e^-$
colliders for future work.

  
\subsection*{Acknowledgements}

\begingroup 
S.H.\ thanks K.~Radchenko for helpful discussions.
The present work has received financial support from the `Spanish Agencia 
Estatal de Investigaci\'on'' (AEI) and the EU
``Fondo Europeo de Desarrollo Regional'' (FEDER) through the project
FPA2016-78022-P and
by the grant IFT Centro de Excelencia Severo Ochoa CEX2020-001007-S
  funded by MCIN/AEI/10.13039/501100011033.
The work of F.A.\ and S.H.\ was also supported in part by the
grant PID2019-110058GB-C21 funded by
MCIN/AEI/10.13039/501100011033 and by "ERDF A way of making Europe".
F.A.\ and M.J.H.\  also acknowledge financial support from the Spanish
``Agencia Estatal de Investigaci\'on'' (AEI) and the EU ``Fondo Europeo de
Desarrollo Regional'' (FEDER) 
through the project PID2019-108892RB-I00/AEI/10.13039/501100011033
and from the European Union's Horizon 2020 research and innovation
programme under the Marie Sklodowska-Curie grant agreement No 674896 and
No 860881-HIDDeN.
The work of F.A.\ was also supported by the Spanish Ministry of Science
and Innovation via an FPU grant with code FPU18/06634. 
\endgroup





\end{document}